\documentclass[a4paper,12pt]{article}
\pdfoutput=1
\usepackage{jheppub} 
\usepackage{amsmath,amssymb}
\usepackage{graphicx}
\usepackage{float}
\usepackage[export]{adjustbox}
\usepackage[utf8]{inputenc}
\usepackage{slashed}
\usepackage{bbold}
\usepackage{tikz}
\usetikzlibrary{arrows,shapes}
\usepackage{subfig}
\usepackage[utf8]{inputenc}

\graphicspath{{figures/}}

\def\bea{\begin{eqnarray}}
\def\eea{\end{eqnarray}}
\def\typeI{type S}
\def\typeII{type V}
\def\TypeI{Type S}

\newcommand{\eqs}[2][0.3]{\includegraphics[width=#1\linewidth, valign=c]{#2}}
\newcommand{\takelimit}{{\cal R}}

\newcommand{\e}{\epsilon}

\title{Locally finite two-loop amplitudes for off-shell multi-photon production in electron-positron annihilation}

\author[a]{Charalampos Anastasiou,}
\author[a]{Rayan Haindl,}
\author[b]{George Sterman,}
\author[a]{Zhou Yang,}
\author[a]{Mao Zeng}

\affiliation[a]{Institute for Theoretical Physics, ETH Zurich, 8093 Z\"urich, Switzerland}
\affiliation[b]{C.N.\ Yang Institute for Theoretical Physics and Department of Physics and Astronomy, Stony Brook University, Stony Brook, NY 11794,
USA}
\emailAdd{babis@phys.ethz.ch}
\emailAdd{haindlr@phys.ethz.ch}
\emailAdd{george.sterman@stonybrook.edu}
\emailAdd{yangzho@student.ethz.ch}
\emailAdd{mzeng@phys.ethz.ch}

\abstract{ 
We study the singularity structure of two-loop QED amplitudes for the production of multiple off-shell photons 
in massless electron-positron annihilation and develop counterterms that remove their
infrared and ultraviolet divergences point by point in the loop integrand.
The remainders of the subtraction are integrable in four
dimensions and can be computed in the future with numerical
integration.  The counterterms capture the divergences of the
amplitudes and factorize in terms of the Born
amplitude and the finite remainder of the one-loop amplitude. They
consist of simple one- and two-loop integrals with at most three
external momenta and can be integrated analytically in a simple
manner with established methods.
We uncover novel aspects of fully local IR factorization,
where vertex and self energy subdiagrams must be modified by 
new symmetrizations over loop momenta, in order to expose their tree-like tensor structures and hence
factorization of IR singularities prior to loop integration.
This work is a first step towards isolating locally the hard
contributions of generic gauge theory
amplitudes and rendering them integrable in exactly four dimensions with numerical methods.
}


\begin{document}
\preprint{YITP-SB-2020-25}
\maketitle 
%\flushbottom
%\newpage
\section{Introduction}
\label{sec:introduction}
Perturbative Quantum Chromodynamics (QCD) has enjoyed rapid progress in the past two
decades. Most remarkably, we have witnessed an ``next-to-leading order (NLO) revolution''~\cite{hep-ph/0609007,arXiv:0708.2398,arXiv:0801.2237, arXiv:0803.4180, Ellis:2011cr,Bredenstein:2010rs}:
all important  hard-scattering LHC processes can be
computed, in an automated way, at next-to-leading-order in the strong-coupling expansion \cite{Hirschi:2011pa, Cascioli:2011va}.
The automation of NLO computations has marked the culmination of years of challenging
research which revealed elegant structures in gauge-theory amplitudes, loop
integration and phase-space integration over real partonic
radiation. Impressive progress on the same fronts (amplitudes, loop
and phase-space integration) has been achieved beyond NLO.
Many $2 \to 2$ processes are by now computable at
next-to-next-to-leading-order (NNLO) \cite{Czakon:2013goa,Cascioli:2014yka,Currie:2013dwa,Czakon:2019tmo,Grazzini:2017mhc,Boughezal:2015ded,DelDuca:2016ily}, while hadron collider processes with simpler kinematics can be
calculated at N$^3$LO~\cite{Anastasiou:2015ema,Duhr:2019kwi,Duhr:2020seh,Cieri:2018oms,Dulat:2018bfe,Dreyer:2018qbw,Dreyer:2016oyx}. 
A first cross-section for a $2 \to 3$ process has been computed at NNLO in the approximation of a large number of colors~\cite{Chawdhry:2019bji}. Here we shall explore a factorization-based approach to that may be useful in the computation of multiloop amplitudes with multiple external lines, and give a first application at two loops in massless quantum electrodynamics.

Currently, theoretical simulations and experimental measurements are in most cases 
of  comparable accuracy.  However, it is expected that with the collection of 15
times more data by the end of the LHC physics program, in the next two
decades, and further advances in data analysis methods, the precision
of many measurements  will surpass the accuracy of theoretical
predictions~\cite{Cepeda:2019klc}.  Theory will therefore be the limiting factor in the
precise extraction of fundamental parameters, in testing the Standard
Model of particle physics and in revealing potential signals of new
phenomena.  It will become important in the future to perform the many more NNLO and N$^3$LO cross-section computations necessary to match the projected experimental uncertainties.

We believe that the biggest stumbling block
towards this goal is the computation of new two-loop and three-loop amplitudes. 
This may be surprising given that we have seen an enormous progress in the development of
methods for multi-loop amplitude computations in recent years. On the analytic front,
powerful algebraic Feynman diagram simplifications with reductions to ``master integrals'' have been
devised~\cite{Laporta:2001dd,Anastasiou:2004vj,Smirnov:2008iw,Smirnov:2014hma,Studerus:2009ye,vonManteuffel:2012np,Lee:2013mka,Maierhoefer:2017hyi,Smirnov:2019qkx,Peraro:2019svx,Frellesvig:2019uqt,Larsen:2015ped,Boehm:2017wjc,Boehm:2018fpv,Kosower:2018obg,Peraro:2016wsq,Mastrolia:2016dhn,CaronHuot:2012ab,Ita:2015tya,Mastrolia:2018uzb, Frellesvig:2019kgj, Mizera:2020wdt, Frellesvig:2020qot}. The master integrals themselves have been objects of an
intense field of research and their evaluation is nowadays made with
powerful mathematical methods which exploit newly discovered 
algebraic properties of polylogarithmic  functions and iterated
integrals~\cite{Remiddi:1999ew,Vollinga:2004sn,Goncharov:2010jf,Duhr:2011zq,Duhr:2012fh,Duhr:2019tlz,Broedel:2017kkb,Ablinger:2018sat,Duhr:2019rrs}. 
On the numerical side, methods such as 
solving master integral differential equations numerically~\cite{Czakon:2013goa,Czakon:2020vql,Abreu:2020jxa,Casconi:2019xof,Francesco:2019yqt,Frellesvig:2019byn}, sector decomposition~\cite{Hepp:1966eg,Roth:1996pd,Binoth:2000ps,Anastasiou:2005pn,Anastasiou:2008rm,Lazopoulos:2007ix,Smirnov:2008py,Borowka:2017idc,Carter:2010hi,Borowka:2016ehy,Borowka:2016ypz},
Mellin-Barnes integration~\cite{Smirnov:1999gc,Tausk:1999vh,Anastasiou:2005cb,Czakon:2005rk,Smirnov:2009up,Gluza:2016fwh,Dubovyk:2018rlg,Dubovyk:2019krd}, and direct parametric integration \cite{Panzer:2014caa} using (quasi-) finite bases \cite{vonManteuffel:2014qoa} have proven to be very powerful. 

In their current form, however, the aforementioned methods are seriously challenged by 
the core of multi-loop amplitude computations needed for the purposes of the future precision 
physics program at the LHC.  Their computational cost 
increases fast  with complexity, measured either by the number of loops or the number of
scattered particles.  Next generation problems, such as three-loop
amplitudes for $2 \to 2$ processes~\footnote{In supersymmetric gauge
  theories, scattering amplitudes  of even higher external state
  multiplicities and at four loops have been computed
  recently, e.g.\ in \cite{dixon:2020cnr}.}
beyond the massless case (for which we note recent spectacular breakthroughs~\cite{Henn:2020lye,Henn:2016jdu,Jin:2019nya}) 
and two-loop amplitudes for $2 \to 3$ processes with two or more mass parameters, appear to have a prohibitive algebraic complexity. Beyond the algebraic obstructions, on the analytic front, the ``basis'' of special functions (elliptic polylogarithms, etc.) which emerge routinely in such computations is not fully known.  There is a realistic chance that the computational complexity cannot be tamed by following these well established and, up to now, very successful paths. New approaches must also be pursued.
In this spirit, new ideas aiming at fundamental aspects of
perturbative computations are emerging~\cite{Catani:2008xa,Bierenbaum:2010cy,Hernandez-Pinto:2015ysa,Buchta:2014dfa,Driencourt-Mangin:2019aix,Baumeister:2019rmh,Buchta:2015wna,Runkel:2019zbm,Aguilera-Verdugo:2019kbz,Runkel:2019yrs,Capatti:2019ypt,Page:2018ljf,Pittau:2012zd}.

We believe that numerical integration directly  in momentum space
can provide a powerful solution to the computational complexity of multi-loop
amplitudes.   In momentum space, Feynman diagrams take the form of
simple rational functions before loop integration.  The evaluation of the multi-loop amplitude
integrands is inexpensive, especially when modern recurrence and other generation methods \cite{Hirschi:2011pa,Buccioni:2017yxi,Cascioli:2011va,Berends:1987me,Kosower:1989xy,Caravaglios:1995cd,Britto:2004ap} are employed.  
The idea of integrating scattering amplitudes directly in
momentum-space is appealing because at a loop order $N_L$ the number
of integrations ($ \leq 4 N_L$)  is independent of the number of
scattered particles. This is very helpful in the effort to automate the
evaluation of amplitudes for generic processes, since the number of
integrations in other  approaches depends strongly on the number of external
particles. This is, for example, one main reason that while analytic
approaches have been very successful in two-loop $2 \to 2$ amplitudes,
the computation of $2 \to 3$ two-loop amplitudes is much more
complicated~\cite{Badger:2017jhb, Abreu:2017hqn, Badger:2018enw, Abreu:2018zmy,
Abreu:2019odu, Hartanto:2019uvl, Abreu:2018aqd, Chicherin:2018yne,
Chicherin:2019xeg, Abreu:2019rpt, Badger:2019djh, Dunbar:2019fcq, Chawdhry:2019bji}. 

However, it is also true that direct momentum-space integration cannot be applied in a straightforward manner either.  
Feynman amplitudes are divergent in $d=4$ space-time dimensions due to
ultraviolet (UV) and infrared (IR) singularities. While UV singularities are canceled by renormalization, IR
singularities cancel when all components of physical cross-sections
(including phase-space integration over real radiation and
convolution with parton densities/fragmentation functions) are
assembled.  In order to make this cancellation manifest and to obtain
convergent representations of the physical remainder, the singularities
need to be explicitly extracted from the integrand of multi-loop
amplitudes.  Similar extractions are of course required for phase-space
integrands~\cite{GehrmannDeRidder:2008ug,Weinzierl:2008iv,Currie:2016bfm,Czakon:2010td,Boughezal:2015dra,DelDuca:2016csb,Somogyi:2006cz,Somogyi:2006da,Somogyi:2009ri,DelDuca:2016ily,Catani:2007vq,Grazzini:2017mhc,Cieri:2018oms,Boughezal:2016wmq,Boughezal:2019ggi,Billis:2019vxg,Moult:2016fqy,Gaunt:2015pea,Cacciari:2015jma,Dreyer:2016oyx,Currie:2018fgr,Caola:2017dug,Herzog:2018ily,Magnea:2018ebr,Magnea:2018hab,Behring:2019quf,Melnikov:2019pdm,Melnikov:2018jxb,Ebert:2020unb,Ebert:2020yqt,Catani:2019nqv,DelDuca:2019ggv,DelDuca:2020vst}.

Rendering amplitudes integrable with removal of IR and UV
divergences followed by appropriate contour deformations 
has been pursued at one loop,  in foundational works of Refs.~\cite{Nagy:2006xy,Soper:1999xk,Gong:2008ww,Becker:2010ng,Assadsolimani:2009cz,Becker:2012aqa,Becker:2011vg, Becker:2012bi}. 
At higher loop orders, the structure of IR and UV divergences is severely entangled and designing
subtractions which remove simultaneously all singularities is
challenging.  In a recent work~\cite{Anastasiou:2018rib}, it was shown
that such singularities can be successfully removed in practice at higher loops by an amplitude-level ``nested subtraction'' approach~\cite{Erdogan:2014gha,Ma:2019hjq,Collins:2011zzd,Zimmermann:1969jj}. 
There has also been related work on expressing amplitude integrands in a basis of integrals which makes the divergence structures manifest, known as the ``local numerator'' approach. This method has been applied  in $\mathcal N=4$ super-Yang-Mills theory \cite{ArkaniHamed:2010kv, ArkaniHamed:2010gh, Bourjaily:2015jna, Bourjaily:2019iqr, Bourjaily:2019gqu} and to less supersymmetric theories, including QCD \cite{Kalin:2018thp, Kalin:2019vjc, Badger:2016ozq, Hartanto:2019uvl}.

In this article, we demonstrate for the first time that two-loop amplitudes for the 
processes $e^-+ e^+ \to \gamma^* \mbox{'s}$ can be freed locally, i.e., at the integrand level,
from all infrared and ultraviolet divergences by a limited number of counterterm subtractions.  The process of annihilation into off-shell photons is chosen for its relative simplicity, and because its IR structure is closely related to annihilation into massive electroweak bosons.
The counterterms remove all types
of soft, collinear and ultraviolet singularities. Infrared counterterms are 
summed into simple universal factors which are easy to integrate analytically, regardless of the number of final-state photons considered. 
Our approach is inspired by the picture of infrared factorization
~\cite{Akhoury:1978vq,Sen:1982bt,Korchemsky:1994jb,Catani:1998bh,Sterman:2002qn,Bauer:2000yr,Aybat:2006wq,Aybat:2006mz,Ferroglia:2009ii,Becher:2019avh,Collins:1989bt,Sterman:1995fz,Dixon:2008gr,Gardi:2009qi,Gardi:2009zv,Becher:2009cu,Becher:2019avh,Feige:2014wja,Collins:2011zzd,Erdogan:2014gha,Ma:2019hjq}
which is established for amplitudes after integration of the loop momenta.
Unlike Ref.~\cite{Anastasiou:2018rib}, we will not attempt to construct counterterms for every individual diagram,
but will construct counterterms for the sum of all diagrams, using a suitable choice of loop momentum routings
to combine all diagrams into a single integrand. The benefit of summing over diagrams is that IR singularities
become factorized, that is, proportional to lower-loop amplitudes, and the needed counterterms are very simple even
when the number of external legs is very large.

The origin of gauge theory infrared divergences in terms of factorized and universal jet and soft functions has been 
demonstrated by a number of methods \cite{Feige:2014wja,Ma:2019hjq}.   In this paper,
we use a consequence of this factorization:  processes with the same set of jet functions and soft infrared functions
can differ only in their short-distance, ``hard" functions.   In this way, a known process can be used as a guide to isolate the hard functions
in a more complex process.   We shall use below the color-singlet annihilation form factor at two loops (here, in QED) as the known function,
which we shall use to determine the much more complex annihilation cross section for $n$ off-shell photons.
The construction of counterterms below will follow the procedures implemented in proofs of factorization, and will 
rely on Ward identities that decouple unphysically polarized vectors from physical processes.   

The efficacy of such a subtraction method depends on the logarithmic nature of infrared divergences in gauge theories.   The
power counting procedure that gives this important result has been known for a long time, and has been described in detail in the literature~\cite{Sterman:1978bi,Libby:1978qf,Collins:1989gx}.   
Key steps in these arguments depend on essential features of self energy and three-point vertex subdiagrams
in massless renormalized perturbation theory.  First, a renormalized fermion self energy diagram takes
the form $\slashed{p}\, \pi_F(p^2/\mu^2)$, with $\mu$ the renormalization scale. This is the inverse propagator times a function, $\pi_F$, of
the invariant mass that is no more than logarithmically divergent.
In the same way, the renormalized vector self energy takes an analogous form, with 
a potentially logarithmically divergent function, $\pi_V$, times the unique transverse tensor, $(g^{\mu\nu}q^2 - q^\mu q^\nu) \pi_V(q^2/\mu^2)$.
Finally, after integration and renormalization, a three-point subdiagram can depend only on its external momenta, and must 
 obey the relevant Ward identities of the theory.

As discussed below, the tensor properties of two- and three-point functions that we have just described are a challenge to any local subtraction
method, because they are only
guaranteed to hold after the internal loop integrals have been carried out.   At fixed values of internal
loop momenta, self energy subdiagrams can produce power divergences in infrared limits.   In addition, we will see that the internal
loop momenta of vertex subdiagrams generically produce polarizations of their external photons
that are in the directions of those loop momenta.   We shall refer to these below as ``loop polarizations".
Loop polarizations, of course, are not necessarily in the directions of external momenta, and, as we
shall see, interfere with the factorization, and hence universality, of infrared enhancements.   All of these complications
begin at two loops, and much of this article will be involved with showing how such obstacles can be overcome. Specifically, we will modify the loop integrand of certain diagrams by adding terms that vanish upon integration, to restore manifest logarithmic singularities that factorize into lower-loop expressions \emph{before} integration is carried out.

After setting the notation in the following section, we discuss the underlying factorization properties of the class of processes under consideration, and 
point out the possibility of realizing this factorization at the level of the integrands.   In Sec.\ \ref{sec:oneloop} we
construct a set of infrared and ultraviolet counterterms for the one-loop amplitude of multiple off-shell photon production in electron-positron annihilation. The infrared counterterms are manifestly proportional to the tree-level amplitude, thanks to Ward identities that reveal the factorized nature of the singularities. We continue to the construction of IR and UV counterterms for the two-loop amplitude, first for diagrams with fermion loops (Sec.\ \ref{sec:fermionloop.tex}), and then for diagrams with loops formed by additional photons (Sec.\  \ref{sec:photonic.tex}). In these constructions, we will modify certain one-loop self energy and vertex subgraphs to make factorization properties of IR divergences manifest before integration. UV subdivergence counterterms for one-loop subgraphs  will be carefully constructed to obey Ward identities even away from the strict UV limit, to prevent spoiling the IR factorization properties of the remaining loop. We will present in Sec.\ \ref{sec:checks} numerical checks to demonstrate that all IR and UV singularities are removed by our subtraction terms. The subtraction terms can themselves be analytically integrated by standard methods, and the detailed calculation will be presented in a forthcoming publication. We conclude with a summary of the procedures developed in this paper, and a brief discussion of the outlook for further progress.
 
\section{Expansions and factorization for integrands}
\label{sec:setup}

We consider the QED Lagrangian in the Feynman gauge,  with $N_f$ 
massless lepton fields which carry  an  electric charge $q_f$, normalized in units of the positron charge $e$,
\begin{equation}
{\cal L} = \sum_{f=1}^{N_f} i \bar \psi_f \left( \slashed \partial -
  i q_f e \slashed A
\right) \psi_f - \frac{1}{4}F_{\mu \nu} F^{\mu \nu} -\frac{1}{2 \xi}
\left(\partial \cdot A \right)^2, \qquad \xi =1.
\end{equation}
In addition to soft singularities, the  theory also exhibits collinear
singularities due to the masslessness of the fermionic
fields.  

We will focus on processes with multiple photons produced 
in electron-positron annihilation, 
\begin{equation}
e^{-}(p_1) + e^+(p_2)  \to \gamma^*(q_1)+ \ldots + \gamma^*(q_n)\, ,
\label{eq:basic-processes}
\end{equation}
where in parenthesis we denote the momenta of the external electron, positron and photons. 
We will consider here the case of off-shell photons, $q_i^2 \neq 0$, focusing on the treatment of initial-state singularities.
Although off-shell, these amplitudes are gauge invariant, and share many of the properties of amplitudes for the production of massive vector bosons.   By restricting ourselves to off-shell photons only, we can concentrate on the IR divergences associated with initial states.  Building on the formalism we develop below, we intend in future work to treat amplitudes with on-shell photons and charged particles in the final state.\footnote{Beginning at two loops, even for outgoing on-shell photons we will encounter ``transient" final-state singularities, which disappear in the fully integrated amplitude, but which require special treatment to achieve local integrability.}

\subsection{Expansions}

The infrared structure of a probability amplitude for a process in
this class, Eq.~(\ref{eq:basic-processes}), is independent of the number of photons. 
We will therefore treat the final state in complete generality. For
the initial state, we define the Mandelstam variable $s  \equiv
(p_1+p_2)^2$.  

As an expansion series in the coupling constant,
a generic $n$-photon production amplitude is given by
\begin{align}
M  &= e^n \left[  
M^{(0)} + e^2 M^{(1)} + e^4 M^{(2)}_{\rm full} + \ldots 
\right],
\label{eq:pert_exp}
\end{align}
where $M^{(0)}$ denotes the tree-level process and $M^{(k)}$ is the
$k$-loop amplitude. 
At two loops, the $n$-photon  production amplitude acquires an explicit dependence
on the electric charges of all fermionic flavors due to closed fermion
loops (as in Fig.~\ref{fig:two-loop-fermionic-k} below).
The two-loop amplitude can  then be  further decomposed as 
\begin{equation}
\label{eq:M2-fermion-loop}
M^{(2)}_{\rm full} =  M^{(2)} + \sum_{c=2}^{n+2} \left( \sum_f q_f^c \right) M_c^{(2)}\, ,
\end{equation}
where $q_f e$ is the charge associated to the fermion species $f$. The term $M^{(2)}$ contains diagrams with no fermion loops.  
The amplitude coefficients $M_c^{(2)}$ originate from Feynman diagrams
with a closed fermion-loop, and $c$ photons  coupled to the loop. At two loops, as many as two of these photons may be virtual.  Since we
are working in QED, due to Furry's theorem, 
these coefficients vanish for odd values of
$c$.

The one-loop, two-loop, etc., orders of the $n$-photon production
amplitude are integrals over loop momenta 
\begin{align}
  &M^{(1)} = \int \frac{d^dl}{(2\pi)^d}   \, \mathcal M^{(1)}(l), \quad M^{(2)} = \int
    \frac{d^d k}{(2\pi)^d}  \frac{d^d l}{(2\pi)^d}  \, \mathcal M^{(2)}(k,l) \,,\ldots
    \label{eq:McalM}
\end{align}
We will adopt the above notation also for counterterms and finite
amplitudes after subtraction;  in all cases, $M$, with various superscripts and subscripts, will be
used to denote integrated quantities while  $\mathcal M$ will be used to
denote unintegrated quantities.   For the tree amplitude, there is no 
loop integration, so $M^{(0)}$ and ${\mathcal M}^{(0)}$ will be used 
interchangeably. The tree-level and one-loop amplitudes, ${\mathcal M}^{(0)}$ and ${\mathcal M}^{(1)}$, respectively, will be written down from diagrams in the Feynman gauge; the same is true for a starting expression of the two-loop ${\mathcal M}^{(2)}$, but a modified version of amplitude ${\mathcal M}^{(2)}$ will be constructed, to make IR factorization properties more manifest without changing the integrated amplitude.

To denote counterterms, we will use the generic notation
\begin{equation}
  \takelimit_{r} \mathcal M \label{eq:tauNotation}
\end{equation}
to represent a local (i.e.\ integrand-level) approximation of some amplitude $\mathcal M$ in a certain IR or UV limit $r$. For example, we will write $\takelimit_{\text{soft}} \mathcal M^{(1)}$ to denote the soft limit of the one-loop amplitude. We stress the inherent freedom and ambiguity left in this notation: $\takelimit_{r} \mathcal M$ only needs to match the divergent part of $\mathcal M$ in the limit $r$, while the finite part is unspecified. The specific choice of counterterms will be a main topic of this paper. Now it's natural to use the notation
\begin{equation}
  \takelimit_{r} M \label{eq:tauNotationIntegrated}
\end{equation}
to denote the result of integrating Eq.~\eqref{eq:tauNotation} over loop momenta.

Due to the electron-positron initial state, the amplitude takes the
form 
\begin{equation}
  \mathcal M=\overline{v}(p_2) \widetilde {\mathcal M} \left (p_1, p_2, \{q_j\} \right ) u(p_1)\, , \ \{ q_j\}\ =\  q_1, \dots , q_n\, ,
\end{equation}
where $\widetilde {\mathcal M}$  is a matrix in spinor space, and where the
dependence of the amplitude on electron and positron spinors is 
shown explicitly. We will adopt a similar notation for
various amplitude components, whereby  $\widetilde { M}^{(X)}$ ($\widetilde {\mathcal M}^{(X)}$) denotes the integrated (unintegrated) component $X$ of the amplitude with external spinors removed.

For future reference, it will be useful to define the Dirac projectors,
\begin{align}
  \mathbf P_1 &\equiv \frac{\slashed p_1 \slashed p_2} {2p_1 \cdot p_2},  
\quad 
  \mathbf P_2 \equiv \mathbb 1 -\mathbf P_1
\label{eq:P1projector}
\end{align}
which satisfy
\begin{equation} 
 \mathbf P_1^2 = \mathbf P_1,  \quad \mathbf P_2^2 = \mathbf P_2, 
\end{equation}
and 
\begin{align}
  &\mathbf P_1\, u(p_1) = u(p_1), \quad \bar v(p_2) \mathbf P_1 = \bar
    v(p_2), 
\quad
\mathbf P_2\, u(p_1) = 0, \quad \bar v(p_2) \mathbf P_2 = 0 \, .
\label{eq:projectorSpinor} 
\end{align}
The first of these projectors, also commonly used in SCET~\cite{Becher:2014oda},
will serve in our definition of infrared counterterms at one and two loops.  

\subsection{Factorization for integrals and integrands}
\label{sec:fact-int-int}

As noted above, for the processes under study, all IR singularities are associated with virtual lines that are either collinear to the incoming
electron or positron, or have vanishing momenta and are attached to one  of the external lines of the pair or to fermion lines
that are collinear to the incoming lines.   All such IR behavior factorizes from the ``hard scattering", in this case the production of off-shell photons.

All such amplitudes enjoy the same pattern of factorization as the (Sudakov) annihilation form factor for massless fermions in QED.   
Let us briefly review the reasons for this result.
 Infrared singularities are generated by the loops of jet functions for the electron and for the positron, and by loops in a soft function, which describes 
the coupling of soft photons to those jets, and their scattering through virtual massless fermion loops.   In fact,
for our discussion, we will not need to construct these functions explicitly.  We need only know that they factor from a hard-scattering, or
short-distance function.   The hard function, in turn, depends only on the momenta of outgoing photons and the incoming lines, and is completely insensitive to collinear
evolution before the hard scattering, as well as to all soft photons.  The factorization is therefore in terms of a product, rather than a convolution.
Product factorization is characteristic of all wide-angle or large-momentum transfer scattering amplitudes in gauge theories \cite{Sterman:2002qn,Ma:2019hjq}.

There are many ways to express the product factorization described above.   For us, it will be convenient to give it as a product in Dirac
space, and in terms of integrands, following the notation of Eq.\ (\ref{eq:McalM}).   (We suppress the subscript ``full" in Eq.~(\ref{eq:McalM}), but the
following discussion applies independently to each of the terms in the expansion in the order of fermion loops, Eq.\ (\ref{eq:M2-fermion-loop}).)    
For the amplitude that describes the annihilation process $e^-(p_1) + e^+(p_2) \to \gamma^*(q_1) + \cdots +\gamma^*(q_n)$, with all photons off-shell,  we will represent the factorized amplitude as
\begin{eqnarray}
M \left( p_1,p_2,\{q_j\} \right) \ &=&\  \left \langle \,0 |\, \bar \psi(0)\, \left [ \mathbf P_1 T\left(p_1,p_2, \{q_j\} \right) \mathbf P_1 \right ] \, \psi(0)\, | p_1,p_2 \right \rangle
\nonumber\\[2mm]
&=&\ F\, \left [ \mathbf P_1 T\left( \{q_j\} \right) \mathbf P_1 \right ]  
\, ,
\label{eq:full-fact}
\end{eqnarray}
where in the first form we define the corresponding (annihilation) form factor as a matrix element.
In the second expression, the matrix element is written in terms of a function
$F$, which represents the all-order sum of form factor diagrams.   Each diagram has a single, local, electron-positron vertex
represented by  $\mathbf P_1 T\left( \{q_j\} \right) \mathbf P_1$, at which the total momentum of the initial state, $p_1+p_2$, flows out of the diagram.  The factor $T$
is itself a Dirac matrix, representing a hard-scattering function that contains all dependence on the final
state momenta $\{q_j\}$.   It appears between the projectors defined in Eq.\ (\ref{eq:P1projector}), whose purpose we discuss at the close of this subsection.
Like $F$, the matrix $\mathbf P_1 T \mathbf P_1$ will be computed from sums of loop diagrams. Its integrals, however, will be completely independent of the loop integrals of $F$, as described below.

The content of Eq.\ (\ref{eq:full-fact}) is that all IR singularities arise from form factor integration, and are
independent of the final state momenta.  Correspondingly, the matrix $\mathbf P_1 T\left( \{q_j\} \right) \mathbf P_1$ is IR finite in four dimensions.   This relation holds to all orders in perturbation theory \cite{Ma:2019hjq} for renormalized amplitudes.

\begin{figure}
\begin{eqnarray}
\label{eq:FF2functional_occ1}
  \mathcal F^{(2)} [\mathbf P_1 T \mathbf P_1]
  &=& 
\eqs[.2]{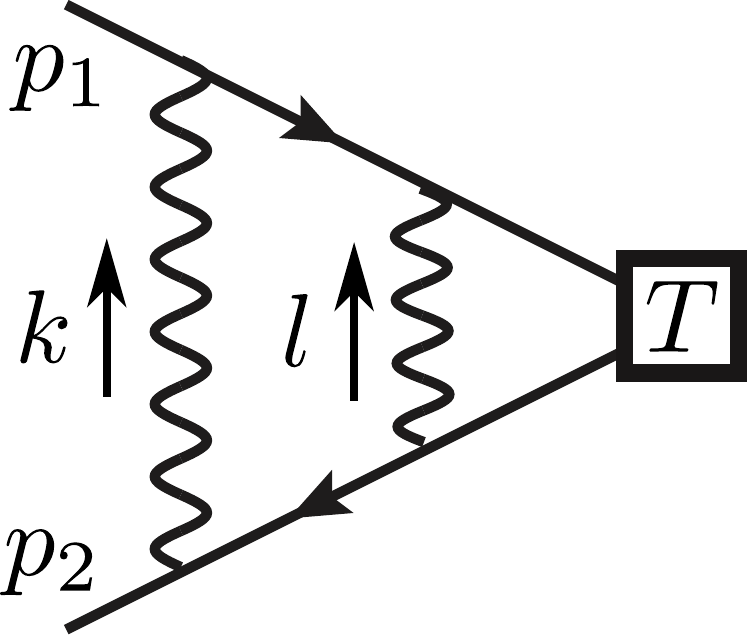} 
+ \eqs[.2]{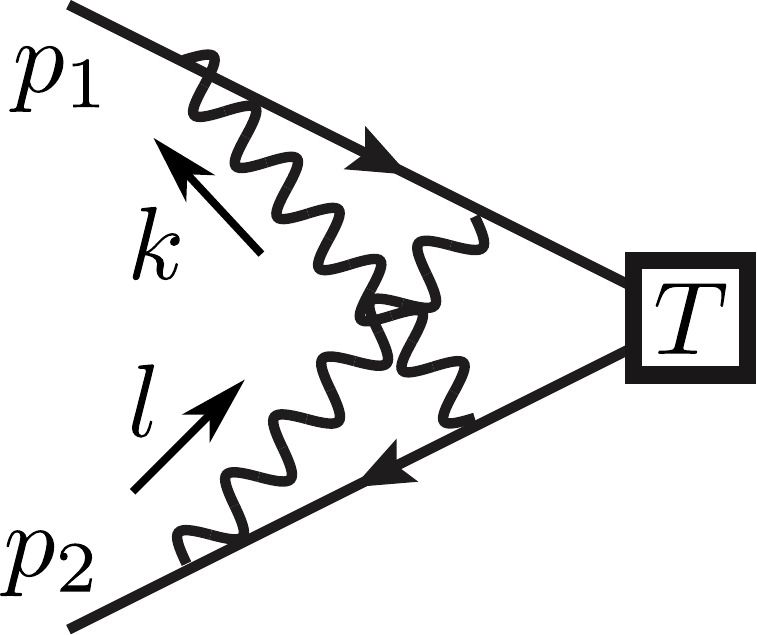} 
+ \eqs[.2]{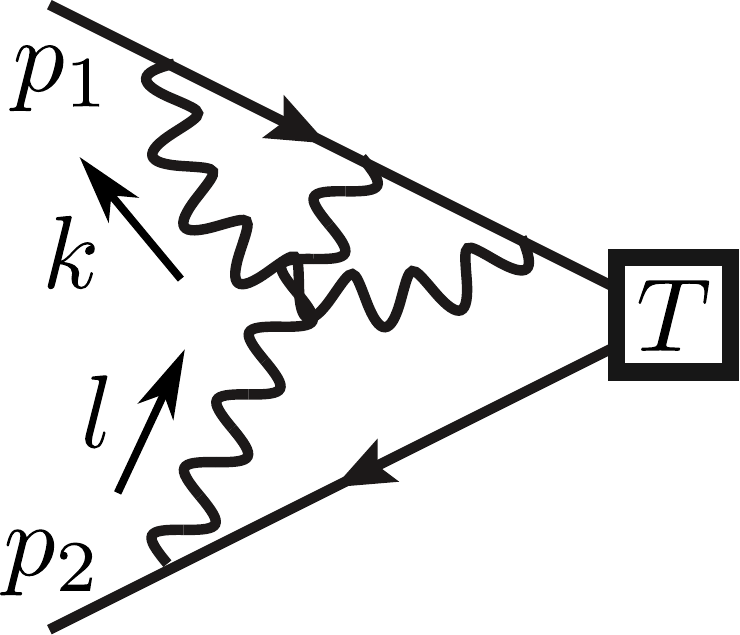}
\nonumber \\
&& + \eqs[.2]{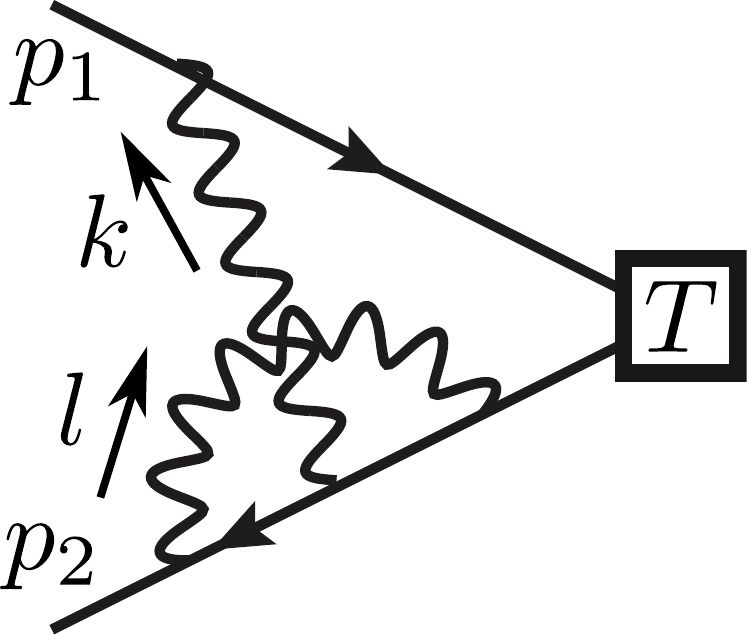} 
+ \eqs[.2]{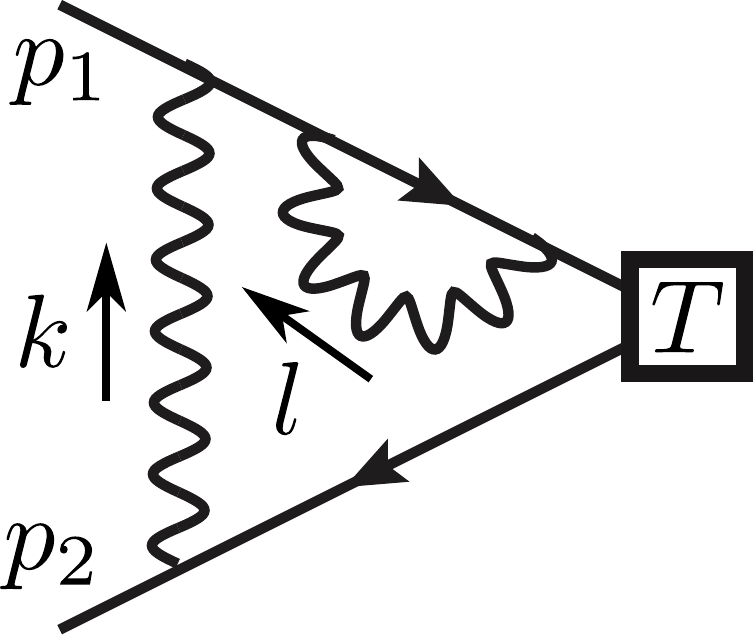} 
+ \eqs[.2]{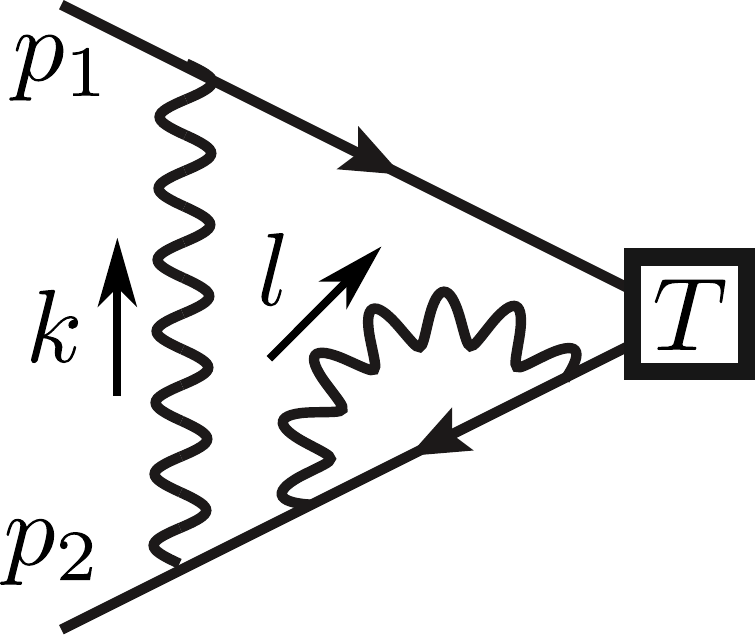} \ ,\nonumber 
\end{eqnarray}
\caption{Diagrams that contribute to ${\cal F}^{(2)}$ for arbitrary order in the local vertex $T$. \label{fig:F2}}
\end{figure}

Equation (\ref{eq:full-fact}) is also a sum over all orders in this fully factorized expression.   To a given power of the coupling, say $e^{2a}$, $a\ge 1$ in the
notation of Eq.\ (\ref{eq:pert_exp}), $M^{(a)}\left( \{q_j\} \right)$ is a function of $a$ loop momenta, while $F$ and $T$ appear at all loop
orders that add up to $a$, as
\begin{eqnarray}
M^{(a)} \left( \{q_j\} \right )\ &=& \int_{k_{1 \ldots a}}\, {\cal M}^{(a)}\left( \{q_j\}, k_1, \dots k_a \right) 
\nonumber\\
 &=&\ 
\sum_{b=0}^{a-1} \int_{k_{b+1\dots a}}\, {\cal F}^{(a-b)} \left [ k_{b+1}, \dots k_a  ; \mathbf P_1 \int_{k_{1\dots b}}\,{\cal T}^{(b)} \left( \{q_j\}, k_1, \dots k_{b} \right) \mathbf P_1 \right ] 
\nonumber\\[2mm]
&\ & \quad +\ {\cal F}^{(0)} \left [ \mathbf P_1 \int_{k_{1\dots a}} {\cal T}^{(a)} \left( \{q_j\}, k_1, \dots k_a  \right) \mathbf P_1 \right ]\, .
\label{eq:expand-fact}
\end{eqnarray}
In this product factorization, the notation implies that the integrals over loops in the form factor $F$ are completely decoupled from those in matrix $T$,
whose integrands are represented by $\cal F$ and $\cal T$, respectively.
The form factor integrand for ${\cal F}^{(2)}$ is illustrated diagrammatically in Fig.\ \ref{fig:F2}.   Also included implicitly are the
UV counterterms that are part of the perturbative expansion in these expressions for renormalized amplitudes.   

Because of the separation of integrations, we can think of Eq.\ (\ref{eq:full-fact}) 
as an effective theory expression for this class of amplitudes, where hard scales, of the order of external invariants, have been integrated
out within $T$.  From this point of view as well, the matrix $\mathbf P_1 T \mathbf P_1$ is a local vertex.   The Dirac structure of
this vertex depends on the nature of the hard scattering.   For our case, with photon production only,
$T$ will be a Dirac vector, dressed by an IR finite coefficient function. For other electroweak processes, potentially including 
heavy lepton flavors in the hard scattering, axial vector, tensor and scalar matrices will all occur in general.    In both
this picture and the factorization formula, it is necessary to regulate UV divergences associated with the separation of scales, and to
renormalize the operator represented here by the vertex.   

We can now exploit the perturbative expansion of the factorized amplitude.
To begin, in the sum over orders in Eq.\ (\ref{eq:expand-fact}), we have  already
separated the case $b=a$, for which the form factor at zeroth order is just the local vertex, evaluated between the external Dirac spinors
$\bar v(p_2)$ and $u(p_1)$.   We then find, using the properties of the projectors in Eq.~(\ref{eq:projectorSpinor}), that at the level of integrands as well as integrals, we may identify
\begin{eqnarray}
\ {\cal F}^{(0)} \left [ \mathbf P_1  \int_{k_{1\dots a}}\, {\cal T}^{(a)} \left( \{q_j\}, k_1, \dots k_a  \right) \mathbf P_1 \right ] \
&\ & \
\nonumber\\
&\ & \hspace{-30mm} =\
\bar  v(p_2)\, \left [ \mathbf P_1  \int_{k_{1\dots a}}\, {\cal T}^{(a)} \left( \{q_j\}, k_1, \dots k_a  \right) \mathbf P_1 \right ] \, u(p_1)
\nonumber\\[2mm]
&\ & \hspace{-30mm} =\
 \ \bar v(p_2)\,  \int_{k_{1\dots a}}\, {\cal T}^{(a)} \left( \{q_j\}, k_1, \dots k_a  \right) \, u(p_1)
\nonumber\\[2mm]
&\ & \hspace{-30mm} \equiv\   \int_{k_{1\dots a}}\, {\cal M}^{(a)}_{\rm finite} \left (\{q_j\}, k_1, \dots k_a \right )
\nonumber\\[2mm]
&\ & \hspace{-30mm} \equiv\ M^{(a)}_{\rm finite} \left (\{q_j\} \right )\, .
\label{eq:M-fin-integrated}
\end{eqnarray}
In the final definitions we recognize from Eq.\ (\ref{eq:expand-fact})  a term from which all IR subdivergences have been subtracted.   The finiteness of this term at all loop
orders is equivalent to the statement of factorization for this amplitude.   Solving for $M^{(a)}_{\rm  finite}$ in (\ref{eq:expand-fact}), we can interpret the additional terms
as subtractions, what we will refer to as ``infrared counterterms" below, which regulate all IR singularities in the integrand of the amplitude at this order.

Given (\ref{eq:M-fin-integrated}), we can now identify, following Eq.\ (\ref{eq:McalM}) for amplitudes without external spinors, that
 \begin{eqnarray}
 T^{(a)}\ =\ \widetilde M^{(a)}_{\rm finite}\, .
 \end{eqnarray}
 This result enables us to solve iteratively for the finite remainder at order $a+1$ in terms of those at lower order,
 \begin{eqnarray}
 M^{(a+1)}_{\rm finite} \ =\ M^{(a+1)}\ -\ \sum_{b=0}^{a}{F}^{(a+1-b)} \left [ \mathbf P_1 \widetilde M_{\rm finite}^{(b)} \mathbf P_1 \right ]\, .
 \label{eq:Mfin-iterative}
\end{eqnarray}
At the lowest orders, we can immediately write down expressions that we will encounter again below,
\begin{eqnarray}
M^{(0)}_{\rm finite}\ &=&\ {M}^{(0)} \, ,
\label{eq:M0-amplitude}
\\[2mm]
{M}^{(1)}_{\rm finite}\ &=&\ {M}^{(1)} - {F}^{(1)} \left [ \mathbf P_1 \widetilde {M}^{(0)} \mathbf P_1 \right ] \, ,
\label{eq:M1-fin-integrated}\\[2mm]
{M}^{(2)}_{\rm finite}\ &=&\ {M}^{(2)} - {F}^{(1)} \left [ \mathbf P_1 \widetilde {M}^{(1)}_{\rm finite} \mathbf P_1 \right ] 
- {F}^{(2)} \left [ \mathbf P_1 \widetilde {M}^{(0)} \mathbf P_1 \right ]\, .
\label{eq:M2-fin-integrated}
\end{eqnarray}
The advantage of this procedure is that all dependence on the multiplicity and masses of the produced photons is isolated in the functions $\widetilde M_{\rm finite}$, which
are infrared and ultraviolet finite after the integrals have been carried out.   For a complex process, however, with high multiplicity in the final state, these integrals may be
difficult to carry out analytically.   

So far, our discussion is still entirely at the level of integrated amplitudes. 
As explained above, our goal is to formulate the $M^{(a)}_{\rm finite}$ so that they may be evaluated numerically.

With this in mind, we turn to the question of how to derive analogs of Eqs.\ (\ref{eq:M1-fin-integrated}) and (\ref{eq:M2-fin-integrated}) at the level of the {\it integrands} for
these quantities, in forms that are at the same time locally integrable at singularities and manifestly convergent at infinity.   
To emphasize these
two aspects, we write the analogs of (\ref{eq:M1-fin-integrated}) and (\ref{eq:M2-fin-integrated}) as
\begin{eqnarray}
{\cal M}^{(0)}_{\rm finite}\ &=&\ {\cal M}^{(0)} \, ,
\\[2mm]
{\cal M}^{(1)}_{\rm finite}\ &=&\ {\cal M}^{(1)}_{\rm UV-finite} - {\cal F}_{\rm UV-finite}^{(1)} \left [ \mathbf P_1 \widetilde {\cal M}^{(0)} \mathbf P_1 \right ] \, ,
\label{eq:M1-fin}\\[2mm]
{\cal M}^{(2)}_{\rm finite}\ &=&\ {\cal M}^{(2)}_{\rm UV-finite} - {\cal F}^{(1)}_{\rm UV-finite} \left [ \mathbf P_1 \widetilde {\cal M}^{(1)}_{\rm finite} \mathbf P_1 \right ] 
- {\cal F}^{(2)}_{\rm UV-finite} \left [ \mathbf P_1 \widetilde {\cal M}^{(0)} \mathbf P_1 \right ]\, .
\nonumber\\
\label{eq:M2-fin}
\end{eqnarray}
In these expressions and below, ${\cal M}^{(a)}$ will denote a sum of $a$-loop integrands  without UV counterterms.
Here the subscript ``UV-finite" represents the relevant set of diagrams with ultraviolet counterterms subtracted.   Schematically, we may write at $a$ loops, for
both ${\cal M}^{(a)}$ and ${\cal F}^{(a)}$,
\begin{eqnarray}
{\cal M}^{(a)}_{\rm UV-finite}\ &=&\ {\cal M}^{(a)}\ -\ \sum_{{\rm UV\ regions}\ i}\ \takelimit_i\, {\cal M}^{(a)}\ +\ \cdots\, ,
\nonumber\\
{\cal F}^{(a)}_{\rm UV-finite}\ &=&\ {\cal F}^{(a)}\ -\ \sum_{{\rm UV\ regions}\ i}\ \takelimit_i\, {\cal F}^{(a)}\ + \cdots
\, ,
\label{eq:M-UV-fin-def}
\end{eqnarray}
where $\takelimit_i$ is an operator that inserts a counterterm
appropriate for UV region $i$.  Terms not shown involve nestings of UV limits and UV limits where more than a single loop momentum diverges.
 We will use this notation, adapted to the specific cases being treated, below.   In the same spirit, the
form factor terms ${\cal F}$ in Eqs.\ (\ref{eq:M1-fin}) and (\ref{eq:M2-fin}), explicitly subtract the singular infrared behavior of the integrands of the original diagrams.

We emphasize that these relations do not generally hold for amplitude and form factor integrands written down directly from the QED Lagrangian.   Beyond one loop, it will be necessary to
modify both the original amplitudes and the IR subtraction terms.   These changes are necessary because factorization properties are expressions of the symmetries,
especially gauge symmetry, of the theory, and such symmetries are not always manifest locally.   We explain the challenges involved in finding a construction that
obeys these relations in the following sections.  We close this section with a brief discussion of counterterms, followed by an explanation of our use of the projection
matrices, $\mathbf P_1$.

In our treatment, the ultraviolet counterterms in Eq.\ (\ref{eq:M-UV-fin-def})
will be written, not directly as a Laurent series in $\epsilon=2-d/2$, but as a set of explicit integrals, whose integrands, once combined with the integrands
of the original diagrams, produce convergent integrals overall.   The counterterm integrals, which can be carried out independently as well,  will depend on arbitrary masses, 
which serve to regularize them in the infrared, and in principle to match them to (massless) QED counterterms in any renormalization scheme.   
The set of counterterms for each diagram that contributes to ${\cal M}$ will include a subset that is
in one-to-one correspondence to the original counterterms of the theory for these diagrams.  We will find, however, that additional counterterms
are necessary to make all integrals converge, even though these counterterms may add up to zero in the final integral.   An example will be found below in light-by-light scattering
diagrams.

Many of the same UV counterterms that appear in ${\cal M}$
will appear in the form factor integrals of ${\cal F}$.  We will also encounter UV counterterms associated with the renormalization of the vertex we have denoted by 
$\mathbf P_1 T \mathbf P_1$.  

 In our discussion, we shall not consider contributions to the amplitude with the self-energies of the
external lines, which we simply truncate and multiply by spinors or photon polarizations.   Correspondingly, we do not carry out a multiplicative renormalization
for our amplitudes.   This process, however, is completely standard and unaltered by our construction.   
We will implement our construction explicitly for the one- and two-loop amplitudes of the full class of processes in Eq.\ (\ref{eq:basic-processes}).   

\begin{figure}
\centerline{\includegraphics[width=0.35\textwidth, valign=c]{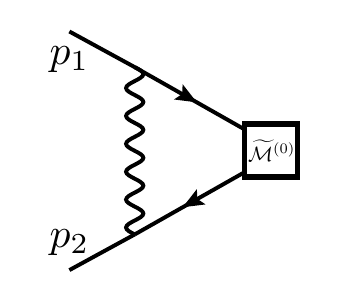}}
\caption{Illustration of ${\cal F}^{(1)}$ at lowest order. \label{fig:for F1}}
\end{figure}

Finally, before going on to the treatment of the one-loop amplitudes, Eq.\ (\ref{eq:M1-fin}), we will address the motivation for the use of the projectors $\mathbf P_1$.
The example of ${\cal M}^{(1)}_{\rm finite}$ shows, at the first nontrivial order, the  essential role of these projections.   
Consider the factor ${\cal F}^{(1)}$, which is illustrated in Fig.\ \ref{fig:for F1}. 
Because the
integral in ${\cal F}^{(1)}$ is fully decoupled from $\widetilde {\cal M}^{(0)}$, the fermion propagators that are contracted with the local vertex in Fig.\ \ref{fig:for F1} are generically off-shell.
This would imply that, without the projectors, the tree-level hard scattering function $\widetilde{\cal M}^{(0)}$ would lose gauge invariance.  At tree level, this means
that it would not be transverse to unphysical polarizations of the external photons.   

To see the loss of gauge invariance explicitly, we consider a generic contribution to $\widetilde{\cal M}^{(0)}$, with $n$ external photons attached in some fixed order $\Pi$, 
with fixed polarizations, $\epsilon^*(q_i)$.  We isolate one
of these, say $\epsilon^{*\nu}(q_e)$, which we will replace by its momentum $q_e^\nu$.   We will refer to such a polarization as ``scalar" or ``longitudinal", 
terms which are equivalent for massless vectors. For massive neutral scalars, $q^\nu_e$ is strictly speaking the scalar polarization.

Let us denote by ${\cal M}^{(0,\Pi)}_\nu$ the sum of all insertions of the scalar-polarized photon of momentum $q_e$ into the fermion line from which the final-state photons emerge, keeping the order of the other photons fixed.
This sum reduces to only two terms, determined by the QED Ward identity.  The result is easily derived in complete generality by repeated use of the lowest order identity, which will be revisited in the next section.   Here, we simply give the result,
\begin{eqnarray}
q_e^\nu\; \sum_{\rm {\it q_e}\ insertions} \widetilde {\cal M}_\nu^{(0,\Pi)}
&=&\ 
\slashed p_2 \left (\slashed p_2- \slashed q_e \right )^{-1} \prod_{c=2,\ne e}^n \slashed \epsilon^*(q_c) \left (\slashed p_1 - \sum_{d=1,\ne e}^{c-1} \slashed q_d \right )^{-1}
\nonumber\\
&\ & \quad +\ \prod_{c=2,\ne e}^n \slashed \epsilon^*(q_c) \left (\slashed p_1 - \sum_{d=1,\ne e}^{c-1} \slashed q_d  - \slashed q_e \right)^{-1}\ \left (\slashed p_1 - \slashed q_e \right )^{-1} \slashed p_1\, .
\nonumber\\
\end{eqnarray} 
If this expression  is inserted in Eq.\ (\ref{eq:M1-fin}) {\it without} the projectors $\mathbf P_1$, both terms will vanish only if both virtual fermion lines in Fig.\ \ref{fig:for F1} are on-shell.   As a result, there will be gauge dependence in all regions except when the virtual photon carries zero momentum.  In particular, in collinear regions for the photon's momentum, gauge dependence is induced.   In contrast, with the projectors in place, this gauge-dependence vanishes identically because $\slashed p_1^2=\slashed p_2^2=0$.   At the same time, in all singular regions, the projectors act as the identity for on-shell lines.

In summary, the result of our procedure will be expressions for the integrands of amplitudes,
as in Eqs.\ (\ref{eq:M1-fin}) and (\ref{eq:M2-fin}), in which all infrared singularities are
subtracted as form factor integrands (the ${\cal F}^{(L)}$ in Eq.\ (\ref{eq:expand-fact})), up to two loops.   All 
process-dependent information is isolated in each $\widetilde {\cal M}^{(a)}_{\rm finite}$, which can in principle be evaluated numerically.   
We anticipate that it will be possible to extend this procedure beyond two loops, and to other theories for a wide class of related processes.

In the following section, we turn to this formalism at one loop.
 
\section{The one-loop amplitude}
\label{sec:oneloop}

The one-loop amplitude $M^{(1)}$  for
the process $e^-(p_1) + e^+(p_2) \to \gamma^*(q_1)+ \ldots \gamma^*(q_n)$  is
a $d$-dimensional integral (with $d=4-2\epsilon$ in dimensional regularization) over the components of
the loop momentum, as in Eq.~\eqref{eq:McalM}.
$M^{(1)}$ is singular in $d=4$ space-time dimensions,
due to IR and UV divergences.
In this section, we will
describe how to remove these singularities systematically from the integrand,
through the introduction of simple counterterms.  
Our approach will be by direct construction of the necessary subtractions.
As we proceed, the general formula for the fully-integrable and process-dependent
result, ${\cal M}^{(1)}_{\rm finite}$ in Eq.\ (\ref{eq:M1-fin}), will emerge from the analysis.   
We emphasize that, even at one loop, we will encounter many individual diagrams
that do not have the structure of Eq.\ (\ref{eq:M1-fin}).

Our aim is to identify a set of counterterms  that are local in momentum space, and which add up to
a single function  ${\cal M}_{{\rm singular}}^{(1)}(l)$  of the loop
momentum that approximates the full one-loop integrand ${\cal M}^{(1)}$  in all 
divergent limits. The integration will then be re-organized into two integrals,
\begin{align}
  M^{(1)} &= \int \frac{d^d l}{(2 \pi)^d}
  \left[
    {\cal M}^{(1)}(l)
- {\cal M}_{{\rm singular}}^{(1)}(l)
\right]
+\int \frac{d^d l}{(2 \pi)^d}  {\cal M}_{{\rm singular}}^{(1)}(l)\, .
\label{eq:M1-IRUV-sub}
\end{align}
The first integral on the right-hand side, which we will identify with $M_{\rm finite}^{(1)}$ in Eq.\ (\ref{eq:M1-fin-integrated}),
will be convergent in four space-time dimensions
and amenable, in principle, to direct numerical integration.
We will not discuss its practical computation here, but note that relevant methods have
been developed in Ref.~\cite{Capatti:2019edf}. The second
integral will contain all divergences in the dimensional regularization parameter
$\epsilon$. We will detail the construction of its integrand
and its evaluation within dimensional regularization.  We emphasize
that the finiteness of the first term in Eq.\ (\ref{eq:M1-IRUV-sub}) is a property of
the sum over all one-loop diagrams, and the construction of counterterms
will reflect properties of the sums of their integrands.  

Generic IR  and UV counterterms at one loop have already been derived in
Refs.~\cite{Nagy:2003qn,Becker:2010ng,Assadsolimani:2009cz}.  The
counterterms that we present in this section 
are equivalent to the ones in the literature, in approximating all
singular limits of one-loop amplitudes.
Their derivation here serves to describe generic features of our methodology and 
as an introduction to the construction of the more complicated
two-loop counterterms that will be presented for the first time, in 
later sections.  Our derivation of local UV counterterms at
one loop is designed, in particular, to respect Ward identities. 
 The full set of counterterms will become important 
at two loops in constructing UV-subdivergence counterterms that do not
spoil IR-factorization properties, which depend crucially on these Ward identities. 
As suggested in Sec.\ \ref{sec:fact-int-int}, we will see that 
the IR and UV counterterms necessary for a process with $n$ photons match those of
the $e^+e^-$ annihilation form factor.
 In this way, we use our knowledge of
the simplest process under consideration, $e^++e^- \to \gamma^*$,
to organize singular contributions to a much larger set of processes.

\subsection{IR divergences of the one-loop amplitude}
\label{subsec:oneloopIR}

In this subsection we will present the subtraction of  IR divergences from $M^{(1)}$,
reflected at the integrand level by the terms ${\cal F}^{(1)}$ in Eq.\ (\ref{eq:M1-fin}).
We will study their UV structure in the following subsection.

We may decompose the one-loop amplitude into four classes,
\begin{equation}
M^{(1)}  \equiv \eqs[.28]{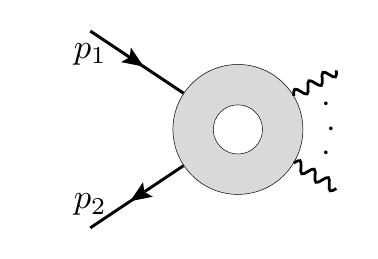} = M^{(1,A)} + M^{(1,B_1)} + M^{(1,B_2)} +M^{(1,C)}   \, ,
\end{equation}
each of which is characterized by the following diagrams:
\begin{itemize}
\item  the virtual photon is adjacent to both incoming fermions,
\begin{equation}
\label{eq:M1Adiags}
M^{(1,A)} \equiv \eqs[.25]{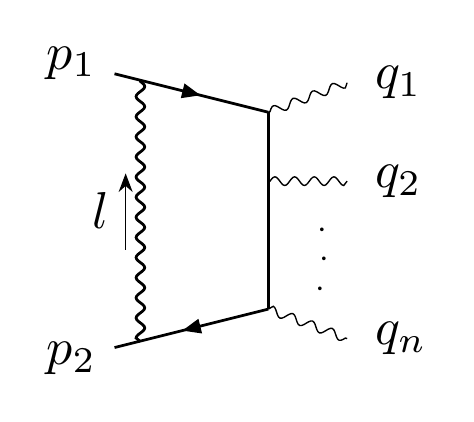}+\,\text{external photon permutations}\,,
\end{equation}
\item the virtual photon is adjacent to the incoming electron but not
  to the incoming positron, 
\begin{equation}
\label{eq:M1B1diags}
M^{(1,B_1)} \equiv \eqs[.25]{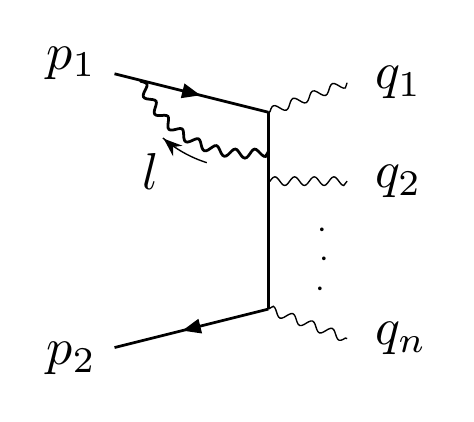}\,+\eqs[.25]{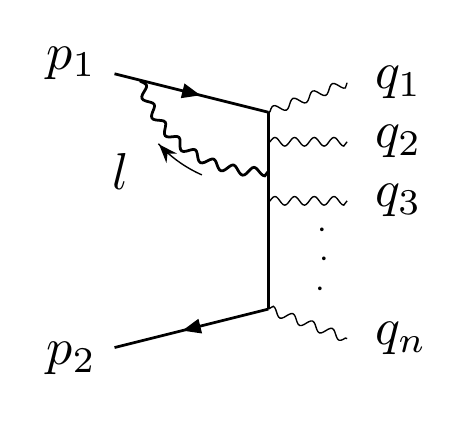}+\, \ldots\,, 
\end{equation}
where the ellipses denote additional photon insertions into the ``hard" subdiagram, consisting of all lines that are off-shell for $l || p_1$, 
as well as permutations among the external photons,
\item the virtual photon is adjacent to the incoming positron but not
  to the incoming electron, 
\begin{equation}
\label{eq:M1B2diags}
M^{(1,B_2)} \equiv \eqs[.25]{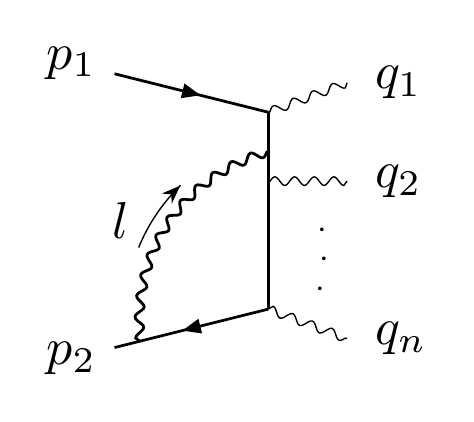}\,+\eqs[.25]{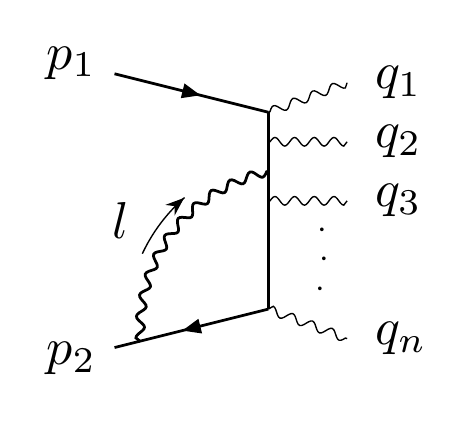}+\, \ldots\,,
\end{equation}
\item and, finally, the virtual photon is not adjacent to either of the incoming fermions
\begin{equation}
\label{eq:M1Cdiags}
M^{(1,C)} \equiv \eqs[.25]{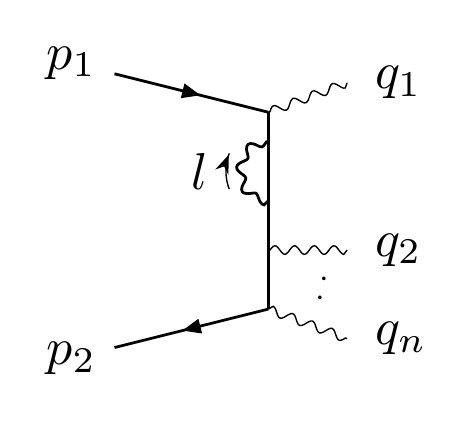}\,+\eqs[.25]{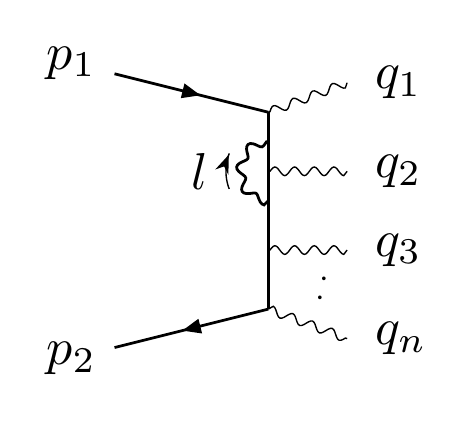}+\, \ldots\,.
\end{equation}
\end{itemize}
We assign a loop momentum label $l$ to all the diagrams, and the loop momentum flow is directed towards the vertex closest to the incoming electron. 
To compute diagrams analytically, it is often convenient to choose different momentum routings 
in different  diagrams, as permitted by the shift-invariance of loop
integrals. However, as will become apparent soon, such loop momentum label
assignments can obscure important cancellations of collinear
singularities in the sum of diagrams. 
A judicious assignment of the loop momentum for all diagrams
is necessary in order to render  these cancellations local in momentum space, and our
momentum assignment achieves this purpose. In addition, the collinear
limits will factorize in terms of the tree-amplitude, again locally at the level of the integrand.

IR singularities are associated with configurations where the loop momentum $l$ either
vanishes in all four components, or is collinear to one of the two incoming lines. 
For the diagrams of $M^{(1)}$, the complete list
of such singularities is
\begin{itemize}
\item a soft singularity $l = 0$ in the graphs of $M^{(1,A)}$, 
\item a collinear singularity $l = -x_1 p_1$, $0\le x_1 \le 1$, in the graphs of
  $M^{(1,A)}$ and $M^{(1,B_1)}$,
\item a collinear singularity $l =  x_2 p_2$, $0\le x_2 \le 1$, in the graphs of
  $M^{(1,A)}$ and  $M^{(1,B_2)}$.
\end{itemize}
We refer to these points or lines as ``pinch surfaces" \cite{Sterman:1995fz,Collins:2011zzd}.
To identify these regions, we recall that diagrammatic integrals are defined in the complex plane for each loop
momentum component, and carried out along a contour that is specified by the ``$i\epsilon$" prescription
for propagator denominators.   In general, these contours can be deformed away from poles where propagator
denominators vanish, unless poles coalesce from opposite sides of the contour, producing a ``pinch".
This happens from two poles of a single denominator, whenever the corresponding line
carries zero momentum in all four components, or from pairs of denominators, when they 
share the light-like momentum of an external particle.   The latter are collinear pinches, the former are soft.  
Note that when the virtual photon line of diagram $M^{(1,A)}$ carries zero momentum, both adjacent internal
fermion lines are automatically on-shell, carrying their respective external momenta, so that in this case
a soft pinch forces three lines to be on shell at once.

The set of collinear and soft  singularities identified above do not exhaust the full set
of pinch surfaces in the amplitudes that we treat here, but they are the only ones that produce
infrared singularities. That is, they are the only pinch surfaces that are not integrable in four dimensions.   The simplest examples of integrable pinch surfaces are
points in loop momentum space where a fermion line carries zero momentum.   All of our diagrams
include such points within their loop momenta, but  they are always integrable, because of the extra
factor of momentum in the numerator of fermion propagators.
We will not reproduce the power counting necessary to show these 
results here, but refer the reader to, for example, Refs.~\cite{Sterman:1995fz,Collins:2011zzd}.    A recent
and very general discussion has been given  in Ref.\ \cite{Collins:2020euz}.
 
An essential feature of the list of divergent pinch surfaces given above is that it is the same for any number of
final-state photons in our basic set of processes, Eq.\ (\ref{eq:basic-processes}).    As we will show, as indicated in Eq.\ (\ref{eq:M1-fin}), that this universality will be
inherited by the IR counterterms necessary to factorize infrared behavior.   Indeed, the full infrared behavior
for every such amplitude is already present in  the case $n=1$ in Eq.\ (\ref{eq:basic-processes}), the annihilation form factor $e^++e^- \to \gamma^*$.

We now seek a function  ${\cal M}_\text{IR}^{(1)}$ that
approximates the one-loop integrand ${\cal M}^{(1)}$  in all 
of the above IR-divergent limits for the general process with an arbitrary number of photons in the final state.
The  construction of such an approximation is complicated somewhat by overlaps
of soft and collinear divergences. For example, the
region $l \sim - x_1 p_1$, which yields a collinear singularity, overlaps
for $x_1=0$ with the $l \sim 0$ region, which yields a soft singularity.
The entanglement of IR divergences takes more complicated
patterns at  higher loop
orders.  However, it has been shown that approximations of the
integrands of arbitrary loop integrals can be constructed systematically with the
method of {\it nested subtractions} ~\cite{Collins:1989gx,Collins:2011zzd,Erdogan:2014gha,Ma:2019hjq}. 
In this approach, the diverse singular regions are ordered according to their extent within the
integration volume. Then, the singularities are removed recursively, subtracting first the ones corresponding to the smallest volumes.
Example applications of the nested subtractions method for one- and two-loop Feynman integrals can be found in Ref.~\cite{Anastasiou:2018rib}.  In this article, we perform an explicit application of the method for physical amplitudes.

Following the nested subtraction approach, we first need to find an approximation
of the integrand in the soft limit, which receives contributions only
from the diagrams
of the class $M^{(1,A)}$. The sum of their integrands can be
written compactly in the form 
\begin{equation}
{\cal M}^{(1,A)} = \frac{-i e^2}{l^2 (p_1+l)^2 (p_2-l)^2} {\cal N}^{(1)}\, ,
\label{eq:M-1}
 \end{equation}
where the numerator ${\cal N}^{(1)}$ reads
\begin{equation}
{\cal N}^{(1)} =  \bar{v}(p_2) \gamma^\mu 
\left(\slashed l  - \slashed p_2 \right) 
\widetilde {\mathcal M}^{(0)} (p_1 + l, p_2 - l, q_1, \dots, q_n)
\left(\slashed p_1+ \slashed l \right) 
\gamma_\mu u(p_1). 
\end{equation}
In the soft limit $l^\mu \to 0$, we neglect $l$ inside $\widetilde {\mathcal M}^{(0)}$, and approximate ${\cal M}^{(1)}$ with
\begin{align}
  {\cal M}^{(1)} \xrightarrow{l \to 0}
  &\takelimit_{{\rm soft}} {\cal M}^{(1)}    \nonumber \\
  & \hspace{-10mm} = \frac{i e^2
  \bar{v}(p_2) \gamma^\mu 
\left(\slashed l - \slashed p_2  \right) 
\widetilde {\mathcal M}^{(0)} (p_1, p_2, q_1, \dots, q_n)
\left(\slashed p_1+ \slashed l \right) 
\gamma_\mu u(p_1)
  }
  {l^2 (p_1+l)^2 (p_2-l)^2}
  \, .
\label{eq:T-soft-1}
\end{align}
In this definition of the soft counterterm, we have left the denominators of the 
outermost electron and positron lines intact, in particular not dropping terms
quadratic in $l$ compared to $p_i\cdot l$, with $i=1,2$. In addition, we keep the exact $l$-dependence in the numerator except inside $\widetilde {\mathcal M}^{(0)}$, where we set $l=0$.   

We now turn our attention to the collinear region $l || p_1$.
Naively, we would have to analyze each diagram individually, but the situation is actually much simpler.  In covariant gauges, collinear divergences appear
when vector particles connect jet subdiagrams to the hard scattering.  In individual diagrams, these contributions are
non-factoring in their momentum dependence.   Such collinear vector lines, however, always carry scalar polarizations (equivalent to longitudinal polarizations for massless particles), contracted with the hard scattering subdiagram.
The Ward identities of the theory (in this case, QED) ensure that the sum of all such contributions factorizes at fixed momenta for the collinear photons \cite{Collins:1989gx}.
This, by now standard, result enables us to introduce local counterterms for the sum of diagrams, avoiding the potentially complicated non-factoring dependence of each diagram
individually.

Only the diagrams in $M^{(1,A)}$ and $M^{(1, B_1)}$ contribute to the collinear divergence in the region $l \parallel p_1$.  We write the sum of their integrands in the form
\begin{align}
\label{eq:M1AplusM1B1}
{\cal M}^{(1,A+B_1)} &\equiv {\cal M}^{(1,A)} + {\cal M}^{(1, B_1)} 
\nonumber\\
& \hspace{-10mm} =  \frac{ -i e}{l^2 (p_1+l)^2}\,  
\bar{v}(p_2)
\eta^{\mu \nu}\, \widetilde {\mathcal M}^{(0,A+B_1)}_\mu (p_1+l, p_2, l;q_1, \dots, q_n) \left( \slashed p_1 + \slashed l \right) \gamma_\nu u(p_1)\,. \nonumber\\
 \end{align}
In this relation, $\widetilde {\mathcal M}^{(0,A+B_1)}_\mu (p_1+ l, p_2,
 l;q_1,  \dots, q_n)$ represents the sum of tree diagrams from the set $A+B_1$ including all vertices at which the final state
 photons  and an
 additional external photon of momentum $l$ attach, but excluding the diagram
 where $l$ is directly attached to $p_1$.  Momentum $l$ flows out of each tree diagram and into
 the vertex adjacent to the electron line that is external to the full
 one-loop diagram, to be consistent with Eq.~\eqref{eq:M-1}.
The polarization vector of the exchanged photon (momentum $l$) as well  as the electron and positron spinors are removed from $\widetilde {\mathcal M}^{(0,A+B_1)}_\mu$.  The momenta of the
additional photon and the incoming electron are generically off-shell,
but they approach the mass shell in the collinear limit $l^\mu \approx x_1 p_1^\mu$.

Using $\eta_1$ as a reference vector, any $q^\mu$ in the direction of $p_1^\mu$, i.e., $q^\mu = x_q p_1^\mu$ can be approximately rewritten as
\begin{equation}
q^\mu \approx \frac{q \cdot \eta_1 }{l \cdot \eta_1}  l^\mu ,
  \label{eq:collprojection1}
\end{equation}
or alternatively,
\begin{equation}
  q^\mu \approx \frac{2 q \cdot \eta_1 }{(l + \eta_1)^2 -\eta_1^2} l^\mu.
  \label{eq:collprojection2}
\end{equation}
Either of these approximations becomes accurate in the diagrams above when the photon of momentum $l$ becomes collinear to $p_1$, with
$q^\mu$ a linear combination of $p_1^\mu$ and $l^\mu$.
While Eq.~\eqref{eq:collprojection1}, with a denominator linear in
$l$, suffices in order to demonstrate the factorization of collinear singularities (see, for example, Ref.~\cite{Ma:2019hjq}), in
this article  we use the equivalent version of
Eq.~\eqref{eq:collprojection2} with a quadratic denominator.

Similarly, in the collinear limit $l \approx - x_1 p_1$, 
the metric tensor can be decomposed in terms of $l^\mu$ and the generic vector $\eta_1^\mu$ as 
\begin{equation}
\eta^{\mu \nu} =\frac{l^\mu \eta_1^\nu}{l \cdot \eta_1} +\frac{l^\nu
  \eta_1^\mu}{l \cdot \eta_1}   -\frac{\eta_1^2}{(l \cdot \eta_1)^2}
l^\mu l^\nu+ \eta_\perp^{\mu \nu}\, ,
\end{equation}
where $\eta_\perp^{\mu \nu} \, l_\nu = \eta_\perp^{\mu \nu} \, {\eta_1}_\nu =0.$ 
Staying always within the collinear approximation, where $l^2 \to 0$, we can equivalently approximate the
metric with
\begin{equation}
\label{eq:metric_decomp_quadratic}
\eta^{\mu \nu} \approx \frac{2 l^\mu \eta_1^\nu}{\left(l
    +\eta_1\right)^2-\eta_1^2} +\frac{2 l^\nu
  \eta_1^\mu}{\left(l +\eta_1\right)^2-\eta_1^2}   -\frac{4\eta_1^2}{ \left( \left (l + \eta_1\right )^2 - \eta_1^2 \right )^2}
l^\mu l^\nu+ \eta_\perp^{\mu \nu}\, .
  \end{equation}
Using anti-commutation relations for Dirac matrices and the Dirac
equation satisfied by the external spinor $u(p_1)$, in the collinear limit $l^\mu \approx - x_1 p_1$,
the rightmost factors in Eq.~\eqref{eq:M1AplusM1B1} can be re-written as, 
\begin{equation}
\left( \slashed p_1 + \slashed l \right) \gamma_\nu \, u(p_1) \approx 2 (p_1+l)_\nu \, u(p_1) \propto p_{1\,\nu} \, . \label{eq:collinearPolarization}
\end{equation}
We now use these considerations to show how collinear singularities factorize on a point-by-point basis in our one-loop diagrams.

\begin{figure}[h]
\begin{center}
\begin{align}
\frac{1}{e} \,\eqs[0.28]{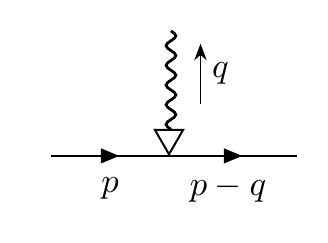}  =
\eqs[0.25]{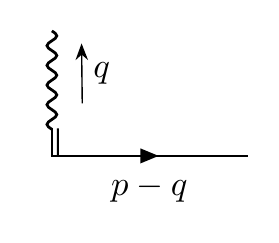} 
-\eqs[0.25]{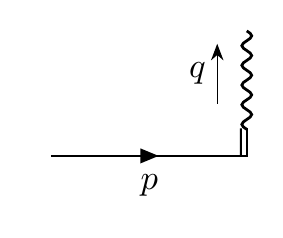} 
\nonumber
\end{align}
\end{center}
\caption{\label{fig:WI} A pictorial representation of a tree Ward
  identity Eq.~\eqref{eq:WI}. Fermion-fermion-photon vertices with arrowed photons
  denote a contraction of the QED triple vertex with the photon
  momentum $-i e \gamma^\mu \to -ie \slashed{q}$, multiplied by
  appropriate extra factors shown on the left-hand side of Eq.~\eqref{eq:WI}. The photon can
be regarded as an external photon with a scalar polarization
vector, equal to its momentum. The double lines represent insertions of
  momenta without a photon-fermion vertex, multiplied by appropriate
  factors as shown on the right-hand side of Eq.~\eqref{eq:WI}.}
\end{figure}

\begin{figure}[!h]
\begin{center}
\includegraphics[width=0.5\textwidth]{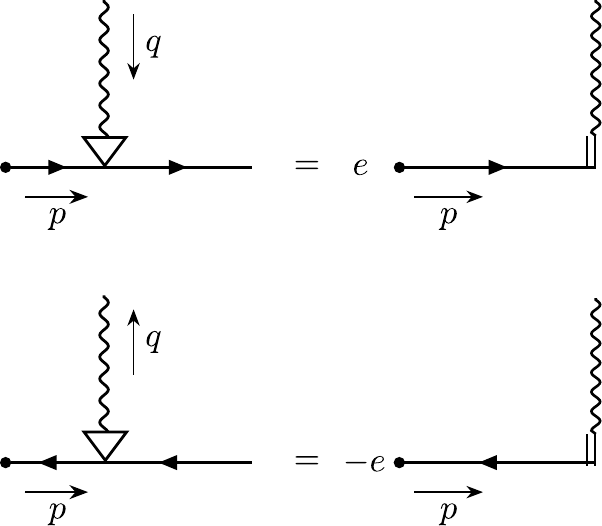}
\end{center}
\caption{\label{fig:WIexternal} A pictorial representation 
of the effect of attaching scalar-polarized photons to external fermion
lines in Eqs.~(\ref{eq:WI-ext1})-(\ref{eq:WI-ext}). The dots represent external fermion legs.}
\end{figure}

\begin{figure}[h]
  \centering
  \begin{align}
  &\eqs[0.25]{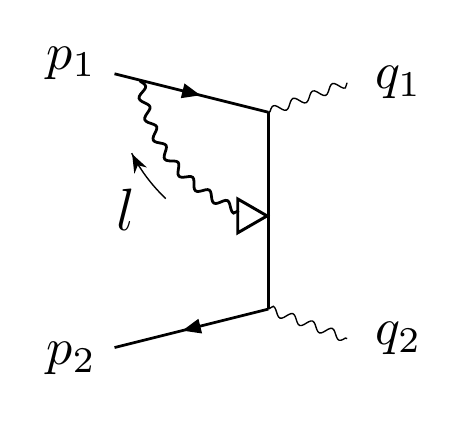} + \eqs[0.25]{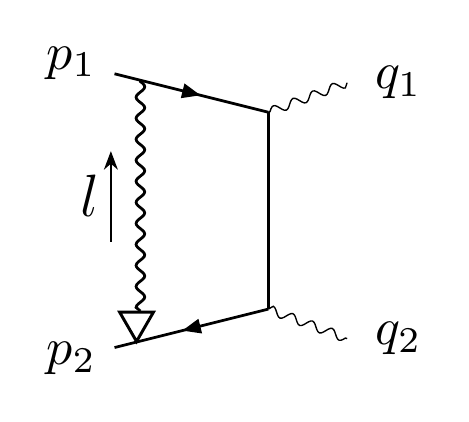} = \nonumber \\
  &-e \left(
  \eqs[0.25]{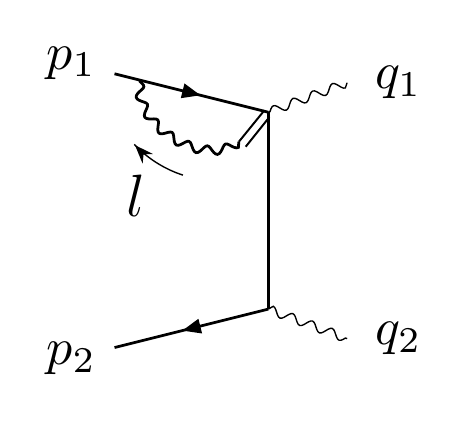}
  - \eqs[0.25]{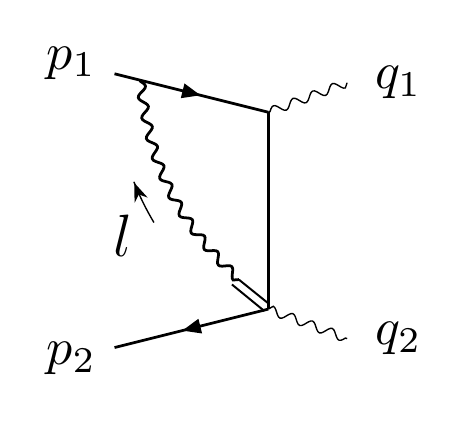}
  + \eqs[0.25]{1L_Ward_diag4} \right) \nonumber \\
  &=-e \eqs[0.25]{1L_Ward_diag3}
  \nonumber
  \end{align}
  \caption{The application of Ward identities to factorize the $l\parallel p_1$ collinear divergence in the case of two-photon production.   This is a graphical
  representation of Eq.\ (\ref{eq:M1AplusM1B1approx}).}
  \label{fig:twoPhotonCollFact}
\end{figure}

By decomposing the  metric tensor $\eta^{\mu \nu}$ in
Eq.~\eqref{eq:M1AplusM1B1} according to Eq.~(\ref{eq:metric_decomp_quadratic}),
we notice that only one term gives a non-vanishing
contribution in the collinear limit.
Thus, in the limit $l \approx -x p_1$  we can make the replacement $\eta^{\mu\nu} \to 2 \eta_1^\nu \, l^\mu / [(l+\eta_1)^2 -\eta_1^2] $ without changing the behavior of ${\cal M}^{(1,A+B_1)}$ in the region $l || p_1$. This allows us to write
\begin{align}
{\cal M}^{(1,A+B_1)} &\equiv {\cal M}^{(1,A)} + {\cal M}^{(1, B_1)} \approx  \nonumber \\
&\frac{-i e}{l^2 (p_1+l)^2}\, \bar{v}(p_2) \left[\frac{2\eta_1^\nu \, l^\mu} {(l + \eta_1)^2  - \eta_1^2}\,
 \widetilde {\mathcal M}^{(0,A+B_1)}_\mu (p_1+l, p_2, l;q_1, \dots, q_n) \right] 
\nonumber \\
&\times\,\left( \slashed p_1 + \slashed l \right) \gamma_\nu u(p_1).
\label{eq:M1AplusM1B1approx0}
\end{align}
We represent the right-hand side of Eq.~\eqref{eq:M1AplusM1B1approx0} graphically by 
\begin{align}
e^2{\cal M}^{(1,A+B_1)}  &\approx \eqs[.25]{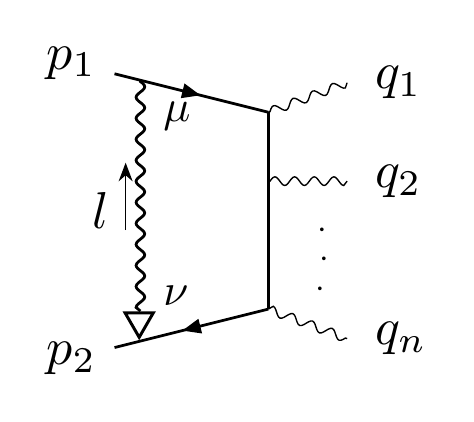}\,+ \eqs[.25]{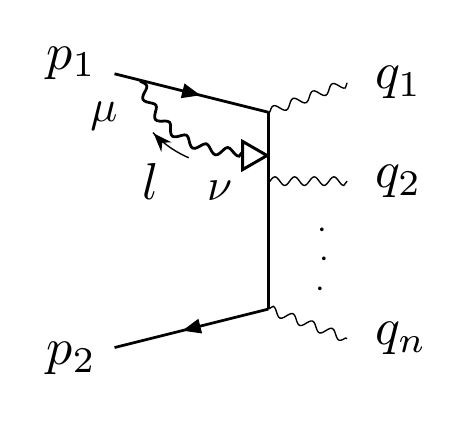} \,\nonumber \\
&+ \eqs[.25]{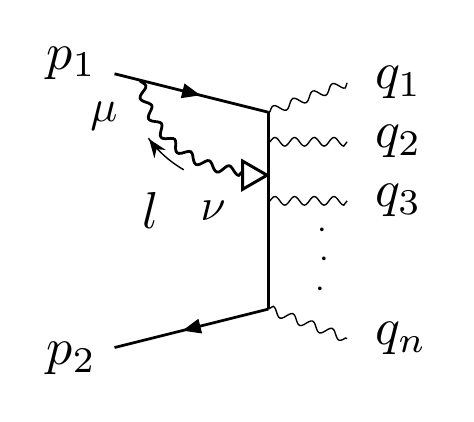}\,+\,\ldots \,. \label{eq:M1AplusM1B1triangle}
\end{align}
The triangle arrow at the lower end of photon lines denotes the following polarization approximation in the photon propagator: 
\begin{equation}
  \eqs[.15]{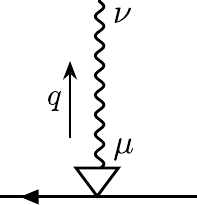} \, : \qquad \frac{-i}{q^2} \eta^{\mu \nu} \rightarrow \frac{-i}{q^2} \frac{2 \eta_1^\nu \, q^\mu} {(q+\eta_1)^2 -\eta_1^2}\,,
  \label{eq:triangleNotation}
\end{equation}
where $\eta_1$ is chosen to have a large rapidity separation from $p_1$ (i.e. it is not collinear to $p_1$). Here we are effectively considering a photon line whose polarization is proportional to its own momentum, a scalar polarization.
Here and below we choose the triangle arrow notation to define a quadratic, rather than linear, denominator in the approximate polarization
tensor. For the opposite collinear limit, we use a gray triangle to denote an analogous approximation, with a reference vector $\eta_2$ chosen to have a large rapidity separation from $p_2$,
\begin{equation}
  \eqs[.15]{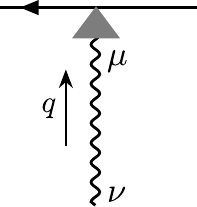} \, : \qquad \frac{-i}{q^2} \eta^{\mu \nu} \rightarrow \frac{-i}{q^2} \frac{-2 \eta_2^\nu \, q^\mu} {(q-\eta_2)^2 -\eta_2^2}.
  \label{eq:triangleNotation2}
\end{equation}
The expression in Eq.~\eqref{eq:M1AplusM1B1approx0} is now ready to be simplified with a partial fractioning
identity, which is a manifestation of the QED Ward identity,
\begin{equation}
\label{eq:WI}
  \frac {2\eta_1^\nu}{(q + \eta_1)^2 - \eta_1^2} \left[ \frac{i}{\slashed{p}-\slashed{q}} (-ie \slashed q) \frac{i}{\slashed{p}} \right]
  = e\,\frac {2\eta_1^\nu} {(q + \eta_1)^2 - \eta_1^2} \left[\frac{i}{\slashed p - \slashed q} - \frac{i}{\slashed p} 
    \right]\, .
\end{equation}
This identity describes the effect of attaching a photon with a
scalar polarization on a fermion line.
We visualize the identity in Fig.~\ref{fig:WI}.
We also note the following identities for insertions of scalar photons in external fermion lines:
\begin{align}
\frac {2\eta_1^\nu} {(q + \eta_1)^2 - \eta_1^2} \left[
  \frac{i}{\slashed{p} + \slashed{q}} (-ie \slashed q) u(p) \right]
  &=e \frac {2\eta_1^\nu} {(q + \eta_1)^2 - \eta_1^2}
    u(p) \label{eq:WI-ext1}\, , \\
\frac {2\eta_1^\nu} {(q+ \eta_1)^2 - \eta_1^2} \left[ \bar{v}(p) (-ie \slashed q)  \frac{i}{\slashed{p} - \slashed{q}} \right] &= -e \frac {2\eta_1^\nu} {(q + \eta_1)^2 - \eta_1^2} \bar{v}(p),
\label{eq:WI-ext}
\end{align}
which are depicted pictorially in Fig.~\ref{fig:WIexternal}.
We now see that repeated applications of Eqs.~\eqref{eq:WI} - \eqref{eq:WI-ext} to
Eq.~\eqref{eq:M1AplusM1B1triangle} result in pairwise cancellations of all terms except one,
\begin{equation}
  e^2{\cal M}^{(1,A+B_1)}
\approx \frac{-i e^2}{l^2 (p_1+l)^2} 
  \bar{v}(p_2) \left[\frac{2\eta_1^\nu} {(l + \eta_1)^2 - \eta_1^2}\,
    \widetilde {\mathcal M}^{(0)} \right]
  \left( \slashed p_1 + \slashed l \right) \gamma_\nu u(p_1)\, ,
   \label{eq:M1AplusM1B1approx}
\end{equation}
where ${\mathcal M}^{(0)}$ is the lowest-order tree diagram, whose external vertices do not include the virtual photon.
The cancellations are illustrated, as an example, for graphs in two-photon production in Fig.~\ref{fig:twoPhotonCollFact}.
With the action of the Ward identity, non-factorizing collinear
singularities cancel, and the one-loop collinear limit is expressed in terms of the lowest order,
on-shell tree amplitude.
The general case of an arbitrary number of photon emissions is proven
analogously, and Eq.~\eqref{eq:M1AplusM1B1approx} holds generally.

Interestingly, if we choose $\eta_1$ to be equal to $-p_2$ in Eq.~\eqref{eq:M1AplusM1B1approx}, then the three 
denominators become identical to those of the soft-photon
approximation given by  the right-hand side of
Eq.~\eqref{eq:T-soft-1}.  After
some simple Dirac algebra, it can be shown that the numerator of Eq.~\eqref{eq:M1AplusM1B1approx} also coincides with that of
Eq.~\eqref{eq:T-soft-1} in the collinear limit $l \parallel p_1$. An
analogous observation can be made for the opposite collinear limit
$l \parallel p_2$. 
Therefore, the right-hand side of
Eq.~\eqref{eq:T-soft-1}, which is simply a form factor whose vertex is
the tree amplitude (with $l$ set to 0), may be used as a ``global'' IR counterterm that simultaneously cancels soft and collinear
singularities of the one-loop amplitude. 
This is an expression of the universality of IR behavior in the complete set of amplitudes under consideration.
As discussed in Sec. \ref{sec:fact-int-int}, we will place the tree amplitude between the Dirac
projectors defined in Eq.\ (\ref{eq:P1projector}).  We arrive at the following global IR counterterm,
\begin{align}
\takelimit_{\rm IR} {\cal M}^{(1)} 
&= i e^2 \frac{
    \bar{v}(p_2) \gamma^\mu \left( \slashed   p_2  - \slashed l \right)
    \mathbf P_1   \widetilde {\mathcal M}^{(0)}(p_1, p_2; q_1, \ldots, q_n)  \mathbf P_1 
    \left( \slashed p_1  + \slashed l \right) \gamma_\mu   u(p_1)
  } {l^2 (p_1+l)^2 (l-p_2)^2} \, , 
  \label{eq:oneLoopFFCT} 
\end{align}
which is equivalent to the soft photon approximation in Eq.\ (\ref{eq:T-soft-1}) because $\mathbf P_1$ acts as the identity in the soft, and collinear, regions.
This is the usual one-loop form factor, which we will denote by $\mathcal F^{(1)}$, but with the QED vertex replaced by the truncated tree-level spinor matrix sandwiched between a pair of projectors, just as in Eq.\ (\ref{eq:full-fact}). Therefore, the one-loop IR counterterm is equivalent to
\begin{align}
\takelimit_{\rm IR} {\cal M}^{(1)}  &\equiv\mathcal F^{(1)} \left[\mathbf P_1  \widetilde {\mathcal M}^{(0)}(p_1, p_2; q_1, \ldots, q_n)  \mathbf P_1 \right]  
\equiv\eqs[0.3]{c19-5} \,,
\label{eq:ffDiagNotation} 
\end{align}
where we have recalled the corresponding diagrammatic notation of Fig.\ \ref{fig:for F1}.

The IR subtraction with the form factor counterterm of Eq.~(\ref{eq:ffDiagNotation}) suffices to remove locally infrared singularities for one-loop amplitudes where the  final-state photons are on-shell as well. 
Even though diagrams such as in $M^{(1,C)}$ have
collinear pinches when two fermion lines share the momentum of a light-like outgoing photon, numerator factors conspire to make them IR finite in physical amplitudes. In contrast, certain two-loop diagrams with on-shell photons are only finite after integration, in their original forms. An extension of our formalism at two loops for on-shell photons will be presented in future work.

%\iffalse
%\else
We comment here on how the $\mathbf P_1$ projector, as defined by
Eq.~\eqref{eq:P1projector}, in Eq.~\eqref{eq:oneLoopFFCT} yields a 
tree-amplitude factor explicitly, in a regularization scheme where
helicities are treated in four dimensions.
Using the spin sum for either $u(p_1)$ or $\bar v(p_2)$, the projector $\mathbf P_1$ has the following two alternative representations,
\begin{align}
  \mathbf P_1 &= \frac{1}{2p_1 \cdot p_2} \slashed p_1 \big( v_L (p_2) \bar v_L(p_2) + v_R (p_2) \bar v_R(p_2) \big), \\
  \mathbf P_1 &= \frac{1}{2p_1 \cdot p_2} \big( u_L (p_1) \bar u_L(p_1) + u_R (p_1) \bar u_R(p_1) \big) \slashed p_2\, .
\end{align}
Without loss of generality, let us consider the case in which the incoming electron is left-handed, and the incoming positron is right handed. We replace $u(p_1)$ by $u_L(p_1)$ and replace $\bar v(p_2)$ by $\bar v_R(p_2)$ in Eq.~\eqref{eq:oneLoopFFCT}. (Due to the chirality-preserving nature of interactions in massless QED, the amplitude vanishes when the incoming electron and positron have the same chirality.) These spinors satisfy,
\begin{align}
  & \frac{1 - \gamma^5}{2} u_L(p_1) = u_L(p_1), \quad \frac{1 + \gamma^5}{2} u_L(p_1) = 0, \\
& \bar v_R (p_2) \frac{1 + \gamma^5}{2} = \bar v_R (p_2), \quad \bar v_R (p_2) \frac{1 - \gamma^5}{2} = 0 \, .
\end{align}
The IR counterterm $\takelimit_{\rm IR} {\cal M}^{(1)} $ in Eq.~\eqref{eq:oneLoopFFCT} is rewritten as
\begin{align}
\takelimit_{\rm IR} {\cal M}^{(1)} \ &=
   \frac{-ie^2}{l^2 (p_1+l)^2 (l-p_2)^2 (2 p_1 \cdot p_2)^2 }  \bar v_R(p_2) \gamma^\mu (\slashed p_2 - \slashed l)
  \slashed p_1 \nonumber \\
  & \quad \times\,  \big( v_L (p_2) \bar v_L(p_2) + v_R (p_2) \bar v_R(p_2) \big)
  \widetilde {\mathcal M}^{(0)}(p_1, p_2, q_1, \ldots, q_n) \\
  & \quad \times\,  \big( u_L (p_1) \bar u_L(p_1) + u_R (p_1) \bar u_R(p_1) \big) \slashed p_2
  \left( \slashed p_1 + \slashed l \right) \gamma_\mu u_L(p_1)\, . \nonumber 
\end{align}
Since an odd number of gamma matrices preserves chirality, the above expression 
reduces to 
\begin{align}
  \label{eq:projectorFactorization}
\takelimit_{\rm IR} {\cal M}^{(1)} \ &=
   \frac{-ie^2}{l^2 (p_1+l)^2 (l-p_2)^2 (2 p_1 \cdot p_2)^2 }  \bar v_R(p_2) \gamma^\mu (\slashed p_2 - \slashed l)
  \slashed p_1  v_R (p_2)\\
  & \quad \times\, \left[ \bar v_R(p_2) \widetilde {\mathcal M}^{(0)}(p_1, p_2, q_1, \ldots, q_n) u_L (p_1)\right]  \bar u_L(p_1) 
  \slashed p_2 \left( \slashed p_1 + \slashed l \right) \gamma_\mu u_L(p_1) \, . \nonumber
\end{align}
The square bracket in this equation is precisely the tree amplitude
contracted with the external spinors $\bar v_R(p_2)$ and $u_L(p_1)$,
so the counterterm of Eq.~\eqref{eq:oneLoopFFCT} is proportional to
the tree amplitude even before integration. 
%This holds in general, and we will sandwich one-loop amplitudes between 
%projectors below, in order to construct counterterms that are manifestly proportional to the one-loop amplitude at the integrand level.   
%% \fi

In summary, in this subsection we have shown
that the following remainder of the one-loop amplitude leads to an IR-finite integration,
\begin{equation}
  \label{eq:M1IRfinite}
  \mathcal M^{(1)}_{\text{IR-finite}}
= \mathcal M^{(1)} - \takelimit_{\text{IR}} \mathcal M^{(1)}\, ,
\end{equation}
with $\takelimit_{\text{IR}} \mathcal M^{(1)}$ given by the explicit subtraction of Eq.\ (\ref{eq:projectorFactorization}), or equivalently, in form factor notation by Eq.\ (\ref{eq:ffDiagNotation}).
Although this procedure provides an integral that is free of infrared singularities, to provide a
numerically computable expression, we must also subtract UV-divergent behavior at the level of the integrand.   We now turn to this procedure.

\subsection{Ward identity-preserving Ultraviolet counterterms}

The IR-finite one-loop integrand, Eq.~\eqref{eq:M1IRfinite}, retains its non-convergent behavior in the UV limit.  It is therefore not yet 
suitable for numerical evaluation.   In order to remove the integral's UV-singularities through a local subtraction, we need to find
an approximating function $\takelimit_{\text{UV}} \mathcal M_{\text{IR-finite}}^{(1)}$ that matches the singular behavior of the integrand in the UV-limit,
\begin{align}
  \takelimit_{\text{UV}} \mathcal M_{\text{IR-finite}}^{(1)} &= \takelimit_{\text{UV}} \left( \mathcal M^{(1)} - \takelimit_{\text{IR}} \mathcal M^{(1)} \right) \nonumber \\
  &= \takelimit_{\text{UV}} \mathcal M^{(1)} - \takelimit_{\text{UV}}
    \takelimit_{\text{IR}} \mathcal M^{(1)}\, ,
\end{align}
with $\takelimit_{\text{IR}} \mathcal M^{(1)}$ given in Eq.\ (\ref{eq:oneLoopFFCT}).   It is at this point that we will encounter UV divergences associated with
the factorized amplitude, Eq.\ (\ref{eq:full-fact}).

At one loop in the process under study, UV divergences occur only in triangle diagrams and fermion self-energies, 
which we will denote by $\Gamma_{ee\gamma}^{(1),\nu}$ and $\Pi_{e}^{(1)}$, respectively.  
Following our convention for the routing of the loop momenta, these
Green's functions have the integrands
\begin{equation}
 e\, \Gamma_{ee\gamma}^{(1),\nu}(p, q, l) \equiv 
\eqs[0.25]{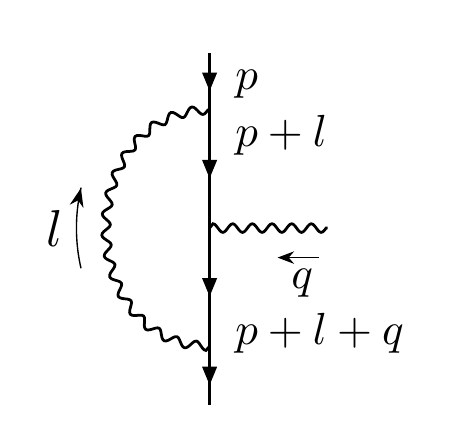} 
= \left(- e^3 \right) \frac{\gamma^\mu(\slashed{p}+\slashed{l}+\slashed{q})\gamma^\nu(\slashed{p}+\slashed{l})\gamma_\mu}{(p+q+l)^2(p+l)^2l^2}\, ,
  \label{eq:oneLoopUVvertex}
\end{equation}
and
\begin{equation}
\Pi_{e}^{(1)}(p, l) \equiv 
\eqs[0.21]{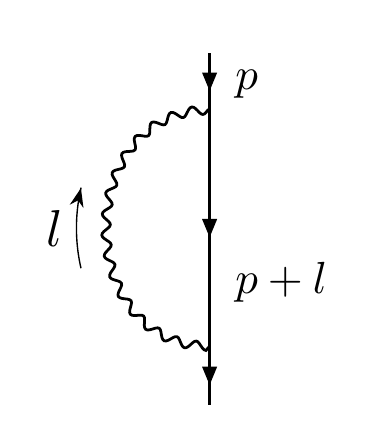} 
= -e^2 \frac{\gamma^\mu(\slashed{l}+\slashed{p})\gamma_\mu}{l^2(l+p)^2}\, ,
\label{eq:oneLoopUVbubble}
\end{equation}
where $p$ is in general off-shell and $q$ can be the momentum of a real or (starting at two loops) virtual 
photon. The Ward identity for these Green's functions takes the form 
\begin{equation}
q_\nu\, \Gamma_{ee\gamma}^{(1),\nu}(p, q, l) 
= \Pi_{e}^{(1)}(p, l) - \Pi_{e}^{(1)}(p+q, l) \, .
\label{eq:oneLoopUVselfen}
\end{equation}
For the vertex diagram, we choose, as suggested above, a UV counterterm that is defined by its integrand, given in this case by
\begin{equation}\label{eq:vtct1}
 e\, \Gamma_{ee\gamma}^{(1,UV),\nu}(l) \equiv
 \eqs[0.25]{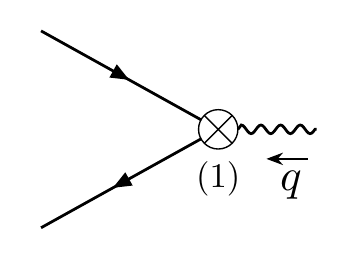}
= \left(- e^3 \right) \frac{\gamma^\mu\slashed{l}\gamma^\nu\slashed{l}\gamma_\mu}{(l^2-M^2)^3} \, ,
\end{equation}
which picks out the only term of the numerator that results in a logarithmic UV divergence, while regulating denominators with a common mass, $M$, which will play the role of a renormalization scale. We stress that in this paper, a ``counterterm'' is always defined at the integrand level, although the graphical notation above draws the counterterm as a local vertex. Our counterterms include all local counterterms in the conventional sense once, but only after, loop integration is carried out.

For the fermion self energy, we must cancel both linear and logarithmic UV divergences, where the latter are linear in the external momentum.   Again regulating denominators with a common mass, the resulting counterterm that we choose is
\begin{eqnarray}
\label{eq:bbct1}
 \Pi_{e}^{(1,UV)}(p, l) &\equiv&
  \eqs[0.12]{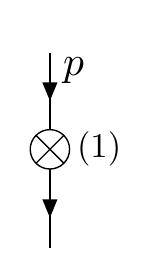}
= \left(-e^2 \right) \left[ 
\frac{\gamma^\mu\slashed{l}\gamma_\mu}{(l^2-M^2)^2} - \frac{\gamma^\mu\slashed{l}\slashed{p}\slashed{l}\gamma_\mu}{(l^2 - M^2)^3}
\right] 
\nonumber \\
&& = -e^2 \frac{\gamma^\mu\slashed{l}\gamma_\mu}{(l^2-M^2)^2} 
-\, p_\nu \Gamma_{ee\gamma}^{(1,UV),\nu}(l) \, .
\end{eqnarray}
In each case, the label $(1)$ indicates that this is a one loop
local counterterm.
Notice that the above counterterms satisfy the Ward identity as well, 
\begin{equation}
q_\nu \Gamma_{ee\gamma}^{(1,UV),\nu}(l) 
=  \Pi_{e}^{(1,UV)}(p, l) -  \Pi_{e}^{(1,UV)}(p+q, l), \label{eq:wardCT}
\end{equation}
for \emph{any values} of $l$ whether or not they are large compared to the scales of external momenta.
Therefore, the renormalized remainders of the above Green's functions
will automatically satisfy  the Ward identity as well.  
Though not essential for the one-loop case, this will turn out to be a particularly useful property at two loops for obtaining
factorized counterterms of collinear limits in the presence of UV sub-divergences. 

Note that the Ward identity for the UV counterterms Eq.~\eqref{eq:wardCT} is by no means guaranteed by the Ward identity for the original diagrams Eq.~\eqref{eq:oneLoopUVselfen}, since we always have the freedom of adjusting non-divergent contributions to the UV counterterms.
For example, an alternative UV counterterm for the self energy diagram, by Nagy and Soper~\cite{Nagy:2003qn}, is
\begin{equation}
-e^2 \frac {\gamma^\mu (\slashed l + \slashed p) \gamma_\mu}{\left[ \left(l + p/ 2 \right)^2 \right]^2} \, .
\end{equation}
This alternative counterterm perfectly matches the UV-divergent behavior of Eq.~\eqref{eq:oneLoopUVvertex} at both linearly and logarithmically divergent orders, but its finite parts would not preserve the Ward identities in conjunction with our UV vertex counterterm, Eq.~\eqref{eq:vtct1}.
We have chosen non-divergent terms so that the UV counterterm for the self energy, Eq.~\eqref{eq:bbct1}, is aligned with our UV counterterm for the vertex, Eq.~\eqref{eq:vtct1}, for both finite and divergent parts, preserving the Ward identity locally.
\footnote{A clean way to arrive at Eqs.~\eqref{eq:vtct1} and \eqref{eq:bbct1} is to perform a series expansion of Eqs.~\eqref{eq:oneLoopUVvertex} and \eqref{eq:oneLoopUVbubble} in the limit of large $l$, and truncate both series at the order that corresponds to a logarithmic divergence after loop integration, before adding a mass regulator to every propagator. The uniform truncation of the series preserves the Ward identity that relates the vertex and the self energy, and the final step of adding an IR mass regulator again preserves the Ward identities.}

We have now specified the local UV-counterterms of the one-loop amplitude,
\begin{align}
\label{eq:UVM1}
e^2\, \takelimit_{\text{UV}} \mathcal M^{(1)} &=
\eqs[0.25]{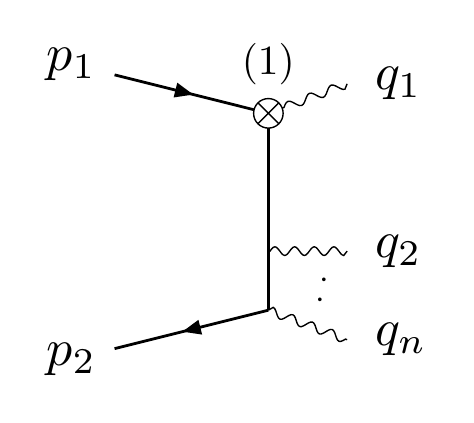} 
+\eqs[0.25]{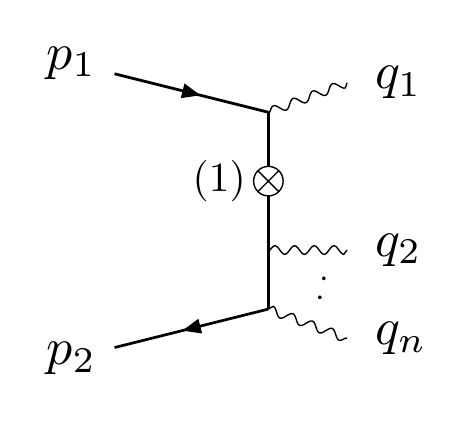}
+\eqs[0.25]{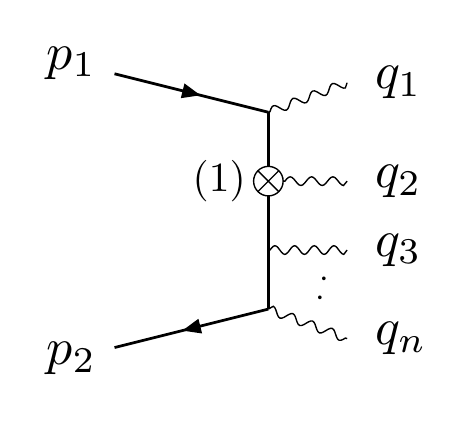} \nonumber \\
&+\ldots 
+\eqs[0.25]{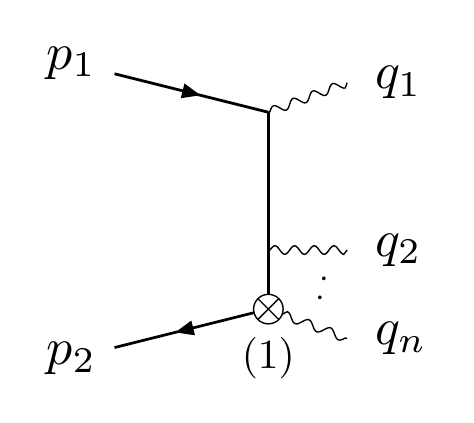} 
+ \text{ external photon permutations}\, .   
\end{align}
Finally, we need to find a UV approximation for the infrared
counterterm, $\takelimit_{\text{IR}} \mathcal M^{(1)}$ in
Eq.~\eqref{eq:oneLoopFFCT}.  We choose
\begin{align}
e^2\, \takelimit_{\text{UV}} \takelimit_{\text{IR}} \mathcal M^{(1)} &\equiv
\bar v(p_2) \left[
\eqs[0.2]{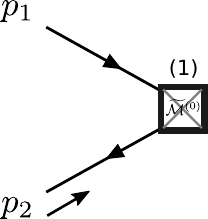} 
\right] u(p_1)
\nonumber \\ 
&= \bar v(p_2)
\left[ 
\left(-i e^2 \right) \frac{\gamma^\mu\slashed{l}
\mathbf P_1
  \widetilde M^{(0)}(p_1, p_2; q_1, \ldots, q_n ) 
\mathbf P_1
  \slashed{l}\gamma_\mu}{(l^2-M^2)^3} 
\right]
u(p_1)
\, . \label{eq:FF1L_UV}
\end{align}
In the above, we have drawn a square vertex with an extra cross to denote the UV limit of the triangle diagram in Eq.~\eqref{eq:ffDiagNotation} with a form factor vertex $\mathbf P_1 \widetilde{\mathcal M}^{(0)} \mathbf P_1$, analogous to how the diagram in Eq.~\eqref{eq:vtct1} denotes the UV limit of the vertex diagram in Eq.~\eqref{eq:oneLoopUVvertex}, with a label ``(1)'' indicating that it is a 1-loop counterterm. Again the projector $\mathbf P_1$ is always omitted in the diagrammatic notation.   In terms
of the relation, Eq.\ (\ref{eq:M1-fin}) that defines ${\cal M}^{(1)}_{\rm finite}$, this counterterm will produce a contribution to ${\cal F}^{(1)}_{\rm UV-finite}$.
  
We have now succeeded in finding an approximation of the one-loop
amplitude integrand in all singular limits.  We can then decompose the
integrand as, 
\begin{equation}
\label{eq:M1separate}
\mathcal M^{(1)}(l) = \mathcal M^{(1)}_{\rm finite}(l) +  \mathcal M^{(1)}_{\rm singular}(l)\, .
\end{equation}
The first term on the right-hand side of Eq.~\eqref{eq:M1separate}, $\mathcal M^{(1)}_{\rm finite}(l)$, is integrable in
four space-time dimensions.  All singularities are now contained in
\begin{equation}
\label{eq:M1singular}
 \mathcal M^{(1)}_{\rm singular}(l) = \takelimit_{\text{IR}} \mathcal M^{(1)}(l) 
+\takelimit_{\text{UV}} \mathcal M^{(1)}(l) 
-\takelimit_{\text{UV}} \takelimit_{\text{IR}} \mathcal M^{(1)}(l).  
\end{equation}
Expressions for  the counterterms on the right-hand side of  
Eq.~\eqref{eq:M1singular} have been given by
Eqs.~\eqref{eq:oneLoopFFCT} for the IR counterterm, Eqs.~\eqref{eq:oneLoopUVvertex}, \eqref{eq:oneLoopUVselfen} and \eqref{eq:UVM1} for the UV subtraction 
and~\eqref{eq:FF1L_UV} for the combined UV/IR term.     
The integration of the counterterms in $\mathcal M^{(1)}_{\rm singular}(l)$ over 
loop momentum $l$ can be performed simply in $d=4-2\epsilon$
dimensions. These results can then be combined with the 
numerical evaluation of $\mathcal M^{(1)}_{\rm finite}(l)$.
In terms of the general relation, Eq.\ (\ref{eq:M1-fin}), we can identify
\begin{align}
{\cal M}^{(1)}_{\rm UV-finite}\ &=\ {\cal M}^{(1)} - \takelimit_{UV} {\cal M}^{(1)} \, ,
\nonumber\\[2mm]
{\cal F}^{(1)}_{\rm UV-finite}\ &=\ \takelimit_{\rm IR} {\cal M}^{(1)} - \takelimit_{\text{UV}} \takelimit_{\text{IR}} \mathcal M^{(1)}
\nonumber\\[2mm]
&=\  {\cal F}^{(1)} - \takelimit_{\text{UV}}  \mathcal F^{(1)}\, ,
\end{align}
where in the second relation for ${\cal F}^{(1)}_{\rm finite}$ we have used Eq.\ (\ref{eq:ffDiagNotation}).  We note that the single IR subtraction, ${\cal F}^{(1)}$,
removed soft and collinear divergences from a large class of diagrams, thus simplifying the application of the method.   
At one loop, our construction was carried out with unmodified
integrands, specified directly by the QED Lagrangian.    At two loops,
we will find again that the set of IR subtractions is much simpler than the full set of diagrams.   To achieve this simplification, however,
it will be necessary to develop modified, but equivalent integrands.

\subsection{Integration of singular counterterms}

Before going on to the two-loop application of out method,
we will complete the discussion of the one-loop case by integrating our one-loop counterterms in dimensional regularization.   This is the lowest-order example
of how the factorization procedure gives a set of closed expression with all singularities of the amplitude, both IR and UV, isolating all dependence on the final state in an
integrable function, specified by Eq.\ (\ref{eq:M1separate}).

Let us consider first the UV counterterms for the one-loop vertex and
propagator, which appear in Eq.~(\ref{eq:UVM1}). 
The vertex counterterm integrates to,
\begin{equation}
\int \frac{d^d l }{(2 \pi)^d}  e\, \Gamma_{ee\gamma}^{(1,UV),\nu}(l)  
= \tilde Z_1^{(1)} \left( -ie \gamma^\nu \right), 
\end{equation}
or, diagrammatically, 
\begin{equation}
\int \frac{d^d l }{(2 \pi)^d}  
\eqs[0.2]{vertexUV_1L}   =
\tilde Z_1^{(1)}  
\eqs[0.2]{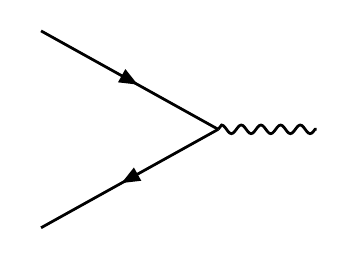} \, ,
\end{equation}
with 
\begin{equation}
\tilde Z_1^{(1)} = \frac{e^2}{(4 \pi)^{\frac d 2}}
\Gamma(1+\epsilon) \frac{ (1-\epsilon)^2}{\epsilon} 
\left(M^2\right)^{-\epsilon}\, .
\end{equation}
For the integral of the electron-propagator counterterm we find 
\begin{equation}
\int \frac{d^d l }{(2 \pi)^d} \frac{i}{\slashed p} 
 \Pi_{e}^{(1,UV)}(p, l) \frac{i} {\slashed p} 
=  - \tilde Z_1^{(1)} \frac{i}{\slashed p},
\end{equation}
or, diagrammatically, 
\begin{equation}
\int \frac{d^d l }{(2 \pi)^d}  \eqs[0.2]{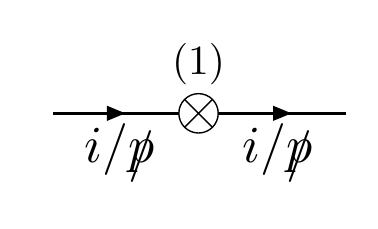}   =
- \tilde Z_1^{(1)}  \eqs[0.2]{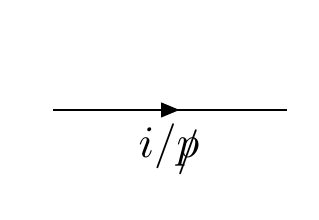}. 
\end{equation}
As expected, every
vertex  or  propagator counterterm in
Eq.~\eqref{eq:UVM1} is proportional to the corresponding tree
diagram. The proportionality factor is equal and opposite for the two
types of counterterms, 
which is a direct consequence of the Ward identity \eqref{eq:oneLoopUVselfen}.
Therefore, in the integral, vertex and
propagator counterterms cancel pairwise.  Since there is one more
vertex than propagators in any such graph, we deduce a very simple
result for the integral of the UV counterterms.  They are proportional to
the tree-amplitude,
\begin{equation}
\label{eq:M1UV}
e^2\, \takelimit_{\text{UV}}  M^{(1)}
\equiv e^2\, \int \frac{d^d l}{(2 \pi)^d} \takelimit_{\text{UV}} \mathcal M^{(1)}
= \tilde Z_1^{(1)} M^{(0)}.
\end{equation}
We will return to the role of the Ward identity in the renormalized amplitude below.
Equally simple is the integration of the UV counterterm of the form factor vertex, the integral of Eq.~\eqref{eq:FF1L_UV},
\begin{align}
\label{eq:M1UVIR}
 e^2\, \takelimit_{\text{UV}} \takelimit_{\text{IR}}  M^{(1)} &
\equiv e^2\, \int \frac{d^dl}{(2 \pi)^{d}}\,  \takelimit_{\text{UV}} \takelimit_{\text{IR}}  
\mathcal M^{(1)}(l)  \nonumber \\
&= \frac{\tilde Z_1^{(1)}}{4 (1-\epsilon)^2} \,
 \bar{v}(p_2)\,\gamma^\mu\, \gamma^\nu \, \mathbf P_1\widetilde M^{(0)} \mathbf P_1\,\gamma_\nu\, \gamma_\mu \,u(p_1)\,.
\end{align}
Meanwhile, the integration of the IR counterterm yields
\begin{align}
\label{eq:M1IR}
\takelimit_\text{IR}\, M^{(1)} &\equiv \int \frac{d^dl}{(2 \pi)^{d}}\
\mathcal F^{(1)}(l)  \\                   
&=
-\frac{1}{(4\pi)^{\frac{d}{2}}}\frac{\Gamma(\e) \Gamma(1-\e)^2}{\Gamma(2-2\e)}\,(-s)^{-\e} 
 \left( \left [ \frac{2}{\e} \, +\, \frac{\e}{(1-\e)} \right]  \, \bar v(p_2) {\mathbf P_1  \widetilde M^{(0)} \mathbf P_1} u(p_1) \right. \nonumber \\
 &\left. \hspace{5mm} - \frac{1}{4(1-\e)}\,
 \bar{v}(p_2) \gamma^\mu \gamma^\nu  \, \mathbf P_1  \widetilde M^{(0)}  \mathbf P_1 \, \gamma_\nu \gamma_\mu u(p_1) \right )\, ,
\nonumber
\end{align}
with $s=(p_1+p_2)^2$.
Combining this result with the ultraviolet counterterm of Eq.~(\ref{eq:M1UV}) and the UV-IR combination counterterm
of Eq.~(\ref{eq:M1UVIR}) to get the full singular contribution to the amplitude in Eq.~(\ref{eq:M1singular}), we find that 
the divergent part of the bare one-loop amplitude is
\begin{align} 
\label{eq:M1singularFinal}
M^{(1)}_{\rm singular} &= 
 -\frac{1}{(4\pi)^{\frac{d}{2}}}\frac{1}{\Gamma(1-\e)} \,(-s)^{-\e} 
 \nonumber \\
 & \quad \times
\bigg\{
\left(\frac{2}{\e^2} + \frac{3}{\e} +\ln\left(\frac{M^2}{-s}\right) + 11 \right)\,  \bar v(p_2) {\mathbf P_1  \widetilde M^{(0)} \mathbf P_1} u(p_1) 
\nonumber \\
&-\frac{1}{4}\left(\ln\left(\frac{M^2}{-s}\right) + 3\right)\,
 \bar{v}(p_2) \gamma^\mu \gamma^\nu  \, \mathbf P_1  \widetilde M^{(0)}  \mathbf P_1 \, \gamma_\nu \gamma_\mu u(p_1)
\bigg\}
+\mathcal O(\e) 
\,.
\end{align}
We note that 
\begin{equation}
\bar v(p_2) {\mathbf P_1
  \widetilde M^{(0)} 
\mathbf P_1} u(p_1) = \bar v(p_2) {
  \widetilde M^{(0)} 
} u(p_1) = M^{(0)}, 
\end{equation}
and the divergent part as $\epsilon \to 0$ is proportional to the
tree-amplitude in agreement with previous expectations~\cite{Catani:1998bh,Sterman:2002qn}.

\section{Two-loop diagrams with fermion loops}
\label{sec:fermionloop.tex}

In the previous section, we exploited factorization to construct local infrared and
ultraviolet counterterms for one-loop amplitudes.  We are now ready to develop counterterms that achieve the same goal
for  two-loop amplitudes.   The result will be a set of integrands that are locally 
integrable in exactly $d=4$ dimensions. 
%and which converge at infinity.   
As mentioned above, to
achieve integrability it will be necessary to develop a modified, but analytically equivalent, 
integrand for this set of diagrams with respect to expressions
obtained by the direct application of Feynman rules.   
The finite remainders found from our modified integrands will satisfy Eq.\ (\ref{eq:M2-fin}), 
thus implementing factorization at the level of the integrand.

In this section, we begin the treatment of two-loop diagrams with the study of those
 diagrams with a fermion loop.  As we shall see, the IR structure of these
 diagrams is closely related to that of the one-loop diagrams treated above.  Nevertheless, we will 
 encounter several obstacles to the construction of local infrared counterterms.   In these diagrams, such issues will involve integrals over the fermion loops.

Once integrated, fermion loops enjoy features associated with the Ward identities of QED, such
as the transversality of the vacuum polarization, and the UV-finiteness of light-by-light scattering.
These are routinely invoked in treatments of the renormalization and IR structure of the full amplitude.
Here, we seek to implement these features on a local basis in momentum space.  

As we shall see, the integrands of diagrams with a one-loop photon vacuum polarization
subgraph exhibit power singularities in addition to logarithmic singularities.  They also give rise to anomalous
polarizations (``loop polarizations" below) that spoil the local factorization in collinear limits that we found at one loop.   
These problems can be sidestepped at the cost of giving up manifest locality,
if we integrate first over the vacuum polarization loop.
This, however, is just what we want to avoid doing here. 
In addition, local collinear factorization is not manifest for individual Feynman diagrams
with four-point or higher fermion loop subgraphs, which, while
they do not contribute to collinear divergences after integration, exhibit
collinear singularities locally.
 
We will solve these problems by  deriving
alternative forms for the amplitude integrands, for which the 
power-counting of singularities is canonical, and whose properties in
collinear regions are locally consistent with factorization.  
For these modified amplitude integrands, we will be able to design
local counterterms that cancel all infrared and ultraviolet singularities, 
point by point in the integration domain.  

\subsection{Diagrams with one-loop photon vacuum polarization subgraphs}

\label{subsec:selfenergy}

\begin{figure}[h]
  \begin{center}
\includegraphics[width=0.3\textwidth]{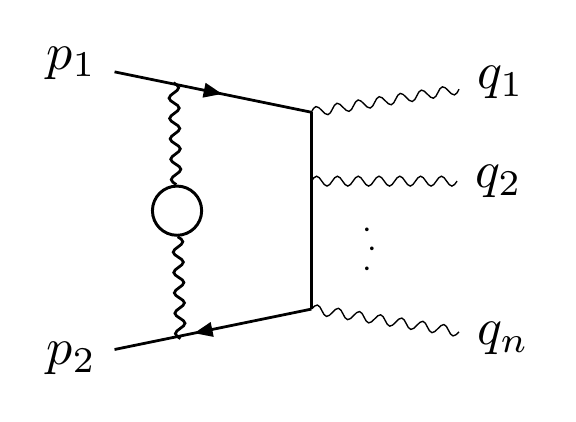}
\includegraphics[width=0.3\textwidth]{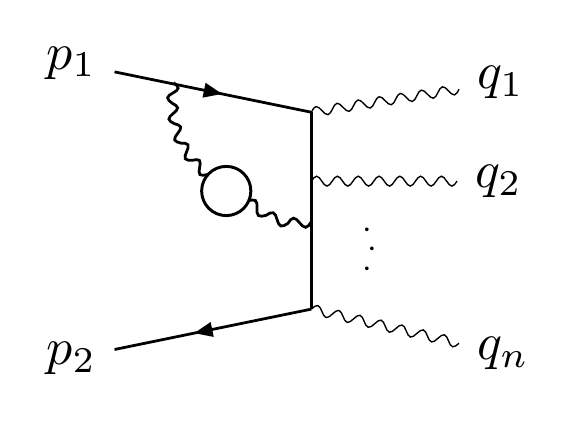}
\includegraphics[width=0.3\textwidth]{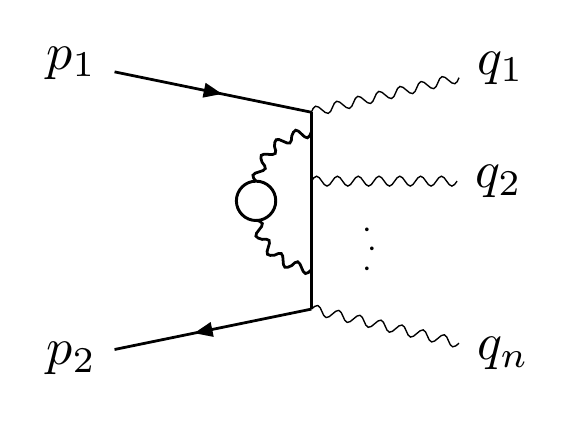}
\caption{ \label{fig:two-loop-fermionic-0} A class of diagrams that
  contribute to ${\cal M}^{(2)}_2$ with a photon vacuum polarization subgraph.} 
\end{center}
\end{figure}

Recalling the notation of Eq.\ (\ref{eq:M2-fermion-loop}), we start
with ${\cal M}^{(2)}_2$, consisting of two-loop diagrams with
a photon vacuum polarization subgraph, 
as in Fig.~\ref{fig:two-loop-fermionic-0}. 
The integrand ${\cal M}^{(2)}_2$ can be generated from the diagrams of
the one-loop amplitude in Eqs.~(\ref{eq:M1Adiags}) - (\ref{eq:M1Cdiags}),
by replacing the free photon propagator with its one-loop correction,
\begin{equation}
  \label{eq:photonprop1L}
  \frac{-i g_{\mu \nu}}{ l^2} \ \to\ 
 \eqs[.3]{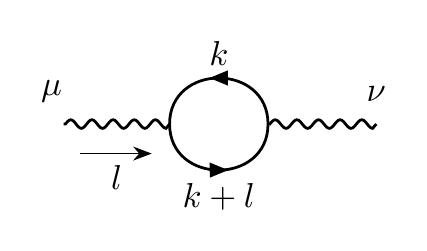} \equiv  \frac{-i }{ l^2} \left( -i \Pi_{\mu \nu }(l) \right)  \frac{-i }{  l^2}\, , 
 \end{equation}
 with
 \begin{equation}
-i \Pi_{\mu \nu }(l) = \ -\, e^2\, \frac{{\rm tr}\left[ \left(\slashed k + \slashed l
    \right)  \gamma_\mu \,  \slashed k \, \gamma_\nu  \right] }{k^2 (k+l)^2} \, .
\end{equation}
We observe that in the single soft limit, $l^\mu \sim  \delta \to 0$, $\mu=0,\dots 3$, at fixed, non-lightlike $k$,
the one-loop photon propagator integrand  scales as  
\begin{equation}
\frac{-i }{ l^2} \left( -i \Pi_{\mu \nu }(l) \right)  \frac{-i }{
    l^2}  \sim \delta^{-4},
\end{equation}
which is more singular than at tree-level by two powers.
This leads to a power-like singularity for the two-loop amplitude in the limit of soft $l$,
and it is no longer possible to rely on the leading-power approximation to write down
a factorized counterterm analogous to Eq.~\eqref{eq:T-soft-1} of the one-loop case.
Similarly enhanced are collinear singularities. For example, in the
collinear limit $l \parallel p_1$ at generic values of the vacuum polarization loop momentum, the divergence in $l$ is linear
instead of logarithmic, simply because of the factor $(l^2)^{-2}$.    At each such point, the polarization of the photon $l$
may be proportional to $k^\mu$, which is arbitrary, and hence non-factorizing in general.  Hence, at the local
level, the two-loop integrand loses important features that we found at one loop.

\begin{figure}
  \centering
  \includegraphics[width=0.8\textwidth]{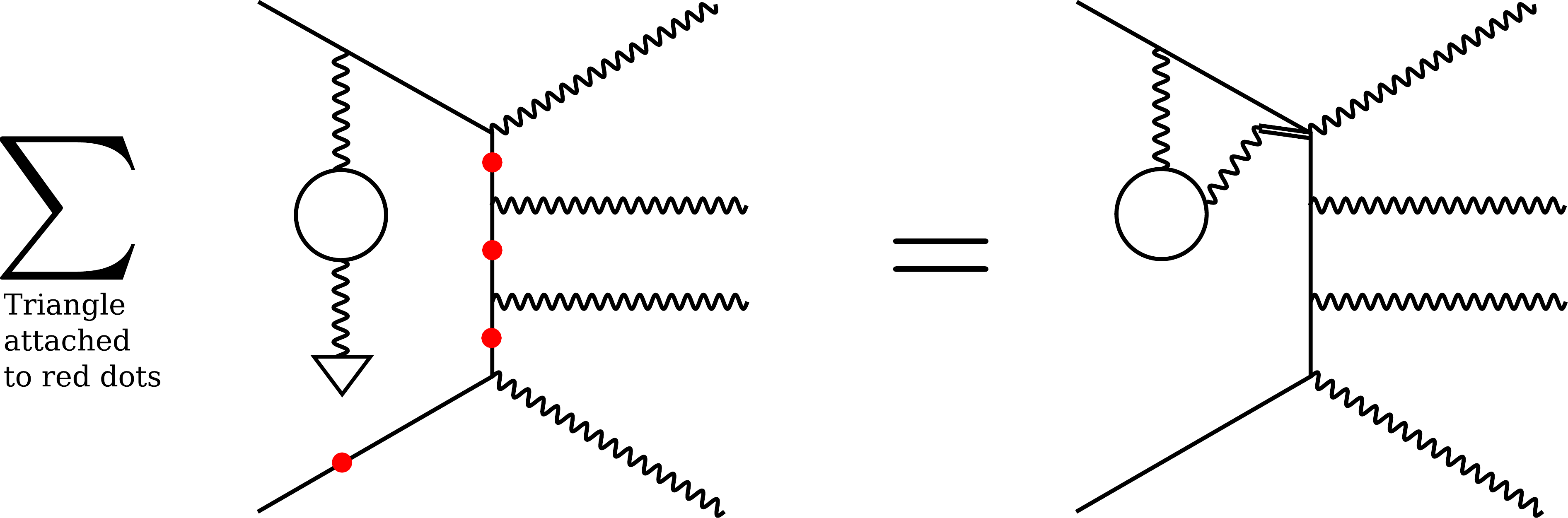}
  \caption{Ward identity for factoring out collinear divergences when $l \parallel p_1$, which relies on the leading-order approximation that the photon polarization is longitudinal or equivalently, scalar.}
  \label{fig:selfEnergyWardId}
\end{figure}

The soft and collinear non-factorizing and power-like singularities of the amplitude 
we have just identified, however,
are only apparent at the local, fully unintegrated level. For
example, after integrating over the fermion loop momentum $k$, a tree-like
$\delta^{-2}$ scaling is restored for the propagator in the soft limit, where all components of $l^\mu$ scale like $\delta$.
However, it is not necessary to integrate the fermion loop to reduce the degree of divergences in $l$ to the 
logarithmic level.   The vacuum polarization loop integral can be rewritten in a fully equivalent manner by, for example, 
Passarino-Veltman  tensor reduction~\cite{Passarino:1978jh}. 
After this reduction, the integrand for the
one-loop photon propagator reads,
\begin{equation}
   \label{eq:photonprop1L_tensor}
   \eqs{photonprop1L}
  = 
  \frac{1}{l^2} \frac{2 (d-2)}{d-1} 
 \, e^2\,
 \left( \eta_{\mu \nu }    - \frac{ l_\mu l_\nu }{l^2} \right)
\frac{1 }
  {k^2 (k+l)^2} \, .
\end{equation}
The result of the $k$ loop integral for this expression is precisely equivalent to the original form,
but in this expression the transversality of the photon is manifest.

The above form can be replaced by an even simpler expression, in
which logarithmic power counting is also regained.
The same gauge-invariant amplitude will be found if the $l_\mu l_\nu/l^2$ term from the
 vacuum polarization is
dropped in every diagram in which it appears. 
In practice, by repeated use of the same Ward identity as in Eq.\ (\ref{eq:WI}), it is easy to show that when
all diagrams are combined these terms produce a fully scaleless integral.
Therefore, the longitudinal tensor gives a zero
contribution to the amplitude integral and it can be
dropped from the propagator.\footnote{Scaleless integrals vanish in dimensional regularization due to cancellation of UV and IR poles. Dropping scaleless integrals mixes UV and IR poles, but this is not a problem because the universal structure of UV and IR divergences of gauge theory amplitudes are known.} After doing so, the tensorial form of the
effective one-loop virtual propagator matches the one of the tree photon-propagator,
\begin{equation}
   \label{eq:photonprop1L_tensor_eff}
   \eqs{photonprop1L}
  \to 
  \frac{\eta_{\mu \nu } }{l^2} \frac{2 (d-2)}{d-1} \,
  \frac 1 {k^2 (k+l)^2} \, .
\end{equation}
Using this result, it is simple to  generate a
suitable integrand for ${\cal M}^{(2)}_2(k,l)$ from the corresponding
one-loop amplitude integrand $ {\cal M}^{(1)}(l) $ as 
\begin{eqnarray}
  \label{eq:M20integrand}
 {\cal M}^{(2)}_2(k,l) 
&=& 
i\, \frac{2 (d-2)}{d-1} 
 \frac{1 }{k^2 (k+l)^2} \, {\cal M}^{(1)}(l)  \, .
\end{eqnarray}
This is our first example of a modified, alternative two-loop integrand, whose integral
is equivalent to the original form, but whose infrared behavior is amenable to 
local subtraction and factorization as in Fig.~\ref{fig:selfEnergyWardId}.

Infrared singularities in Eq.~(\ref{eq:M20integrand}) are
approximated by the counterterm
\begin{eqnarray}
  \takelimit_{\rm IR} {\cal M}^{(2)}_2(k,l)\
  &=& \
i\, \frac{2 (d-2)}{d-1} 
 \, \frac{1 } {k^2 (k+l)^2}  \; 
\takelimit_{\rm IR} {\cal M}^{(1)}(l)
\nonumber\\[2mm]
&\equiv&\ {\cal F}^{(2)}_2(k,l)\, ,
\end{eqnarray}
where the ``form factor'' diagram $\takelimit_{\rm IR} {\cal
  M}^{(1)}(l) = {\cal F}^{(1)}(l)$  is given by Eq.~(\ref{eq:oneLoopFFCT}). 
  The second identify defines the modified two-loop form factor subtraction
  for the contributions of this set of diagrams to the integrable two-loop integrand in Eq.\ (\ref{eq:M2-fin}).
  This modified form factor subtraction would emerge from exactly the same reasoning 
  as above applied to the vacuum polarization in its diagrams.   

We now turn our attention to ultraviolet divergences, subtracting 
the UV subdivergences of $\mathcal M^{(2)}_2 - \takelimit_{\rm IR} \mathcal M^{(2)}_2$
as $k \to \infty$ or $l \to \infty$, as suggested in Eq.\ (\ref{eq:M-UV-fin-def}).
 Instead of
giving separate but straightforward derivations of the UV counterterms, it is equivalent in this case simply
to present the following full remainder after the subtraction, whose integrals are 
free of infrared and ultraviolet singularities, 
\begin{eqnarray}
\label{eq:M22}
{\cal M}^{(2)}_{2,\,\rm finite} 
&=& 
i\, \frac{2 (d-2)}{d-1} 
 \, 
\left[ 
\frac{1 }
  {k^2 (k+l)^2}
- 
\frac{1 }
  {\left( k^2-M^2\right)^2}
\right] 
{\cal M}^{(1)}_{\rm finite}(l)\, .
\end{eqnarray}
Here, the finite one-loop result is specified by Eqs.\ (\ref{eq:M1separate}) and (\ref{eq:M1singular}),
\begin{equation}
\label{eq:M1fin}
{\cal M}^{(1)}_{\rm finite}(l) = {\cal M}^{(1)}(l) 
- \takelimit_{\rm UV} {\cal M}^{(1)}(l) 
- \takelimit_{\rm IR} {\cal M}^{(1)}(l)
+ \takelimit_{\rm UV} \takelimit_{\rm IR} {\cal M}^{(1)}(l)\, .
\end{equation}
Notice that ${\cal M}^{(2)}_{2,\,\rm finite}$ in Eq.\ (\ref{eq:M22}) is not singular in the
simultaneous $k,l \to \infty$ limit, so it is not necessary to
subtract an additional counterterm for the ``double-UV'' limit.
In Eq.\ (\ref{eq:M1fin}), the one-loop IR subtraction produces an integrand
that vanishes like a power when $l^\mu$ or $l^2$ vanishes, and thus controls logarithmic enhancements in momentum $l^\mu$
from IR limits of the vacuum polarization integral over $k$.   A straightforward expansion of the terms in Eq.\ (\ref{eq:M22}), using (\ref{eq:M1fin}),
shows that it is precisely of the form anticipated in Eq.\ (\ref{eq:M2-fin}) in terms of our modified amplitude and form factor integrands.

In summary, we have decomposed the ${\cal M}^{(2)}_2$
coefficient to the two-loop amplitude integrand into a remainder  that can 
be integrated directly in four dimensions,  and a simple singular part 
that can be integrated  in $d=4-2\epsilon$ dimensions,     
\begin{equation}
{\cal M}^{(2)}_2 ={\cal M}^{(2)}_{2,\,\rm  finite}+{\cal M}^{(2)}_{2,\,\rm singular}  \, ,
\end{equation}
with ${\cal M}^{(2)}_{2,\,\rm  finite}$ given by Eqs.\ (\ref{eq:M22}) and (\ref{eq:M1fin}).

\subsection{Other fermion-loop contributions}

\begin{figure}[h]
\begin{center}
  \includegraphics[width=0.5\textwidth]{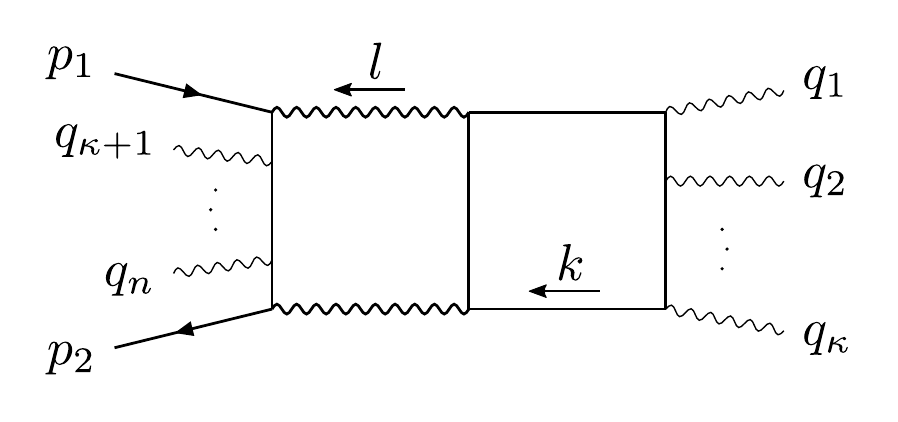}
  \caption{ \label{fig:two-loop-fermionic-k} The class of diagrams that contributes to the ${\cal M}^{(2)}_{\kappa+2},  \kappa>0$ two-loop amplitude coefficient.}
\end{center}
\end{figure}

We will now treat the remaining classes of fermion-loop diagrams, with 
$c > 2$, where  $\kappa \equiv c-2$ final-state photons are emitted from the fermion loop, as in Fig.~\ref{fig:two-loop-fermionic-k}. The loop momentum $l$ is always chosen to be the photon momentum flowing to a vertex adjacent to the $p_1$ electron, and the other loop momentum $k$ is always chosen to be one of the fermion lines in the fermion box.\footnote{It is not important which one of the four lines in the box is chosen as $k$, and the choice does not have to be aligned between different diagrams.}
Due to Furry's theorem, amplitude coefficients for odd values of $\kappa$ vanish.  We do not consider diagrams in which a single virtual
photon decays into an odd number of final-state photons ($\kappa=c-1$), because the IR divergences of this set of diagrams are identical to those of
a one-loop diagram with the emission of $n-\kappa+1$ photons, at least one of which is off-shell.

\begin{figure}[h]
\begin{center}
  \includegraphics[width=0.6\textwidth]{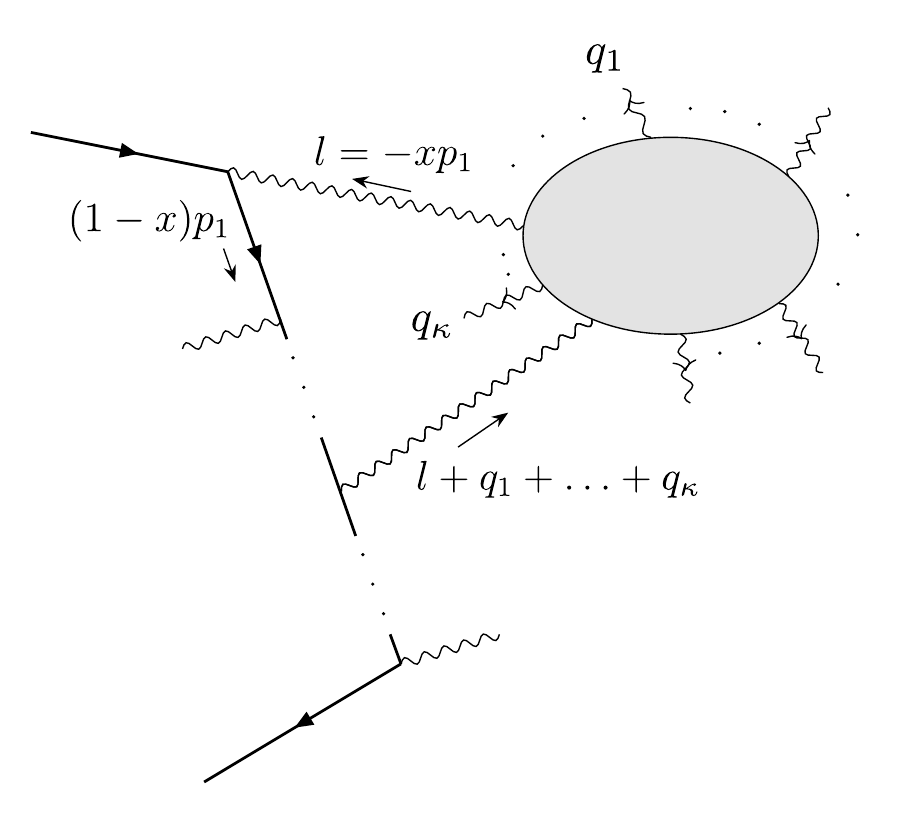}
  \caption{ \label{fig:collinearsing_floop} Illustration of a single collinear
    singularity in graphs with fermionic loops as $l$ becomes parallel
    to $p_1$. The gray blob denotes all permutations of the real and virtual photons
    attached to the fermion loop.
} 
\end{center}
\end{figure}  
Some diagrams that contribute to $M^{(2)}_{\kappa+2},\,  \kappa>0$, illustrated by 
Fig.~\ref{fig:collinearsing_floop}, develop collinear singularities
when a virtual photon that is attached at one end to the incoming electron or positron and at the other to
the fermion loop becomes parallel to the incoming fermion.  
The functional form of all such diagrams that contribute to the
collinear singularity from $l \parallel p_1$ is  
\begin{eqnarray}
\label{eq:Mcfermionloop}
{\cal M}^{(2)}_{\kappa+2} &=& \frac{1}{l^2
  (p_1+l)^2}  \bar{v} (p_2) A_{\nu}(q_{\kappa+1}, \ldots, q_n)
\left(\slashed p_1 +\slashed l \right)
\gamma_\mu u(p_1)
\nonumber \\ 
&& \times 
{\cal M}_{\rm photons}^{\mu \nu} \left(q_1,\, \ldots,\,
  q_\kappa,\, l,\, -\sum_{j=1}^{\kappa}q_j-l \right) .
\end{eqnarray}
In the collinear limit $l \approx - x p_1$, the last few factors times the Dirac spinor $u(p_1)$ become
\begin{equation}
  \left(\slashed p_1 +\slashed l \right) \gamma_\mu u(p_1)\ \approx \ 2(1-x) p_{1 \mu} \, u(p_1).
\end{equation}
Since this  is proportional to $p_{1\mu}$, we can apply the approximation of Eqs.~\eqref{eq:collprojection2} and \eqref{eq:triangleNotation} with $q$ replaced by $l$ to rewrite Eq.~\eqref{eq:Mcfermionloop} as
\begin{eqnarray}
{\cal M}^{(2)}_{\kappa+2} \Big|_{l \parallel p_1} & \to & \frac{1}{l^2
  (p_1+l)^2}  \bar{v} (p_2) A_{\nu}(q_{\kappa+1}, \ldots, q_n)
\left(\slashed p_1 +\slashed l \right)
\, \frac{2 \slashed \eta_1} {(l+\eta_1)^2 - \eta_1^2} u(p_1)
\nonumber \\ 
&& \times \ l_\mu\,
{\cal M}_{\rm photons}^{\mu \nu} \left(q_1,\, \ldots,\,
  q_\kappa,\, l,\, -\sum_{j=1}^{\kappa}q_j-l \right)  .
\end{eqnarray}
The one-loop $(\kappa+2)$-photon amplitude ${\cal M}^{\mu\nu}$ can be simplified further with the use of Ward identities. The two virtual photons that attach to these loops have momenta $l$ and $-l-\sum_{i=1}^\kappa q_i$ flowing out of the loops. Of the two, the photon with momentum $l$ is scalar-polarized. The action of the Ward identity is illustrated below for the case of two photons in the final state. We have, in the notation of Fig.~\ref{fig:WI} and Eq.~\eqref{eq:WI}, 
\begingroup
\allowdisplaybreaks
\begin{align}
&\frac{2 \slashed \eta_1} {(l+\eta_1)^2 - \eta_1^2} l_\mu {\cal M}_{\rm photons}^{\mu \nu} \left(k, q_1, q_2, l, \tilde l \right) \nonumber \\
& \hspace{0.5cm} =  \eqs[0.25]{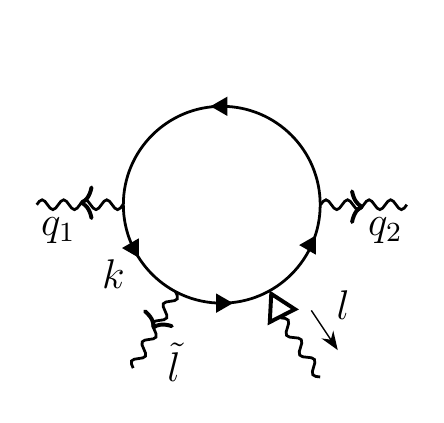} + \eqs[0.25]{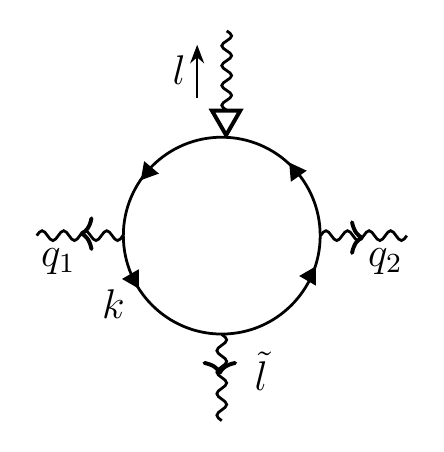} + \eqs[0.25]{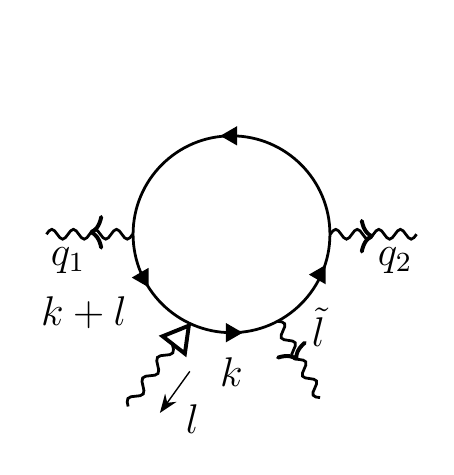} 
\nonumber \\ 
& \hspace{0.5 cm} + \mbox{  symmetric terms} 
\nonumber \\
& \hspace{0.5cm}
= e  \left( \eqs[0.25]{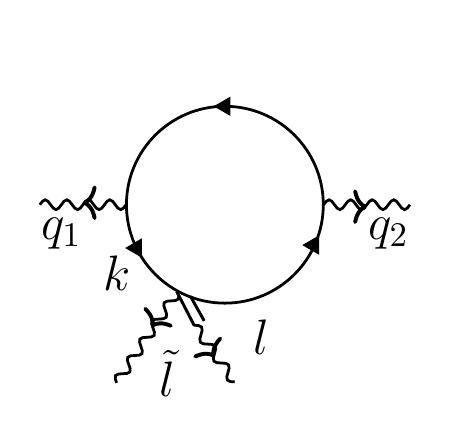} - {\eqs[0.25]{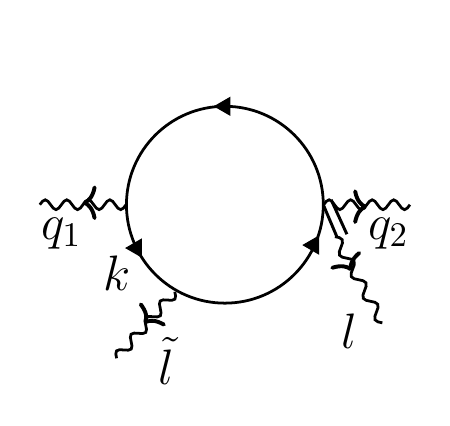}} \right) \nonumber \\
& \hspace{0.5cm}
+ e \left( \eqs[0.25]{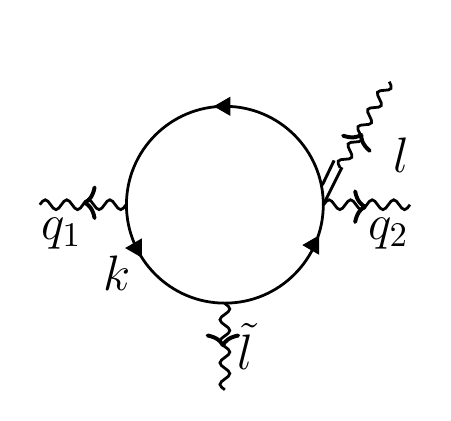} - \eqs[0.25]{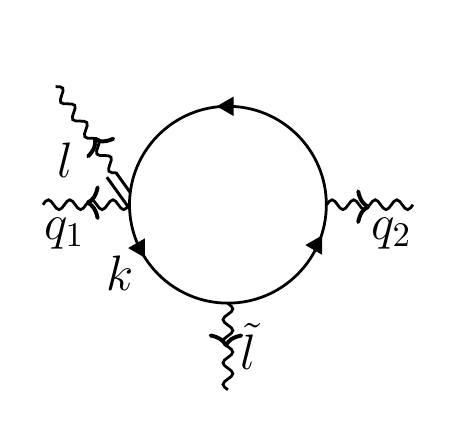} \right)
\nonumber \\
&  \hspace{0.5cm}
+ e \left( \eqs[0.25]{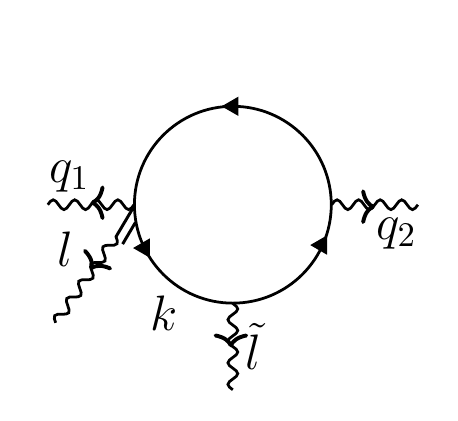} - \eqs[0.25]{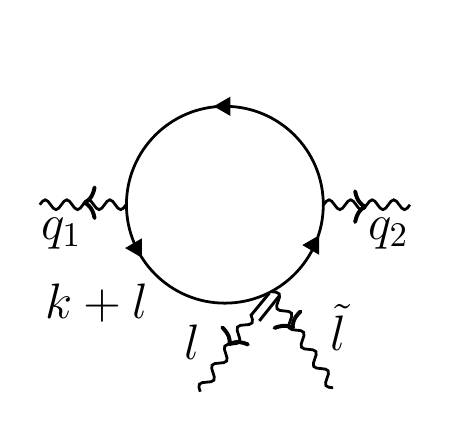} \right)
+ \mbox{  symmetric terms} 
\nonumber \\ 
& \hspace{0.5cm}
 = e \left( \eqs[0.25]{LbL2p_RH} - \eqs[0.25]{LbL3m_RH} \right)
+ \mbox{  symmetric terms} \, .
\label{eq:WI-loop}
\end{align}
\endgroup
We have used the shorthand notation $\tilde l=-l-q_1-q_2$, while ``symmetric terms'' denotes diagrams
with other permutations of external photons $q_i$.
In the final equality, we find the difference of integrands for the same 
three-photon amplitude evaluated at two different values of the loop
momentum, 
\begin{eqnarray}
\frac{2 \slashed \eta_1} {(l+\eta_1)^2 - \eta_1^2} l_\mu {\cal M}_{\rm photons}^{\mu \nu} \left(k, q_1, q_2, l, \tilde l
\right)  &=& \frac{2 \slashed \eta_1} {(l+\eta_1)^2 - \eta_1^2} \left[  
{\cal M}_{\rm photons}^\nu  \left( k, q_1, q_2, -q_1-q_2\right)
 \right. \nonumber \\ 
&&
\left. 
- {\cal M}_{\rm photons}^\nu \left( k+l, q_1, q_2, -q_1-q_2\right)
\right]\, ,
\end{eqnarray}
where ${\cal M}_{\rm photons}^\nu$ is defined in Eq.\ (\ref{eq:WI-loop}).
This difference integrates to zero in dimensional
regularization, which respects momentum-shift invariance. Therefore
\begin{equation}
\label{eq:WIfloop}
\int \frac{d^d k} {(2\pi)^d} \frac{2 \slashed \eta_1} {(l+\eta_1)^2 - \eta_1^2} \, l_\mu {\cal M}_{\rm photons}^{\mu \nu} \left(k, q_1, q_2,\ldots , l,
  \tilde l \right) = 0.  
\end{equation}
We again encounter a situation, however, in which integration over a loop momentum removes singularities
which are present locally, and need to be removed for local integrability.

As for the vacuum polarization diagrams above, we will achieve the removal of local collinear singularities  by a
modification of the integrand, 
adding to it terms that vanish upon integration,  as follows.   Again, gauge invariance will make this possible.
Without changing the integrated value of the amplitude, we can 
exploit Eq.~(\ref{eq:WIfloop}) and modify the integrand of Eq.~(\ref{eq:Mcfermionloop}) as, 
\begin{eqnarray}
\label{eq:Mcfermionloop_mod}
{\cal M}^{(2)}_{\kappa+2} &=& \frac{1}{l^2
  (p_1+l)^2}  \bar{v} (p_2) A_{\nu}(q_{\kappa+1}, \ldots, q_n)
\left(\slashed p_1 +\slashed l \right)
{\hat{\gamma}_\mu\left(l, \eta \right)} u(p_1)  
\nonumber \\ 
&& \times 
{\cal M}_{\rm photons}^{\mu \nu} \left(q_1, \ldots,
  q_\kappa, l, -\sum_{j=1}^{\kappa}q_j-l \right) \, .
\end{eqnarray}
Here, we have modified the vertex adjacent to the incoming electron line, replacing $\gamma_\mu$ by
\begin{equation}
   \label{eq:gamma-hat}
\hat{\gamma}_\mu\left(l, \eta_1 \right) 
\equiv 
\gamma_\mu - l_\mu \frac{2 \slashed{\eta}_1}{(l+\eta_1)^2 - \eta_1^2}\, , \qquad
\eta_1^2 \neq 0\, ,
\end{equation}
where $\eta_1$ is chosen so that it produces no new pinches from the extra denominator.
In the collinear limit, $l= -x p_1$, we have 
\begin{equation}
\left(\slashed p_1 +\slashed l \right)
{\hat{\gamma}_\mu\left(l, \eta_1 \right)} u(p_1)  
\xrightarrow{l = -x p_1}  0. 
\end{equation}
An analogous modification will be made for diagrams which develop a singularity when a photon is collinear 
to the positron.   For diagrams which can 
develop both types of collinear singularities, we perform these
modifications simultaneously. We write, 
\begin{align}
&\eqs[0.6]{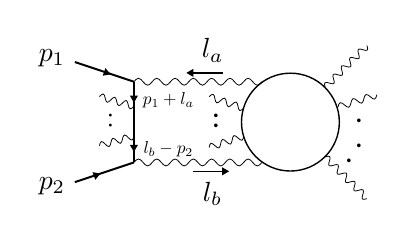}  \nonumber \\
&\hspace{0.5cm}= \overline{v}(p_2) \hat{\gamma}_\nu(l_b, -\eta_2) 
\left( \slashed{l}_b - \slashed{p}_2 \right) 
\ldots
\left( \slashed{p}_1 + \slashed{l}_a \right) 
\hat{\gamma}_\mu(l_a, \eta_1) \,
{\cal M}_{\rm photons}^{\mu \nu}\,.
\end{align}
The above expression integrates to the same value as with
$\hat{\gamma}_\mu \to \gamma_\mu$, and  is free of collinear
singularities. We have chosen the reference vector for the $p_2$-collinear
divergence to be $-\eta_2$, in accordance with Eq.~\eqref{eq:triangleNotation2}.
We note that in this case, there is no corresponding modification to a
form factor subtraction term in Eq.\ (\ref{eq:M2-fin}).   The modified 
contributions are now infrared-integrable on their own, without further subtraction.

In summary, we have modified the integrand of ${\cal M}_{\kappa+2}^{(2)}$, to make each of its diagrams
regular in all its collinear limits, without changing its integral.   This was done with a simple
modification of the fermion-photon vertex when it occurs adjacent to an
incoming leg. This class of diagrams has no further infrared divergences. 

\subsection{UV counterterms for $c=4$}
\label{sec:c=4}

In QED,  the fermion loops in ${\cal M}^{(2)}_{c}$ for $c = 4$ are UV finite after
summing over diagrams that permute the photons connected to the loop, but only after the fermion loop
integral has been carried out.    The complete result must be finite, simply because there is no gauge-invariant local four-photon vertex.
Nevertheless, individual diagrams for $c=4$  (light-by-light scattering) are UV divergent 
in $d=4$. 
To preserve local convergence, we must construct  additional counterterms for these diagrams.   
Analogous counterterms were developed for QCD in Ref.\ \cite{Nagy:2003qn}.  Unlike QCD, however,
for QED such counterterms have no counterpart in the renormalization of the theory.  

\begin{figure}
  \centering
  \includegraphics[width=0.6\textwidth]{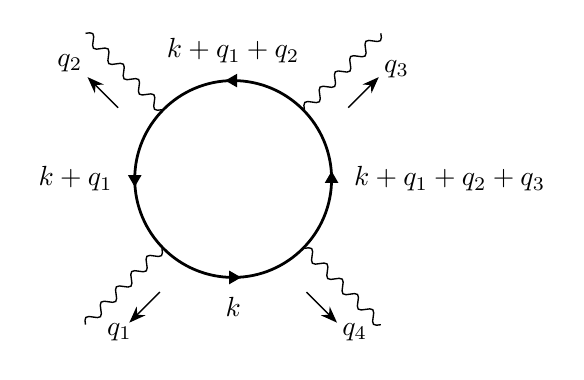}
  \caption{The four-photon subdiagram with a fermion loop, which has unphysical UV divergences diagram by diagram.}
  \label{fig:fourPhoton}
\end{figure}

The general form of a four-photon subdiagram is
\begin{align}
\label{eq:c=4-1234}
  &\quad V^{\mu_1 \mu_2 \mu_3 \mu_4} (q_1, q_2, q_3, -q_1-q_2-q_3)\nonumber \\
  &= e^4 \frac
  {(-1) \operatorname{tr}[\slashed k \gamma^{\mu_1} (\slashed k + \slashed q_1) \gamma^{\mu_2} (\slashed k + \slashed q_1 + \slashed q_2) \gamma^{\mu_3} (\slashed k + \slashed q_1 + \slashed q_2 + \slashed q_3) \gamma^{\mu_4}]} {k^2 (k+q_1)^2 (k+q_1+q_2)^2 (k+q_1+q_2+q_3)^2}\, ,
\end{align}
as illustrated in Fig.~\ref{fig:fourPhoton}.
In the UV limit, we write down the counterterm just as for the vacuum polarization in Eq.\  (\ref{eq:M22}), by dropping all external momenta 
and IR-regulating denominators with a mass $M$,
\begin{equation}
\label{eq:c4-ctr-1234}
\takelimit_{\rm UV} V^{\mu_1 \mu_2 \mu_3 \mu_4} = -e^4 \frac{{\rm tr}
\left( 
\slashed{k} \gamma^{\mu_1}
\slashed{k} \gamma^{\mu_2}
\slashed{k} \gamma^{\mu_3}
\slashed{k} \gamma^{\mu_4}
\right)
}{
\left( k^2-M^2 \right)^4
} \, .
\end{equation}
The counterterm in this expression corresponds to the specific subdiagram of 
Eq.\ (\ref{eq:c=4-1234}).   There are six such counterterms, corresponding to 
the number of non-cyclic permutations of indices, and hence to distinguishable attachments of the two virtual and two real photons to the loop.
Our UV-convergent contribution to ${\cal M}_{4,\,\rm finite}^{(2)}$ is just the sum of these six distinguishable
diagrams, each minus its counterterm.   Each one
of these combinations specifies a locally integrable two-loop integral, which is 
also UV convergent.   

\begin{figure}
  \centering
  \includegraphics[width=0.7\textwidth]{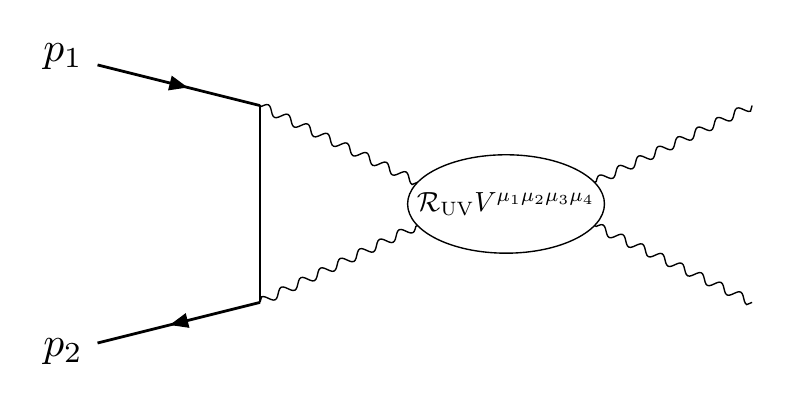}
  \caption{The four-photon UV counterterm inserted into a one-loop diagram.}
  \label{fig:fourPhotonVertexInsertion}
\end{figure}

The corresponding contribution to  ${\cal M}_{4,\,\rm singular}^{(2)}$ is just the sum of the six counterterms,
Eq.\ (\ref{eq:c4-ctr-1234}) plus permutations.   Their sum can be combined into a single, local vertex 
embedded in the one-loop diagram in Fig.~\ref{fig:fourPhotonVertexInsertion}.
Using tensor reduction, we obtain for
this sum, which is necessarily symmetric in the indices $\mu_1 \dots \mu_4$,
an integrand that is proportional to $\epsilon=2-d/2$,
\begin{eqnarray}
&& 
\sum_{\rm permutations} \takelimit_{\rm UV} V^{\mu_1 \mu_2 \mu_3 \mu_4} = 
8 e^4 \frac{\epsilon (1-\epsilon)}{(2-\epsilon) (3-\epsilon)} 
\frac{\left(k^2\right)^2}{
\left( k^2-M^2 \right)^4
} \nonumber \\ 
&&
\hspace{40mm}\times 
\left[ 
\eta^{\mu_1 \mu_2} \eta^{\mu_3 \mu_4}
+ 
\eta^{\mu_1 \mu_3} \eta^{\mu_2 \mu_4}
+ 
\eta^{\mu_1 \mu_4} \eta^{\mu_2 \mu_3}
\right]\, .
\end{eqnarray}
 The integral of loop momentum $k$ has a pole, so that the sum of counterterms is finite in $d=4$, 
\begin{align}
\sum_{\rm permutations} \int \frac{d^d k}{(2 \pi)^d}  
\takelimit_{\rm UV} V^{\mu_1 \mu_2 \mu_3 \mu_4}
=\frac{i e^4}{12 \pi^2} \left[ 
\eta^{\mu_1 \mu_2} \eta^{\mu_3 \mu_4}
+ 
\eta^{\mu_1 \mu_3} \eta^{\mu_2 \mu_4}
+ 
\eta^{\mu_1 \mu_4} \eta^{\mu_2 \mu_3}
\right]\, .
\end{align}
As noted above, this finite result does not correspond to a counterterm of QED renormalization.  In fact,
it is not by itself gauge invariant, simply because it does not vanish when index $\mu_i$ is
contracted with the corresponding photon's momentum.   This is not a problem, however,
because the integral of the diagram in Fig.\ \ref{fig:fourPhotonVertexInsertion} 
is locally IR finite and UV convergent.   Figure \ref{fig:fourPhotonVertexInsertion} simply produces finite
shifts in ${\cal M}^{(2)}_{4,\,\rm finite}$ and ${\cal M}^{(2)}_{4,\,\rm singular}$, with the same magnitudes but with opposite
signs.   Its only effect is to enable us to express  ${\cal M}^{(2)}_{4,\,\rm finite}$ as
a sum of locally integrable and UV convergent diagrams.

To summarize the results of this section, we have identified integrable 
expressions for the sets of two-loop diagrams with fermion loops, $\{{\cal M}^{(2)}_c\}$.  For those with
a vacuum polarization, ${\cal M}^{(2)}_2$, the integrable remainder is found by subtracting singular terms that are closely
related to one-loop IR and UV singularities.    The remainder is given in convenient form by Eq.\ (\ref{eq:M22}), which is equivalent
to the subtraction specified in Eq.\ (\ref{eq:M2-fin}), in terms of the one-loop integrable remainder function. 
 The sums of diagrams with  $c>2$, with final-state photons attached to the
fermion loop, are finite, even though individual diagrams have collinear singularities, and in the case $c=4$, ultraviolet
divergences.   For all such diagrams, modifications of the integrands themselves, as in Eq.\ (\ref{eq:Mcfermionloop_mod}), lead to expressions that are locally integrable on a 
diagram-by-diagram basis, and which taken together give the original
integral.  For the $c=4$ light-by-light case, the introduction of a
set of UV counterterms whose sum is finite  provides a convergent
integrand, while giving the same finite result as in the original 
form after integration.   In the following section, we turn to the
remaining two-loop diagrams, without fermion loops.

\section{Two-loop photonic contributions}
\label{sec:photonic.tex}
\newcommand{\colorboxed}[1]{\colorlet{saved}{.} \color{red}
  \boxed{\textcolor{saved}{#1}}}

\def\figscale#1#2{\epsfxsize=#2\epsfbox{#1}}
\def\nn {\nonumber}
\def \psla{\rlap{$p$}{/}}
\def \ksla {\rlap{$k$}{/}}
\def \lsla {\rlap{l}{/}}
\def \qsla {\rlap{$q$}{/}}
\def \epsla {\rlap{$\epsilon$}{/}}
\def \Qsla {\rlap{$Q$}{/}}

In this section, we will describe the construction of subtraction
terms for the ultraviolet and infrared singularities in the ${\cal M}^{(2)}$
integrand that consists of two-loop Feynman diagrams with no fermion
loops. All of these diagrams will have two virtual photons.   As for the
fermion loop diagrams, the finite remainders found from our modified integrands will satisfy Eq.\ (\ref{eq:M2-fin}), 
again implementing factorization at the level of the integrand.

For the removal of infrared divergences, we will adopt the same
form factor approach as we used for the one-loop amplitude, where 
the local integrand of a $2 \to 1$ form factor 
with an appropriately defined vertex served as the infrared counterterm for the amplitude of
a generic $2 \to n$ process. 
For this approach to succeed, collinear photons must factorize locally from the off-shell subdiagram in
all singular limits, that is, prior to any integration.   

As we shall see, the factorization of a collinear photon in certain two loop diagrams is less obvious than the one loop, 
precisely because of the presence of the extra loop momentum.
The procedure we will describe below entails two steps: 
\begin{itemize}
\item first generating the two-loop integrand with an
appropriate routing of loop momenta,
\item and then replacing some terms
in the integrand with equivalent ones ({i.e.,} which integrate to the same
value) that are locally integrable in collinear singular limits.
\end{itemize}
The following  three subsections detail and summarize these steps, which will bring the infrared singularities of this set of diagrams under control. 

Once we have derived a locally integrable integrand for the two-loop amplitude,
we will  still need two sets of ultraviolet counterterms, just as we did at one loop.   The elements of the first set are
in one-to-one correspondence with the counterterms of the QED Lagrangian, but presented in integral form.
They are formulated so that, when added to the full integrand, they give a fully convergent integral in all UV limits of the original amplitude.
The second set of UV counterterms is required because 
form factor-based infrared counterterms induce additional UV divergences.   We emphasize again that the latter counterterms are a 
familiar feature of many factorization techniques.   Subsection \ref{subsec:UV-photon-2} will deal with the construction of
both sets of UV counterterms.   

The outcome of the construction described below will be a reorganized expression for the full two-loop photonic contributions to the amplitude.  It will be the sum of a set of ``simple" but singular integrals, which can be carried out in dimensional regularization for any of the $2 \to n$ processes in question, plus an integral that depends on all the details of the process, but which is numerically integrable in four dimensions. The number of IR and UV counterterms generated by this procedure will depend on the process, but will not be qualitatively different than for the original set of diagrams with their QED counterterms.

\subsection{Generation of the two-loop integrand}
\label{subsec:mod}

We construct the two-loop integrand starting from the usual
application of QED Feynman rules in the Feynman gauge.
This construction is not unique, given the invariance of
individual Feynman diagrams under shifts of the loop momenta. 
To combine all diagrams into a single integrand, we need to make an unambiguous choice of momentum labeling for every diagram.
As at one loop, we aim to label loop momenta in a way which allows the
application of Ward identities for a local cancellation of
non-factorizing collinear singularities.

We label the momenta of the two virtual photons with $l$ and
$k$, both directed away from the vertex closest to the
incoming positron. It is unimportant to specify which of the two photons carries
momentum $k$ and which carries momentum $l$, due to a symmetrization over momentum
routings, which we will discuss shortly.
This initial integrand is illustrated by,
\begingroup
\allowdisplaybreaks
\begin{eqnarray}
\label{eq:M2a_start}
{\cal M}^{(2)}_0(k, l) &=& 
\eqs[.25]{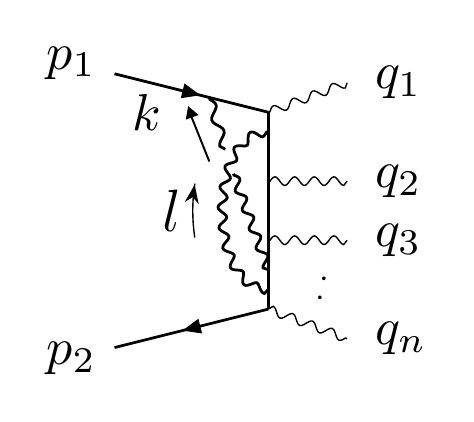} 
+\eqs[.25]{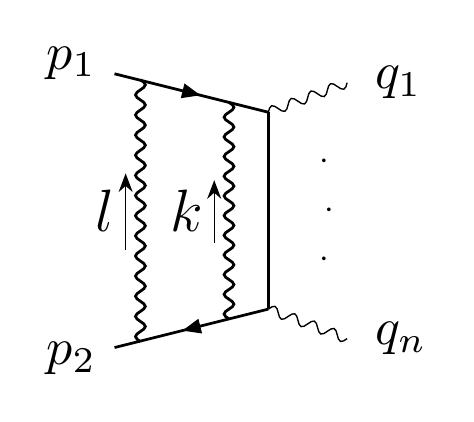} 
+\eqs[.25]{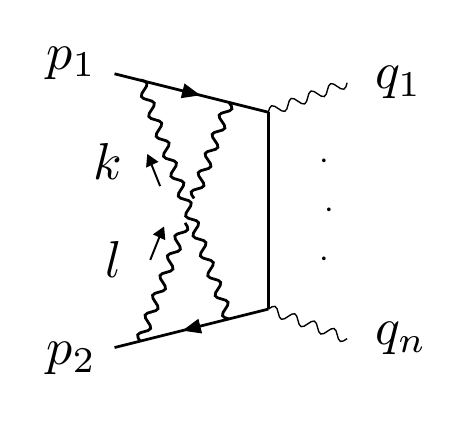} 
\nonumber \\
&& \hspace{-1.5cm}
+\eqs[.25]{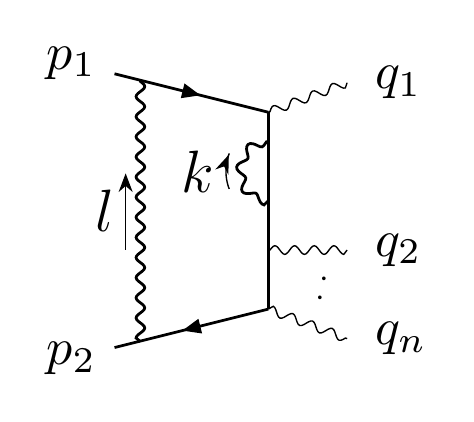} 
+\eqs[.25]{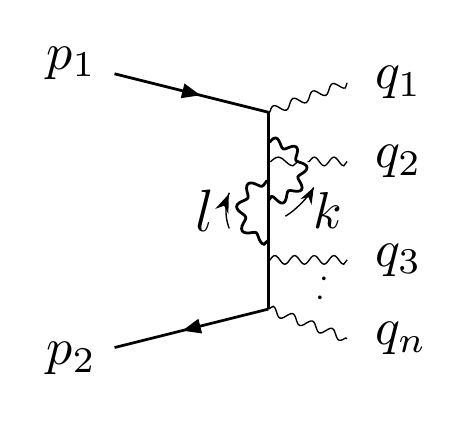} 
+\eqs[.25]{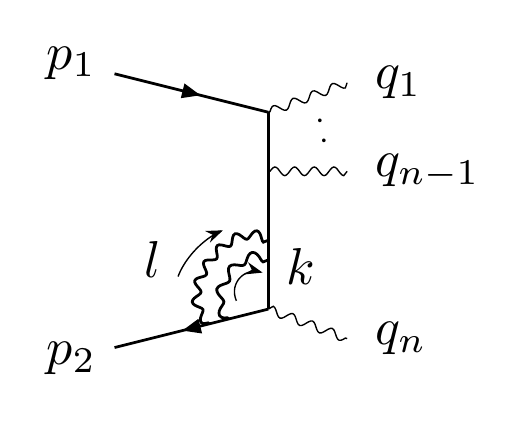}
\nonumber \\
&& \hspace{-1.5cm} 
+\eqs[.25]{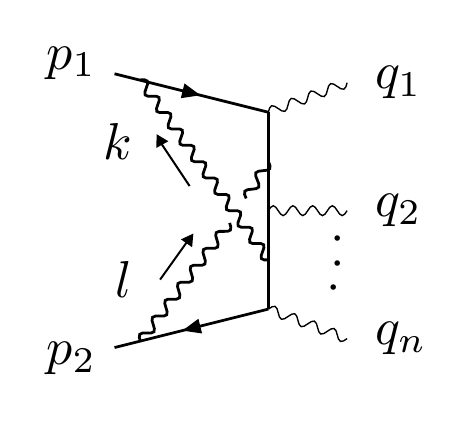} 
+\eqs[.25]{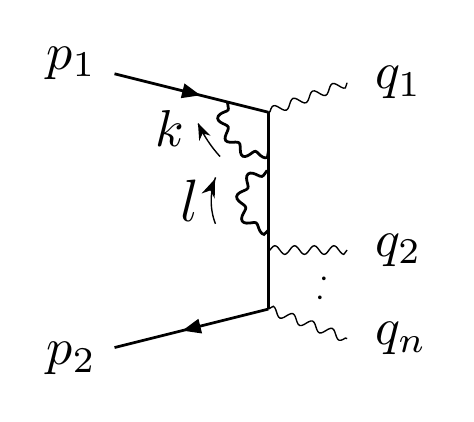} + \ldots\, .
\label{eq:integrand0}
\end{eqnarray} 
\endgroup
We will need to manipulate this form further, in order to produce a modified integrand, which we will denote as
\begin{equation}
  {\cal M}^{(2)} (k, l)\, ,
\end{equation}
which will factorize in all singular collinear limits.  Our manipulations
will include symmetrizations of the loop momenta, and modifications of integrands that do not change integrated results. We detail these manipulations in the
following subsections.  
 
\subsubsection{Global amplitude symmetrizations}
As we will soon explain, we need to perform the four-fold
symmetrization generated by
\begin{equation}
k\leftrightarrow l, \quad (k_\perp, l_\perp) \leftrightarrow (-k_\perp, -l_\perp), \label{eq:fourFoldSym}
\end{equation}
where $k_\perp$ and $l_\perp$ are the components of the loop momenta $k^\mu$ and $l^\mu$ that are orthogonal to both the incoming electron momentum, $p_1$, and the incoming positron momentum, $p_2$.
This symmetrization is applied globally, to all diagrams of the
amplitude.  The symmetrized integrand, 
\begin{eqnarray}
\label{eq:M2symm}
{\cal M}^{(2)}_{\rm sym}(k, l) &\equiv& 
\frac{1}{4} \left[ 
 {\cal M}_0^{(2)}(k, l)
+{\cal M}_0^{(2)}(l, k)
+ {\cal M}_0^{(2)}(k^*, l^*)
+{\cal M}_0^{(2)}(l^*, k^*) \right]\, ,
\end{eqnarray}
with $l^*=l(l_\perp \to -l_\perp)$, $k^*=k(k_\perp \to -k_\perp)$, 
exhibits more complete factorization properties than the individual terms that make it up
on the right-hand side.   Here, the notation ${\cal M}_0^{(2)}$ refers to the integrands
written down directly from the QED Lagrangian, which are symmetrized as shown to define ${\cal M}_{\rm sym}$. 

To see the need for the $k\leftrightarrow l$  symmetrization, let us
examine, as an  example, two-loop diagrams of the $e^-(p_1) +e^+(p_2)
\to \gamma^*(q_1) + \gamma^*(q_2)$
amplitude that are singular in two different
collinear limits.  First, we consider a loop momentum $k$ collinear to
the positron momentum $p_2$, with the second loop momentum $l$ in the
hard region. 
Analogous to the analysis of collinear singularities at one loop using Ward identities, a sum of diagrams which factorize in the collinear $k \parallel p_2$ singularity are,
\smallskip

\begin{align}
\label{eq:M2p2}
{\cal M}_0^{(2)}(k, l) &\xrightarrow{k \parallel p_2}  
\eqs[.17]{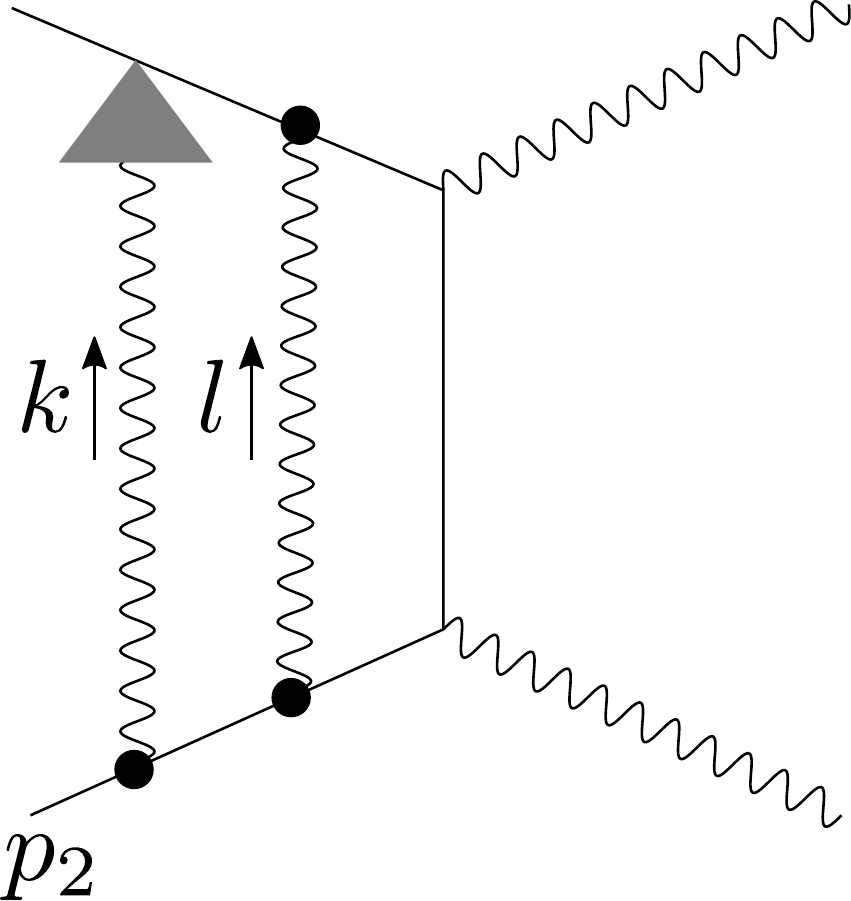}
+\eqs[.17]{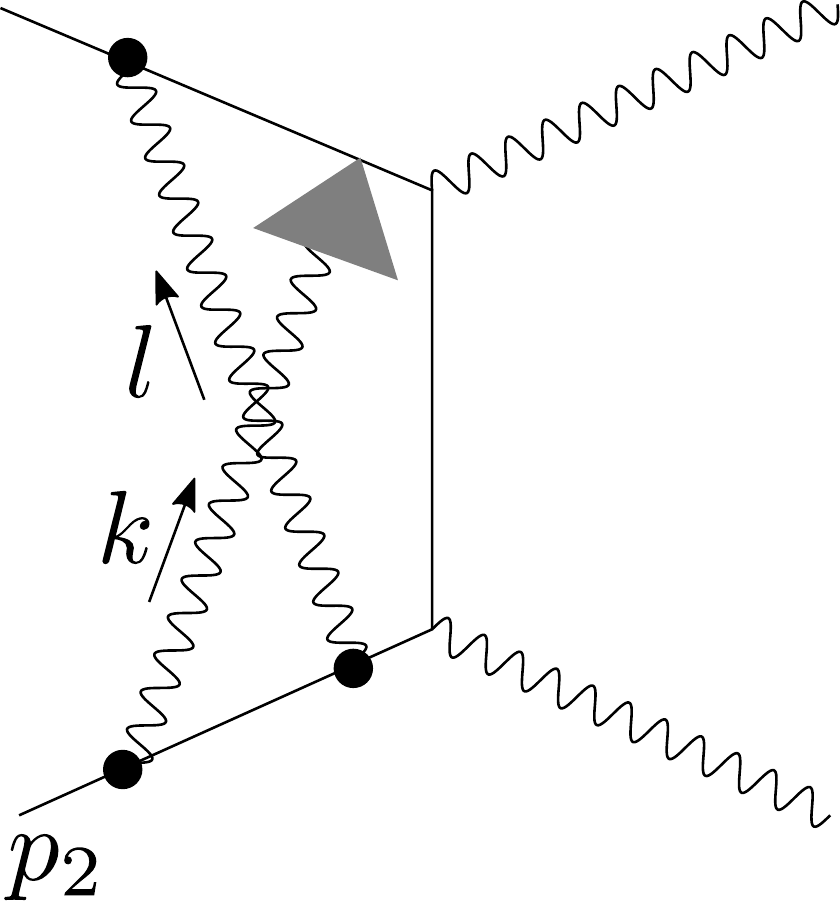}
+ \eqs[.17]{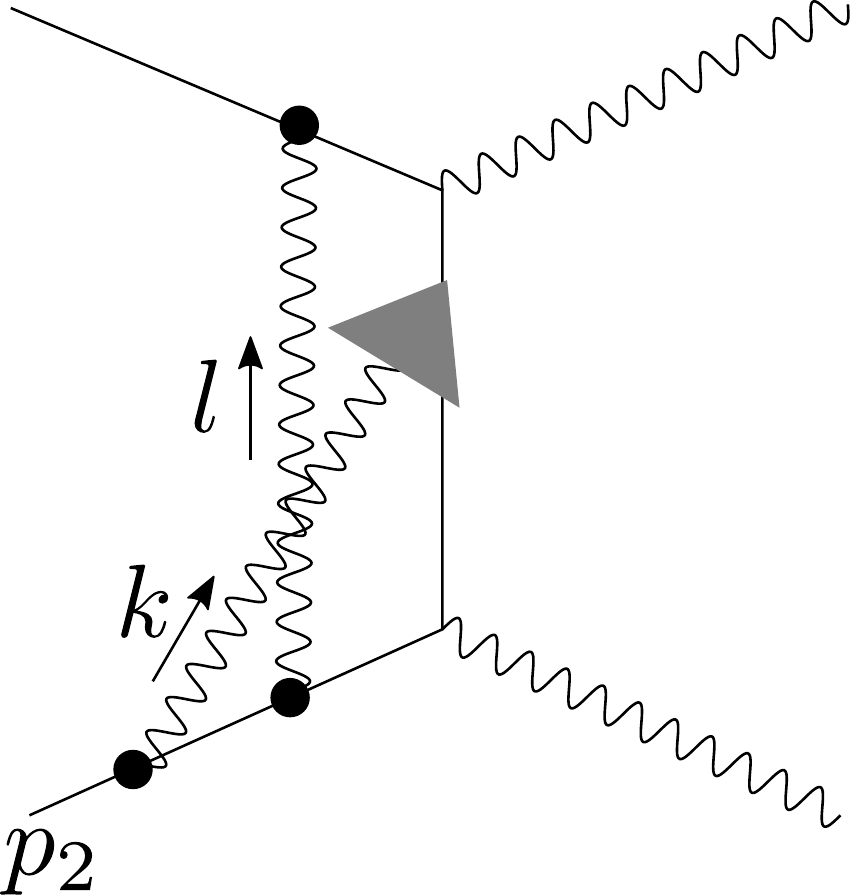}
+ \eqs[.17]{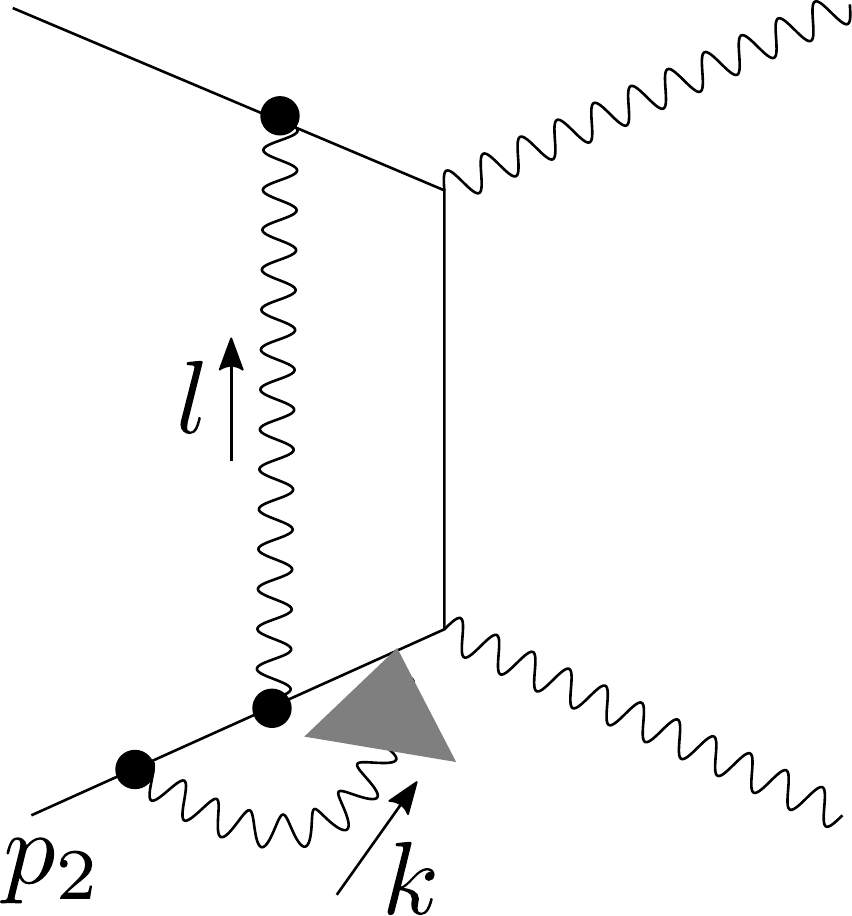} \nonumber\\
& \qquad + \dots \nonumber \\[1em]
&= \eqs[.17]{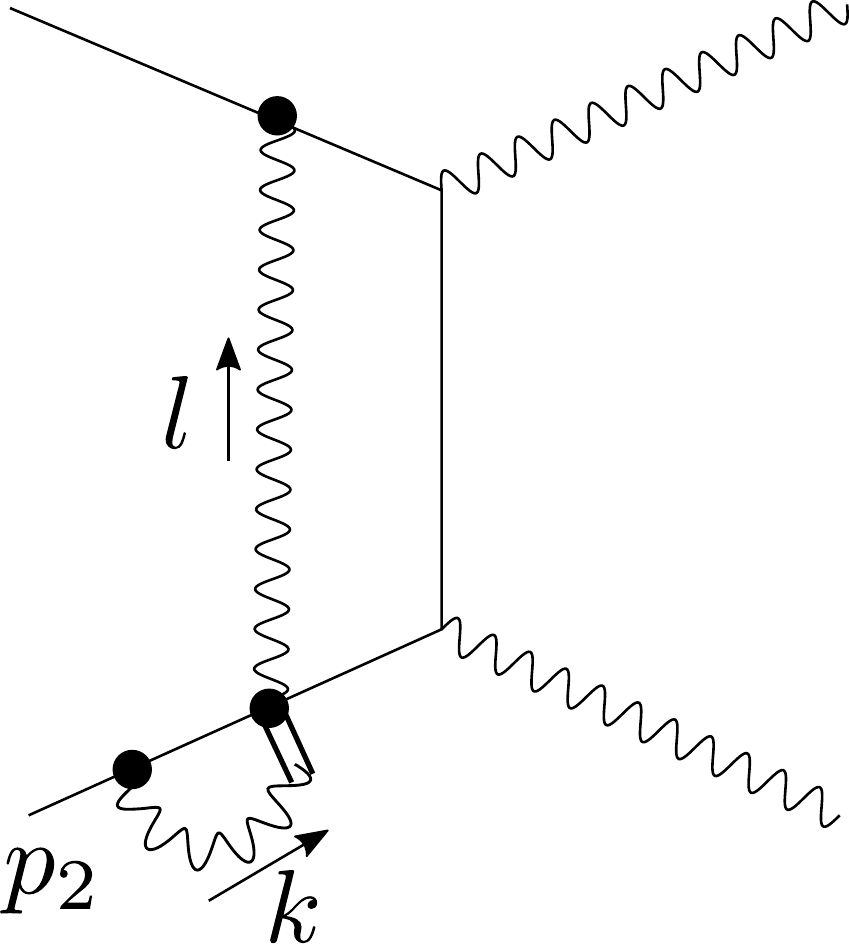} + \dots
\end{align}
In the above,  grey triangles denote the collinear approximation for photon propagators collinear to $p_2$, defined in Eq.~\eqref{eq:triangleNotation2}.
The identity follows, as at one loop, by successive application of the lowest order Ward identity, Eq.\ (\ref{eq:WI}).
In Eq.~\eqref{eq:M2p2}, we have included only a typical class of diagrams where the momentum $k$ is adjacent
to the collinear leg $p_2$ and attaches to all propagators of the same one-loop subgraph.
The dots denote omitted diagrams that can also be grouped into
similar classes. In the collinear limit that we consider here,  Ward
identities (leading to factorization) can be applied to each class independently.

Let's now look at  a distinct limit, where the photon of momentum $k$ becomes collinear to the incoming electron,
$k \parallel p_1$,
\begin{align}
\label{eq:M2p1}
{\cal M}_0^{(2)}(k, l) &\xrightarrow{k \parallel p_1}
\eqs[.17]{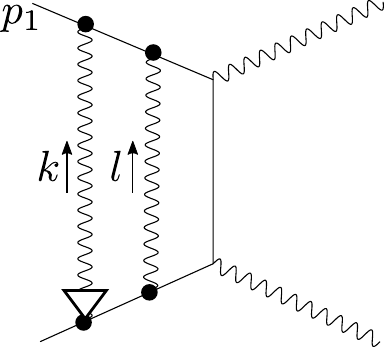}
+\eqs[.17]{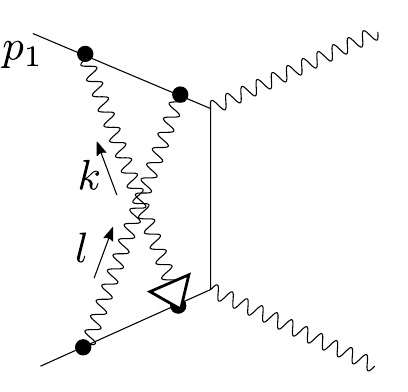}
+ \eqs[.17]{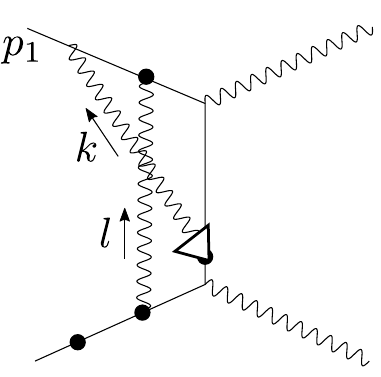}
+ \eqs[.17]{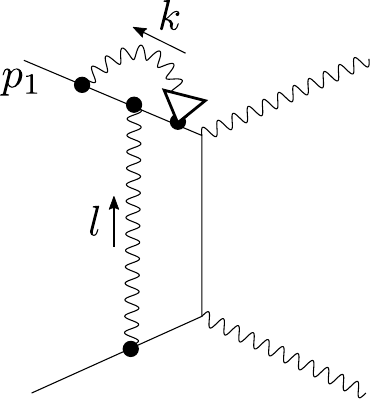}  
\nonumber \\
& \qquad + \dots \nonumber \\[1em]
&= \eqs[.17]{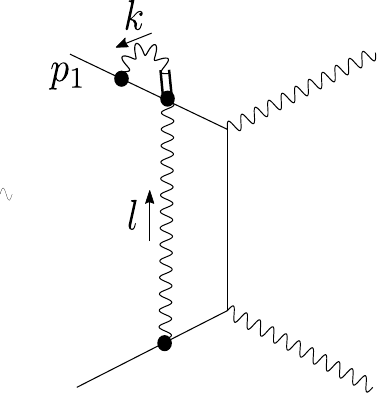} + \dots \, .
\end{align}
In Eq.~(\ref{eq:M2p1}), we show the analogous set of Feynman diagrams
that are needed as a sum for factorization, using white triangles to denote collinear approximations defined in Eq.~\eqref{eq:triangleNotation}. Once again, 
the collinear singularity occurs when the loop momentum $k$ is adjacent to the 
external leg, which carries momentum $p_1$ this time.  We now observe that
the second Feynman diagram on the right-hand
side of Eqs.~(\ref{eq:M2p1}) is actually also needed in Eq.~(\ref{eq:M2p2}), but
with the momenta $l$ and $k$ exchanged. If we make  a single 
assignment of loop momenta that is consistent with factorization of
the $k \parallel p_1$ limit, then  we will spoil the factorization of the
$k \parallel p_2$ limit, and vice versa.  
In effect, the assignment of momenta can be thought of as an
instruction for how to combine the integrands of the diagrams shown in these
figures.  With fixed assignments of line momenta as shown
in the figures, the association of the points in momentum space
where the Ward identity acts would not be automatic, but would require a change of variables.   This conflict is  resolved simply by
symmetrizing over the two momentum assignments, and both collinear limits 
factorize for the combination ${\cal M}_0^{(2)}(k, l) + {\cal M}_0^{(2)}(l, k)$ without a change of variables in momenta.  
Once the momenta are averaged in this manner, each momentum assignment will find
its matching assignment in every diagram. 

The additional reflections $(k_\perp,l_\perp) \leftrightarrow (-k_\perp, -l_\perp)$ in
the transverse plane in Eq.\ (\ref{eq:M2symm}) are required for the factorization of
collinear singularities in propagators which receive hard self energy
and vertex corrections. 
\begin{figure}[h!]
  \centering
  \subfloat[\label{fig:Is}]{\includegraphics[width=.3\linewidth]{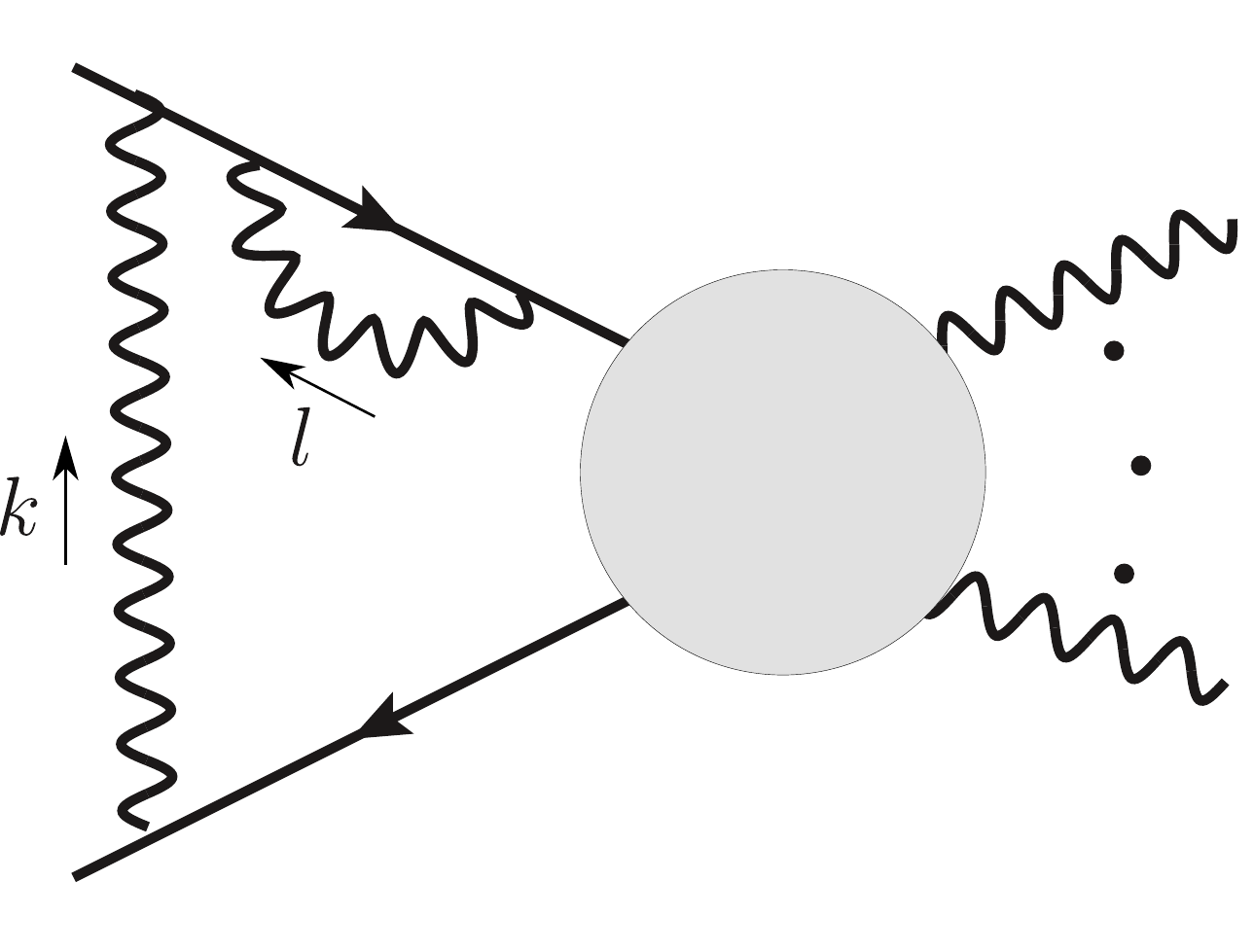}}
  \hspace{.1\linewidth}
  \subfloat[\label{fig:Ic}]{\includegraphics[width=.3\linewidth]{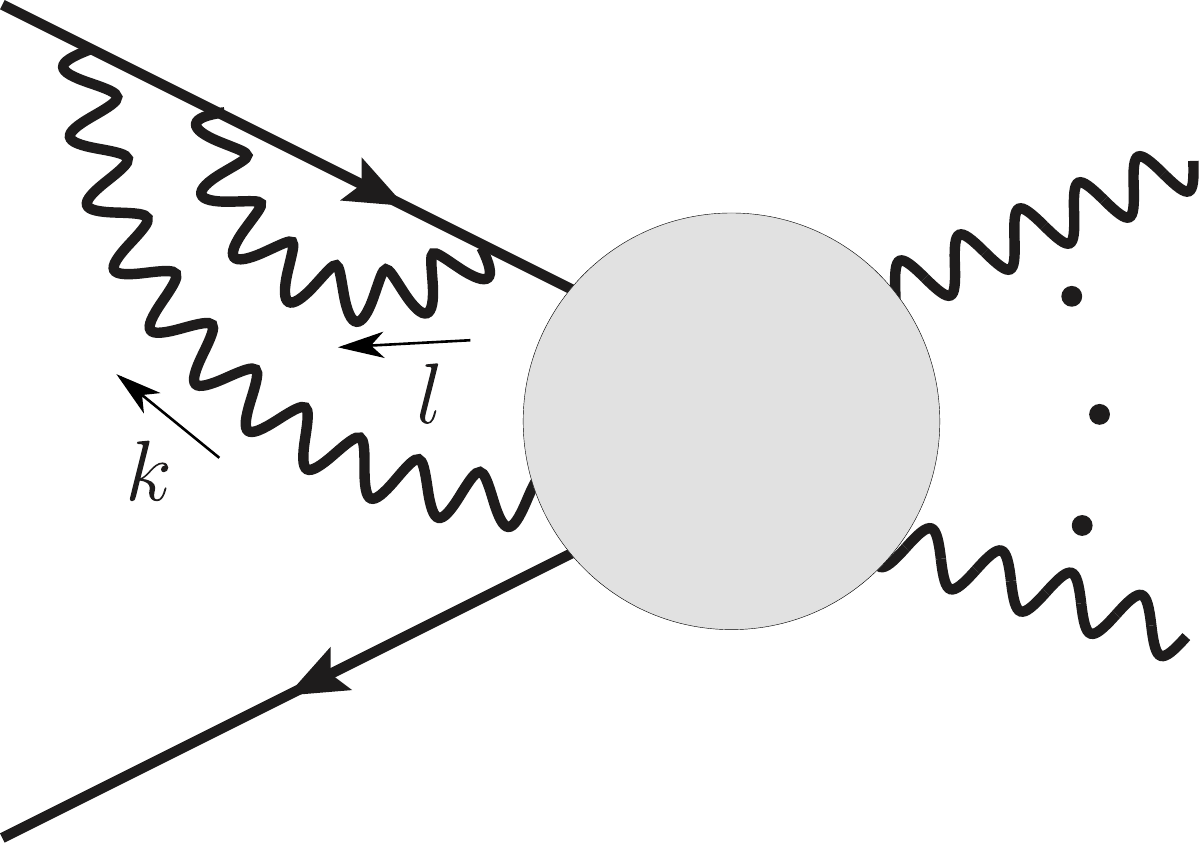}}\\
  \subfloat[\label{fig:IIs}]{\includegraphics[width=.3\linewidth]{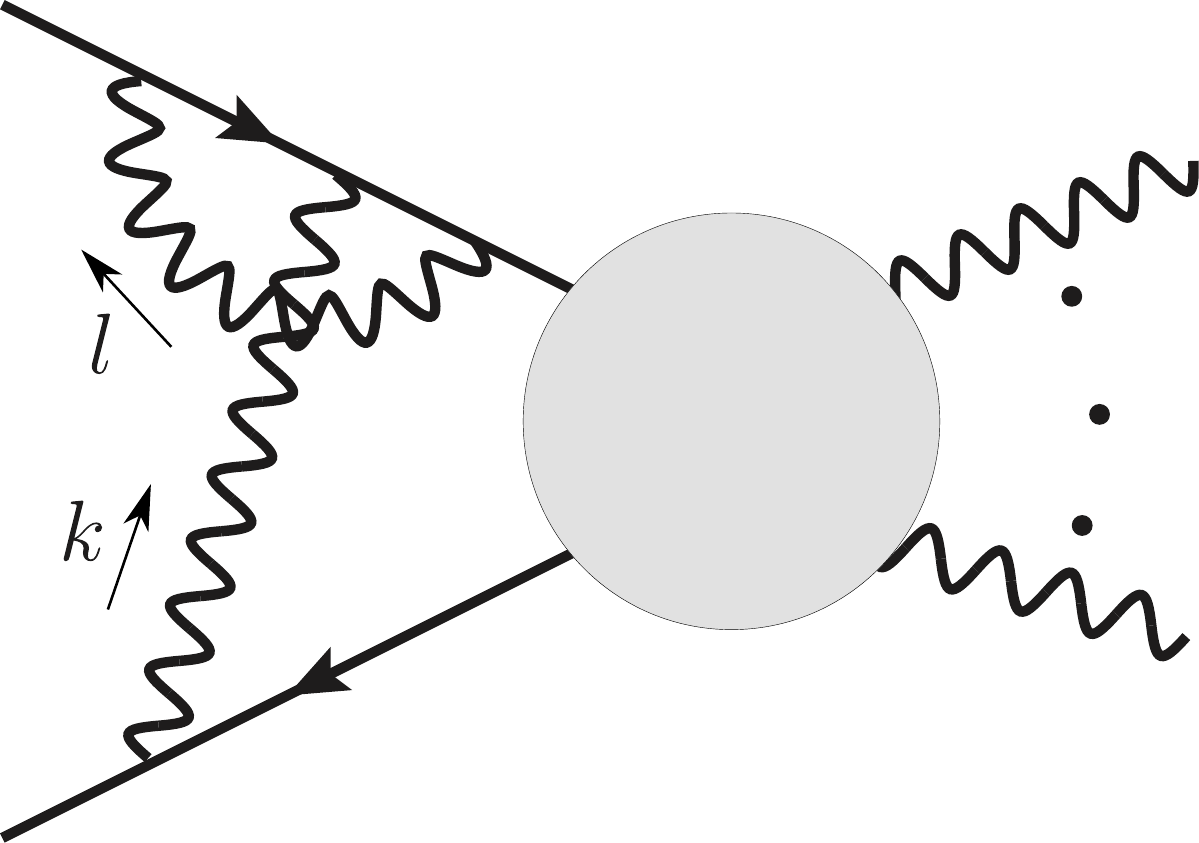}}
  \hspace{.1\linewidth}
  \subfloat[\label{fig:IIc}]{\includegraphics[width=.3\linewidth]{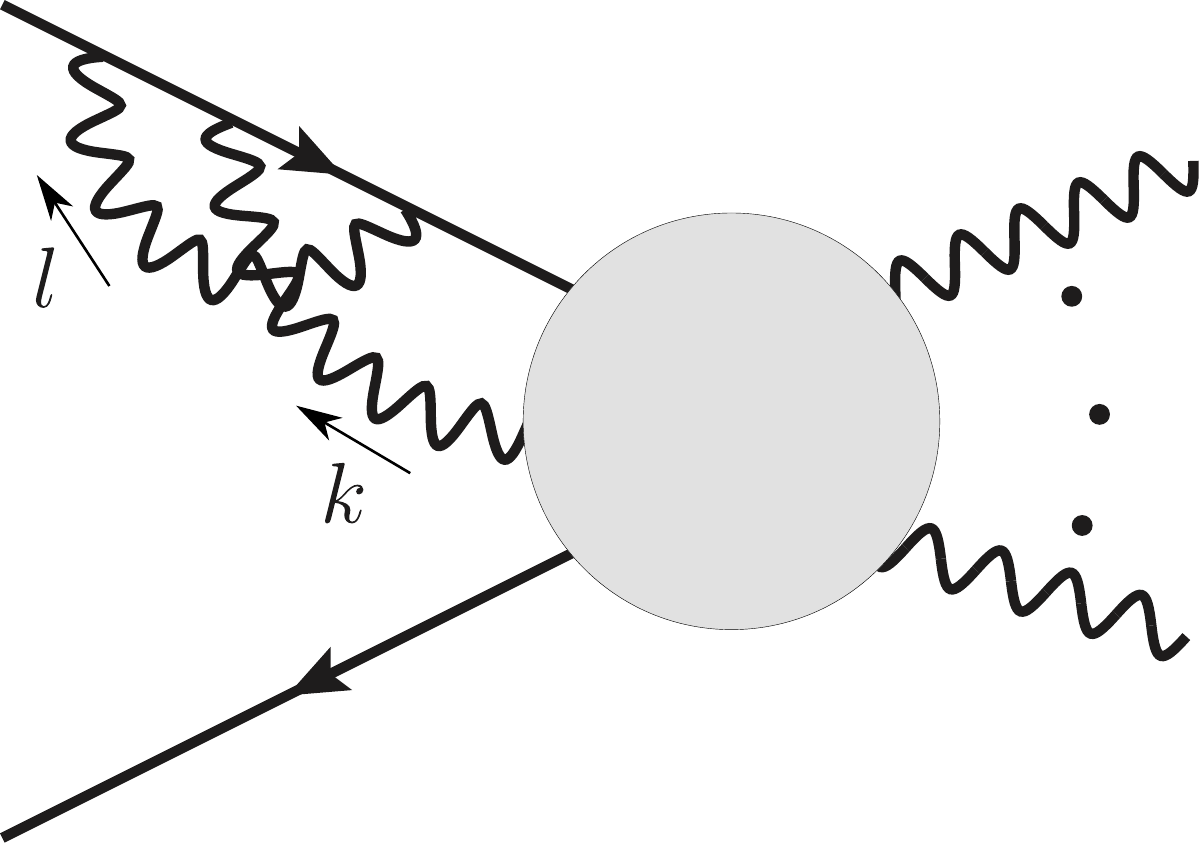}}
  \caption{Diagrams requiring special treatment. We refer to the
    diagrams of the first
    line as \typeI, and of the second line as \typeII\ . 
%\draftnote{Are the labels
%      $k$ and $l$ too small? There's a "2" floating around beneath the
%      k loop in fig. (a). Labels $p_1$ and $p_2$ are missing}
}
  \label{fig:typeInII}
\end{figure}
The diagrams in question are shown in Fig.~\ref{fig:typeInII}, for the case
when the self energy or vertex corrections are on the electron side. The case when
they appear on the positron side is completely analogous and is omitted from the
figure as well as much of the later discussions.   We will see below the role of 
reflections in the transverse planes when we study collinear limits involving these diagrams.

We will refer to the diagrams with an electron or positron self energy correction
outside the gray off-shell subdiagram, from which the final-state photons emerge, as \textbf{\typeI}, 
and the diagrams with a QED vertex correction outside the gray
photon emission blob as \textbf{\typeII}.
For concreteness, we will assign $l$ to be the momentum of the virtual photon of the
self energy or vertex sub-loop in \typeI\ and \typeII\ diagrams as in
figure \ref{fig:typeInII}. This choice is simply a convention, since the integrand will be
symmetrized according to Eq.~(\ref{eq:M2symm}) in the end. 
We call all the other diagrams \textbf{regular diagrams}, decomposing the two-loop integrand as a sum of contributions from these three classes,
\begin{equation}
\label{eq:M0-types}
  \mathcal M_{\rm sym}^{(2)}\ =\ \left [ \mathcal M^{(2)}_{\text{\typeI}} + \mathcal M^{(2)}_{\text{\typeII}} + \mathcal M^{(2)}_{\text{regular}} \right ]_{\rm sym}\, ,
\end{equation}
where ${\cal M}_{\rm sym}^{(2)}$ denotes the full initial 
 integrand with fixed momentum assignments and symmetrizations in Eq.\ (\ref{eq:fourFoldSym}) for each of the terms on the right.
We will not alter the integrand of $\mathcal M^{(2)}_{\text{regular}}$
beyond these momentum symmetrizations, while $\mathcal
M^{(2)}_{\text{\typeI}}$ and $ \mathcal M^{(2)}_{\text{\typeII}}$ will
require a somewhat more elaborate treatment. 

\subsubsection{Collinear regions in  \typeI\ and \typeII\ diagrams}

The diagrams in $\mathcal M^{(2)}_{\text{\typeI}}$ and  $\mathcal
M^{(2)}_{\text{\typeII}}$ of Fig.~\ref{fig:typeInII}, until altered, 
do not exhibit factorization in all infrared singular limits  locally.
In this discussion, we demonstrate that  local
factorization can be achieved by overcoming three challenges, as summarized below and then elaborated upon.

\begin{enumerate}
\item Stronger than logarithmic divergences due to doubled fermion propagators,  
\item apparent contributions of unphysical ``loop polarizations''  that are not compatible with factorization in collinear limits,
\item and, finally, scaleless integrals appearing as
  self energy corrections to external legs.
\end{enumerate}

\begin{figure}[h]
  \begin{center}
\includegraphics[width=0.75\textwidth]{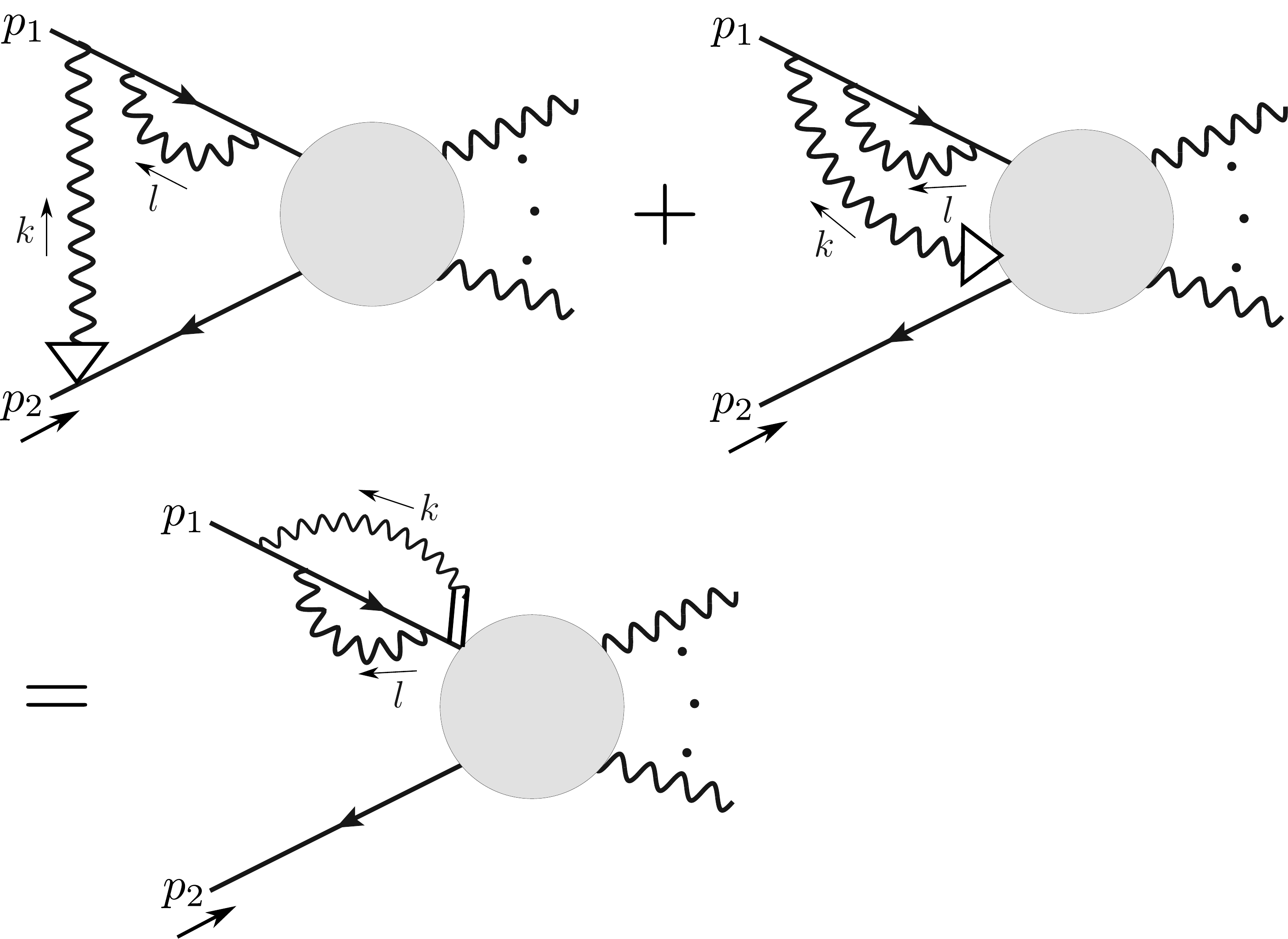}
\end{center}
  \caption{Illustration of the desired factorization property of
    \typeI\ diagrams in the collinear $k \parallel p_1$ limit. Prior
    to integration, the propagator denominator $(p_1+k)^2$ appears
    twice, giving rise to power divergences in the collinear
    limit. Appropriate modifications of the diagrams are needed to
    reduce the divergence to a logarithmic one, so that we can apply
    the collinear approximations, in particular the polarization
    approximations Eq.~\eqref{eq:triangleNotation} with $q=k$
    indicated by the triangle arrow.}
  \label{fig:typeIproblem}
\end{figure}

The first challenge, which we encountered for photon self-energies in the previous section, is apparent in diagrams of \typeI, where self energy corrections to
a fermionic propagator adjacent to an external electron or positron are inserted.
These diagrams possess single-soft $k \to 0$ and collinear $k \parallel p_1$   singularities, whose
strength is exacerbated by the presence of the self energy correction
and the double occurrence of a denominator~\footnote{A related discussion of
  double propagators can be found in Ref.~\cite{Baumeister:2019rmh}.}. 
A straightforward power counting shows power divergences, instead of tamer logarithmic divergences.
A naive subtraction of the infrared divergences would require counterterms for both the leading and next-to-leading singular terms, while the factorization of $k \parallel p_1$ collinear singularities  with the application of Ward identities, schematically illustrated in Fig.~\ref{fig:typeIproblem}, would apply only to the leading terms.
Of course, if we perform the integral over $l$, the amputated self energy diagram gives a result proportional to $\slashed p_1 + \slashed k$ which combines with the identical $\slashed p_1 + \slashed k$ from the numerator of the adjacent fermion propagator to cancel the extra $(p_1+k)^2$ denominator, but this would not achieve our goal of \emph{fully local} factorization.

The second challenge posed by ``loop polarizations'' emerges, for
example, in the  calculation of the collinear limit $k \to - x p_1$
in \typeII\ diagrams. Let us focus  on the vertex subgraph given by
Eq.~(\ref{eq:oneLoopUVvertex})  multiplied with a common fermion
propagator and a spinor factor,  as they appear in \typeII\ diagrams,
\begin{align}
  \label{eq:typeIIs10}
  &
\eqs[.35]{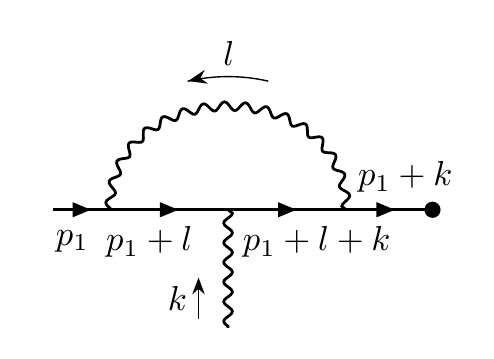}
  = -i \, e^3 \,  \frac{(\slashed{p}_1 + \slashed{k})\gamma^\mu (\slashed{p}_1+\slashed{l}+\slashed{k})\gamma^\nu(\slashed{p}_1+\slashed{l})\gamma_\mu u(p_1)}{(p_1+l+k)^2(p_1+l)^2l^2}
  \nonumber \\
& \xrightarrow{ k = - x p_1}  
                 -i\, e^3 \, 2\, (2-d) 
\frac{1+ \frac{2 k \cdot p_2}{s}}{
(p_1+l+k)^2
} \left[
\frac{l^\nu}{l^2}-
\frac{l^\nu+p_1^\nu}{(l+p_1)^2}
\right] u(p_1)\, .
\end{align}
The tensor structure of the subgraph in the collinear limit  is
crucial for factorization.
Analogous to the one-loop case in Eq.~\eqref{eq:collinearPolarization}, the term with the vector $p_1^\nu$
in the numerator corresponds  to a photon with a scalar
polarization in the collinear limit. This term, when all Feynman diagrams of \typeII\ are
summed and cancellations due to Ward identities are carried out, yields a factorized contribution in this limit.
The $l^\nu$ vector, on the other hand, corresponds to a photon with a random polarization,
which we call ``loop polarization'', and it is problematic. 
Ward-identity cancellations are not present for these terms.  
Before going on, we note that both loop polarization terms in (\ref{eq:typeIIs10}) have one fewer propagator
than the original diagram, and  appear in integrals
characteristic of self energies, rather than the full vertex.   This feature will be used to advantage below.

\begin{figure}
  \centering
  \includegraphics[width=0.75\textwidth]{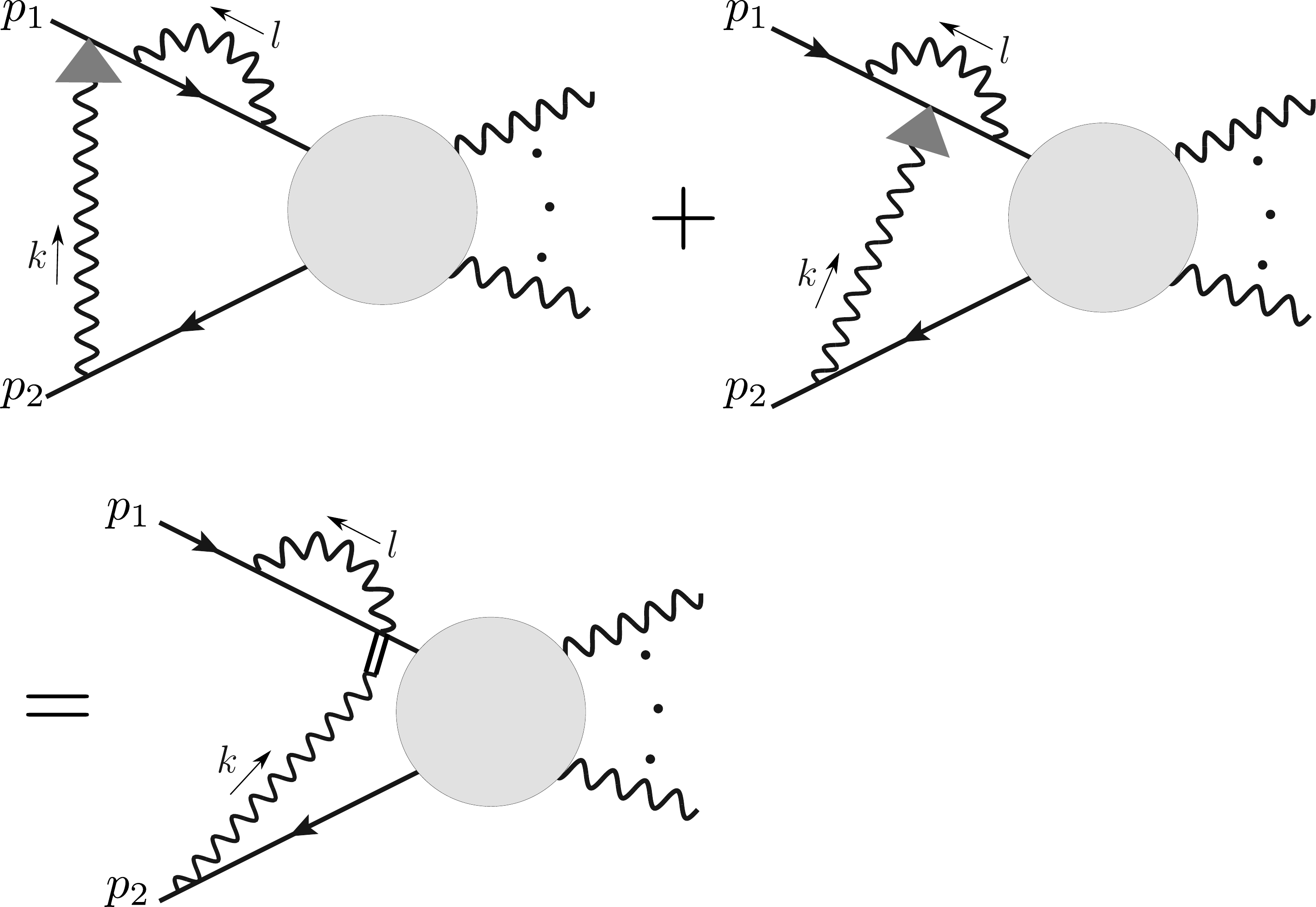}
  \caption{Illustration of the $k\parallel p_2$ collinear
    singularities when summing \typeI\ and \typeII\ diagrams. The
    sum is a non-factorized collinear singularity at the integrand
    level, and vanishes only after integration.}
  \label{fig:region2fact2}
\end{figure}

\begin{figure}
  \centering
  \begin{align}
  \eqs[0.4]{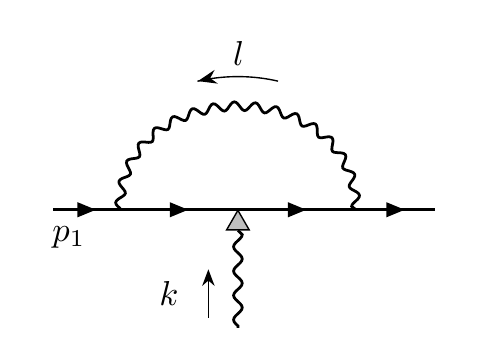} + 
  \eqs[0.57]{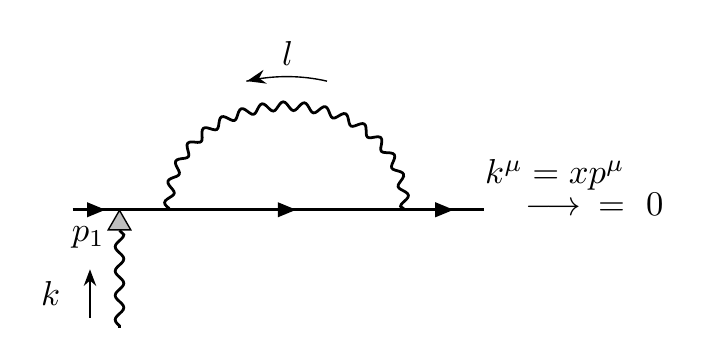}
\nonumber
  \end{align}
  \caption{The exchange photon Ward identity for the modified integrands. This figure should be contrasted with Fig.~\ref{fig:region2fact2} for the unmodified integrands, where the right-hand side is  nonzero locally.}
\label{fig:exchange-WI}
\end{figure}

The third challenge concerns the factorization of $k\parallel p_2$
collinear singularities. 
Regular diagrams (i.e., not of \typeI\ or \typeII) factorize
straightforwardly by Ward identities after using the collinear
polarization approximation of Eq.~\eqref{eq:triangleNotation2} with
$q=k$. But the sum of \typeI\ and \typeII\ diagrams, through Ward
identities, becomes an expression which does \emph{not} factorize into
lower-loop amplitudes times an one-loop correction on the incoming
positron ($p_2$) leg, as illustrated in Fig.~\ref{fig:region2fact2}.
Of course, this problem goes away if we do not attempt to formulate
factorization locally; after integrating over the 
sub-loop $l$, the self energy correction to the incoming electron line
produces a factor of $\slashed p_1$, which annihilates the massless external
spinor $u(p_1)$, and the right-hand side of
Fig.~\ref{fig:region2fact2} vanishes. 
However, in order to be able to construct local counterterms for the
amplitude, we need to exhibit factorization
at the integrand level.  It is therefore our aim  to rewrite the
\typeI\ and \typeII\ diagram expressions so that the sum in
Fig.~\ref{fig:region2fact2} vanishes locally in the limit $k\parallel p_2$.
Later we will construct the modified vertex integrand, Eq.~\eqref{eq:VmuS-average} and modified self energy integrand, Eq.~\eqref{eq:Smu-symm}, which exactly satisfy the desired Ward identity,
\begin{equation}
\frac{\eta_2^\nu k_\mu}{(k-\eta_2)^2 - \eta_2^2} V_{\rm mod}^\mu +
\frac{\eta_2^\nu k_\mu}{(k-\eta_2)^2 - \eta_2^2} S_{\rm mod}^\mu
= 0, \quad \text{when } k \parallel p_2, \label{eq:region2wardid}
\end{equation}
as illustrated in
Fig.~\ref{fig:exchange-WI}.

\subsubsection{Strategy for factorizable vertex and self energy corrections}
\label{sec:strategy}

In the following, local factorization is made possible by specifying
modified one-loop vertex and self energy integrands for subdiagrams
that appear adjacent to the incoming lines.
The modified integrands are designed to leave the integrated amplitudes
unchanged in dimensional regularization. 
These modifications will apply to generic
$e^-(p_1) +e^+(p_2) \to \gamma^*(q_1)+ \ldots +\gamma^*(q_n)$ amplitudes of an arbitrary
photon multiplicity, $n$, in the final state,   as well as to the two-loop 
$2 \to 1$ form factor. 
We will see that suitably modified integrands of the
latter can serve as local IR counterterms of all other
amplitudes in the class, providing a non-trivial two-loop
generalization of the one-loop IR counterterm 
of Eq.~\eqref{eq:oneLoopFFCT}.

Each integral in the \typeI\ and \typeII\ diagrams of
Fig.~\ref{fig:typeInII}  is characterized by an internal loop momentum for the
vertex or self energy, which we will denote as $l$, and the momentum
of the exchanged photon, which we denote by $k$.   As noted above, the loop
labels will be symmetrized, but the arguments here will not require that.

In these diagrams,
the singular regions can be denoted by seven pairs $(A,B)$.   
The first index, $A$, refers to the loop momentum $k$ and takes the values
$A=1,2,S,H$ when $k$ is collinear to $p_1$, collinear to $p_2$, soft
or hard respectively. Here ``hard'' means the typical hard momentum transfer scale of the process, for example, the energy of
final-state photons, which are taken to be all of the same order.  The second, $B$, refers to the
loop momentum $l$, and can take the values $B=1,S,H$ only for the singular regions of these diagrams.

For all leading singular regions of the form $(A,B)$ where neither $A$ nor $B$
equals $H$,  there are no hard sub-integrals by definition.   In these regions,
$(1,1)$, $(2,1)$, $(S,1)$,
on-shell lines  are attached to tree rather than one-loop subgraphs and the
necessary Ward identities, based as always on Eq.\ (\ref{eq:WI}), work point-by-point in loop momentum space.  
The singularities factorize locally, and  do not require
any alterations of  the functional form of the integrand.   
We can thus focus on ``single limits", $(1,H)$, $(2,H)$, $(S,H)$,
$(H,1)$, where only one of the loop momenta becomes soft or collinear.
We will now show how to treat the single-collinear limits.  The region $(S,H)$ may be treated similarly to $(1,H)$.

\begin{figure}[!h]
  \centering
  \begin{align}
  &\eqs[0.4]{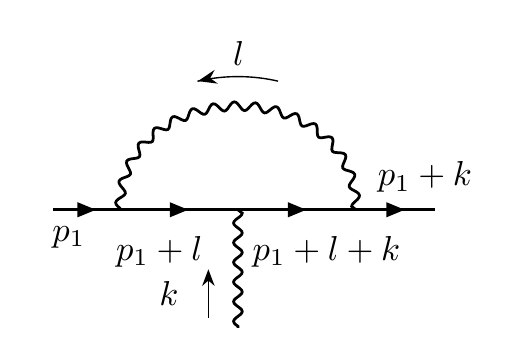}  
\underset{l \to -(1-x) p_1}{\longrightarrow}\nonumber \\
& e \, \frac{2 x}{1-x} \,   \left[ 
\eqs[0.3]{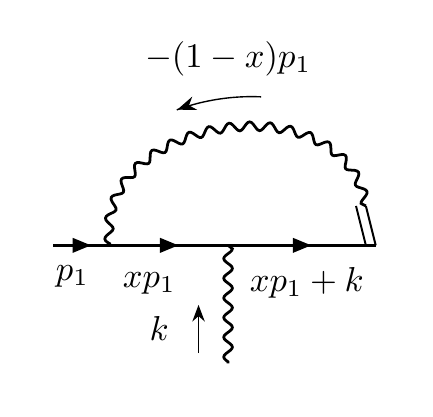} 
   - \eqs[0.36]{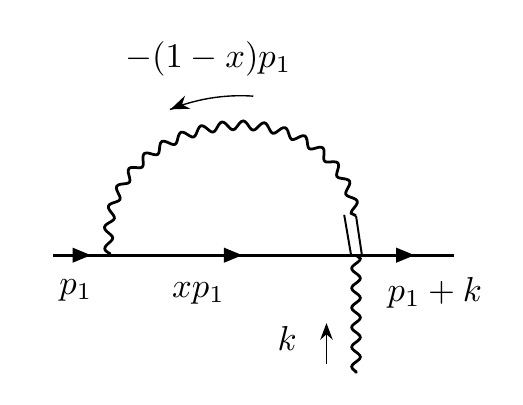} \right]
\nonumber
  \end{align}
  \caption{ Result of the Ward identity in
    the $l$ collinear to $p_1$ limit. 
  \label{fig:H1-WI}
}
\end{figure}

To factorize the amplitude in the three single-collinear limits, where a collinear line attaches to a vertex
carrying hard loop momentum,   we will try to construct integrands
that have four necessary and sufficient properties:

\begin{itemize}

\item In region $(1,H)$ (exchange photon $k$ collinear to $p_1$ and
  loop momentum $l$ hard and fixed) there should be no ``loop
  polarizations" in singular terms.   Leading singular
  behavior in this region should arise only from terms in which the
  photon with momentum 
  $k^\mu$ is scalar-polarized,  i.e., with a polarization vector proportional to $p_1^\mu$ or $k^\mu$,
  on the index that is contracted with the hard scattering subdiagram.   This polarization is
  captured by the polarization approximation introduced in Eq.~\eqref{eq:triangleNotation}.

\item In the same region $(1, H)$, power-like singularities coming from doubled
  propagators in \typeI\ diagrams, shown in
  Fig.~\ref{fig:typeInII}(a)-(b),  should be reduced to no stronger
  than logarithmic.

\item In region $(2,H)$ ($k$ collinear to $p_2$, $l$ fixed), singular terms should obey the modified Ward identity for the scalar polarized photon of momentum $k$ when it attaches to the \typeII\ vertex diagram on the $p_1$ line or is adjacent to the \typeI\ self energy diagram, as in Fig.\ \ref{fig:exchange-WI}.

\item  The Ward identity for a scalar-polarized photon carrying loop momentum
  $l$ in the region $(H,1)$ ($k$ hard and $l$ collinear to $p_1$),
  illustrated in Fig.~\ref{fig:H1-WI}, is responsible for factorizing
  the $l \parallel p_1$ singularity 
  into a product of a singular collinear function and a one-loop
  amplitude with loop momentum $k$.   This identity is 
  respected by the original Feynman diagram expressions
  and must remain valid after any modifications that we perform on \typeII\ diagrams.
  In practice this means that the modification of the \typeII\ diagram, which
  appears on the left-hand side of Fig.~\ref{fig:H1-WI}, should not change its singular
  behavior in the $l \parallel p_1$ limit.
\end{itemize}

\subsubsection{Modification of \typeII\ integrands}

For \typeII\ diagrams in Fig.~\ref{fig:typeInII}(c)-(d), i.e., QED 
vertex diagrams on an external fermion leg, with $k$ the exchanged
photon and $l$ the loop momentum, the $l$ sub-loop integrand can be
written as 
\bea
\eqs[.35]{type2v_RH}  &\equiv& 
V_0^\mu(l,k,p_1)\ 
\nonumber \\ 
&&\hspace{-3.5cm} 
 =-i e^3\left [ \frac{ ( \slashed p_1 + \slashed k )\,  \gamma_\alpha  ( \slashed p_1 + \slashed k + \slashed l )  \gamma^\mu ( \slashed p_1  +\slashed l )\gamma^\alpha }
  {(p_1+k)^2 (p_1+k+l)^2 l^2 (p_1+l)^2} \right]\, u(p_1)\, .
\label{eq:V0}
\eea
Note that we have included the adjacent $(\slashed p_1 +
\slashed k) / (p_1+k)^2$ fermion propagator in the 
expression for the vertex.
The subscript in $V_0$ indicates that we will later modify this factor
of the integrand.

After some Dirac algebra in $d=4-2\epsilon$ dimensions, using identities such as $\gamma_\alpha \gamma^\mu \gamma^\nu \gamma^\rho \gamma^\alpha = 2\epsilon \gamma^\mu \gamma^\nu \gamma^\rho - 2 \gamma^\rho \gamma^\nu \gamma^\mu$, we find 
 \bea
 V_0^\mu(l,k,p_1)\ &=&\ \frac{ 2 \, i \, e^3 }{(p_1+k)^2}\, \Bigg \{ (1-\epsilon) \Bigg ( \frac{ 2(l^\mu + p_1^\mu) }{l^2 (p_1+l)^2}  \left[ 1 - \frac{k^2 + (\slashed p_1 + \slashed l)  \slashed k }{(p_1+k+l)^2}  \right ]
 \nn\\[2mm]
&\ & \quad 
-\  \frac{2(p_1^\mu + l^\mu)}{(p_1+l)^2 (p_1+k+l)^2}\  -  \frac{ 2(p_1+k)^\mu - \gamma^\mu \slashed k }{l^2 (p_1+k+l)^2} \, \Bigg )
\nn\\[2mm]
&\ & \quad 
-  \frac{ (\slashed p_1 + \slashed k ) \left[ -(\slashed p_1 + \slashed l) \gamma^\mu \slashed k\,  + \, \epsilon \slashed k \gamma^\mu (\slashed p_1 + \slashed l) \right ] }{l^2 (p_1+l)^2 (p_1+k+l)^2} 
\Bigg \}\, u(p_1)\, .
\label{eq:Vmu-expand}
  \eea
  This factor includes all the dependence on loop momentum, $l^\mu$.
The overall denominator $(p_1+k)^2$ produces a pinch for $k$ collinear
to $p_1$ in combination with the denominator $k^2$ of the
  propagator for the photon emerging from the one-loop vertex.  
Similarly, a pinch singularity is produced for $l$ collinear to $p_1$ due to 
the denominators $l^2$ and $(l+p_1)^2$.  Crucially, we observe that
these two collinear singularities are disentangled.  In particular,
the terms in the first two lines of Eq.~(\ref{eq:Vmu-expand}) that are
singular for both $k \parallel p_1$ and $l \parallel p_1$ cancel in the $l \parallel p_1$
limit.   Similarly, the final term of Eq.~(\ref{eq:Vmu-expand}), proportional to $\slashed p_1 +\slashed k$, is suppressed 
in the $k \parallel p_1$ limit and is singular only when $l \parallel p_1$.
This natural separation of the $(1, H)$ and $(H, 1)$ collinear regions
allows us to modify terms of the integrand which contribute to the
$k \parallel p_1$ limit without affecting the
singular behavior in the $l \parallel  p_1$ region. 
 
 Based on this, we propose to replace $V_0^\mu$ in Eq.\
 (\ref{eq:Vmu-expand}) by an equivalent form, 
in which loop polarizations proportional to $l^\mu$ are eliminated from terms that
are singular when the exchanged photon momentum  $k$ becomes collinear
to $p_1$.   
This is done by averaging the full 
expression  
on the first line of Eq.~(\ref{eq:Vmu-expand}), containing the factor
$(p_1^\mu + l^\mu) /[l^2(p_1+l)^2] $ , 
over a change of variables from $l$ to $-p_1-l$, 
and by averaging over an analogous change of variables in the
$1/[(p_1+l)^2(p_1+k+l)^2]$ term of the second line, exchanging $p_1+l$ with $-(p_1+k+l)$.  
Finally, anticipating our treatment of region $(2,H)$, we explicitly write down the average $k_T \leftrightarrow -k_T, \, l_T \leftrightarrow -l_T$, as required by the global symmetrization of Eq.~\eqref{eq:fourFoldSym} applied to both regular diagrams and \typeI/\typeII\ diagrams, which we denote by a subscript $\langle T \rangle$. Note that in either limits $k \parallel p_1$ or $k \parallel p_2$, $k_T$ is close to zero and the symmetrization is essentially just $l_T \leftrightarrow -l_T$.

The resulting expression, only slightly more complicated than (\ref{eq:Vmu-expand}) is 
 \bea
 V_{\rm mod}^\mu(l,k,p_1)\ &\equiv&\ \  \frac{ 2i e^3 }{(p_1+k)^2}\, \Bigg \{ \frac{(1-\epsilon) }{l^2(p_1+l)^2}\, \Bigg[ p_1^\mu  - \frac{ (p_1^\mu + l^\mu) (k^2 + (\slashed p_1  + \slashed l) \slashed k )}{(p_1+k+l)^2} 
 \nn\\
 &\ & \hspace{5mm}  +\ \frac{l^\mu ( k^2 - \slashed l \slashed k )}{(k-l)^2} \, \Bigg ] +\  (1-\epsilon)\frac{k^\mu}{(p_1+l)^2(p_1+k+l)^2}\
 \label{eq:VmuS-average}  \\[2mm]
  &\ & 
\hspace{-25mm}    -   \    (1- \epsilon) \frac{(\slashed p_1 + \slashed k)\,  \gamma^\mu   }{l^2\,  (p_1+k+l)^2}\   \ 
   -   \    \frac{(\slashed p_1 + \slashed k)\, [-  (\slashed p_1 + \slashed l) \,  \gamma^\mu \slashed k \, +\,  \epsilon \slashed k \,  \gamma^\mu (\slashed p_1 + \slashed l) ] }{l^2\, (p_1+l)^2\, (p_1+k+l)^2}  \  \Bigg \}_{\langle T\rangle} \, u(p_1)\, .
   \nn
  \eea
  On the right, we can readily confirm the absence of factorization-breaking loop polarizations for singular contributions in the region $(1,H)$.    Explicit factors of $l^\mu$ are suppressed in this region by the factor of $\ksla$ acting on the Dirac spinor, or by an explicit factor of $k^2$.   We emphasize that although $V_{\rm mod}^\mu$ 
  in (\ref{eq:VmuS-average}) is not equal to $V_0^\mu$, its integral over $l$ gives exactly the same result, so that it can be used in place of $V_0^\mu$ in the calculation of the amplitude.
  
\subsubsection{Modification of \typeI\ integrands}

The \typeI\ self energy integrand is treated in a manner analogous to
the \typeII\ vertex.  We seek a functional form 
that can be combined with the expressions for the vertex in the Ward identity relevant to the region $(2,H)$, Fig.~\ref{fig:exchange-WI}, when $k$ is collinear to $p_2$ and $l$ is hard.   Here we shall simply present the form that will provide the correct combination, confirming that it is equivalent to the defining form of the self energy integrand, after integration.

We start with the integrand of the self energy subdiagram,
\begin{align}
&\eqs[.4]{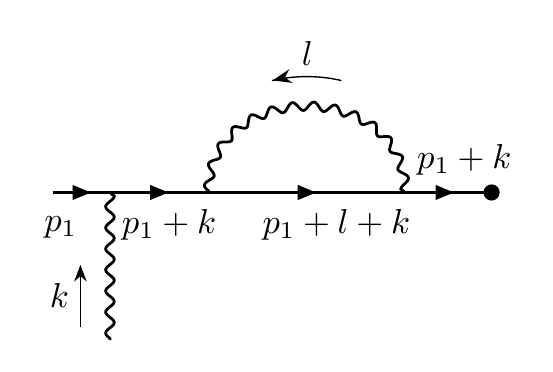}   \equiv S_0^\mu(l,p_1,k)  \nonumber \\
&\hspace{1.5cm}= -i e^3 \, \frac{ ( \slashed p_1+ \slashed k )}{\left ( p_1 +k \right )^2 }\, 
 \frac{\gamma^\alpha ( \slashed p_1 + \slashed k + \slashed l ) \gamma_\alpha}{l^2 (p_1+k+l)^2  }
 \frac{( \slashed p_1 +\slashed k )}{\left ( p_1 +k \right )^2 }\, \gamma^\mu u(p_1)
% \nonumber\\ & \hspace{1.5cm} \equiv S_0^\mu(l,p_1,k)
 \, ,
 \label{eq:Smu-original}
\end{align}
which defines the original integrand, $S_0^\mu(l,p_1,k)$.
Again we have included the adjacent $(\slashed p_1 + \slashed k) / (p_1+k)^2$ fermion propagator in the expression for the self energy, just as we did for the vertex.
In this standard form, the factor $S_0$ has a double pole at $(p_1+k)^2=0$, times the truncated self energy, which, after integration over $l$, leaves a single pole in the combination.

 To get started, we easily symmetrize the integral over a change of variable,
\begin{align}
  \int d^d l\, S_0(l,p_1,k)\ &= \int d^d l \,\frac{1}{2}\left[ S_0(l,p_1,k) + S_0(-p_1-k-l,p_1,k) \right] \nonumber \\
  & = i e^3 \, (1-\epsilon) \frac{(\psla_1 + \ksla )}{(p_1+k)^2}\, \gamma^\mu\, u(p_1)\, \int d^dl\, \frac{1}{l^2(p_1+l+k)^2}\, ,
\label{eq:S_0-int}
\end{align}
in which the single pole is explicit, and proportional to the scalar self energy integral.

Using the symmetries of the integral and the Dirac equation, there are many ways of finding an integrand for the $l$ integral that is equivalent to the ones found from (\ref{eq:Smu-original}) and (\ref{eq:S_0-int}).    We shall find an expression that preserves the Ward identity for the exchange vector $k$ in regions $(1,H)$ and $(2,H)$.    (We do not demand that the Ward identity be exact in other, finite regions.)

In our next step, we provide a form equivalent to (\ref{eq:S_0-int}), which is a simple combination of two terms, related by a change of variables,
  \bea
S_1^\mu(l,p_1,k)\ &\equiv&\  -e^3 \frac{i (1-\epsilon) }{\left ( p_1 +k \right )^2 }\, \Bigg [
 \frac{\slashed k - \slashed l }{(p_1+l)^2 (k-l)^2} \ -\
 \frac{\  \psla_1 + \ksla - \slashed l  }{ l^2 (p_1+k+l)^2 }\, \Bigg ] \gamma^\mu u(p_1)\, .
 \nn\\
 \label{eq:S1-symm}
\eea
To verify that the integral of $S_1^\mu$ equals that of $S_0^\mu$, it is only necessary to change variables to $l'=l+k$ in the first term of (\ref{eq:S1-symm}).    It is straightforward to check that this expression, together with the modified vertex expression, Eq.~\eqref{eq:VmuS-average}, obeys the modified Ward identity in region $(2,H)$ as in Fig.~\ref{fig:exchange-WI}.

As it stands, however, Eq.\ (\ref{eq:S1-symm}) is not quite what we want, because it introduces a new problem of loop polarizations, previously encountered in the unmodified \typeII ~diagrams, in region $(1,H)$, where $k$ is parallel to $p_1$.  These arise  from the components of $l^\mu$ in both the $p_2$ and transverse directions. 

To solve this problem, we will analyze the loop momentum in the
numerators of Eq.~(\ref{eq:S1-symm}) in terms of
components parallel and transverse to the incoming momenta $p_1$ and
$p_2$,
\begin{equation}
l^\mu = \frac{l \cdot p_2}{p_1 \cdot p_2} \, p_1^\mu
+\frac{l \cdot p_1}{p_1 \cdot p_2} \, p_2^\mu
+{l}_T^\mu\, .
\end{equation}
We will perform a  symmetrization only for
the $p_2$ light-cone component in the two numerators  of 
Eq.~\eqref{eq:S1-symm}.   We leave
the  $p_1$ light-cone component and the transverse components intact. 
This partial tensor reduction introduces modified versions of the loop
momentum vector, $\bar l_+^\mu$ and $\bar l_-^\mu$, defined by
\bea
\bar l_\pm^\mu \ =\ \frac{l\cdot p_2}{p_1\cdot p_2}\, p_1^\mu\, \pm \, \frac{1}{2}\, \frac{k\cdot p_1}{p_1\cdot p_2}\, p_2^\mu \, + \, l_T^\mu 
\label{eq:l-mod}
\eea
in the two numerators. 
It is a simple exercise to show that replacing $\slashed l$ by $\overline{ \slashed l}_+$ (or $\overline{\slashed   l}_-$) in the first (or second) term in
the square bracket of Eq.~\eqref{eq:S1-symm} does not change the
result after integration over $l$. 
The resulting expression is, 
\bea
S_{\rm mod}^\mu(l,p_1,k)\ &\equiv&\  -ie^3 \, \frac{(1-\epsilon) }{\left ( p_1 +k \right )^2 }\, \Bigg [
 \frac{\slashed k - \overline{\slashed l}_+ } {(p_1+l)^2 (k-l)^2} \ -\
 \frac{\ \psla_1 + \ksla - \overline {\slashed l}_- } { l^2 (p_1+k+l)^2 }\, \Bigg ]_{\langle T \rangle} \gamma^\mu u(p_1) \, .
 \nn\\
 \label{eq:Smu-symm}
\eea
We are again using $\langle T \rangle$ to denote the transverse part
of the global symmetrization Eq.~\eqref{eq:fourFoldSym}. 
This is our final form for the \typeI\ self energy diagram of Fig.\
\ref{fig:typeInII}(a)-(b). 

Eliminating the $p_2$ light-cone component of $l$ removes its
contributions to ``loop polarizations''. 
The transverse component of $l$, which also gives rise to loop
polarizations, is eliminated in the $k \parallel p_1$ limit by
averaging with a  global reflection of the amplitude on the transverse 
plane in Eq.~\eqref{eq:fourFoldSym}.  This reflection is implicitly applied to
every term of the amplitude integrand. 
Under this transverse reflection and averaging, the denominators in 
Eq.~\eqref{eq:Smu-symm} are unchanged, but the 
$\slashed l_T$ parts of the numerators cancel.  

The transverse symmetrization is specifically designed to eliminate
the ``loop polarization'' problem of the \typeI. Although this appears
to be of a limited scope,  the transverse symmetrization needs to be
applied globally, i.e.\ to all diagrams 
as in Eq.~\eqref{eq:fourFoldSym}, to preserve the interconnected web of
Ward identity relations that enable factorization of divergences 
in a variety of other collinear limits.

\subsection{Summary and validation of  modified amplitude integrand in
  single-collinear limits}

We have now arrived at an alternative expression for the two-loop
amplitude which solves the challenges of the original form.  
We first specified the assignments of loop momenta
in all Feynman diagrams of the two-loop integrand ${\cal M}^{(2)}_0 (k, l)$, 
derived in the Feynman gauge,  in Eq.~\eqref{eq:integrand0}.  
Then we manipulated the integrand in the following manner.

 \begin{enumerate}

\item For \typeII\ diagrams, i.e.\ vertex corrections adjacent to
  external fermion lines as illustrated in
  Fig.~\ref{fig:typeInII}(c)-(d), we modified the integrand expression
  $V_0^\mu$ in Eq.~\eqref{eq:V0} into $V^\mu_{\rm mod}$ in
  Eq.~\eqref{eq:VmuS-average} which is identical after integration. 

\item For \typeI\ diagrams, i.e.\ self energy corrections adjacent to
  external fermion lines as illustrated in
  Fig.~\ref{fig:typeInII}(a)-(b), we modified the integrand expression
  $S_0^\mu$ in Eq.~\eqref{eq:Smu-original} into $S^\mu_{\rm mod}$ in
  Eq.~\eqref{eq:Smu-symm}. 
\item For analogous \typeI\ and \typeII\ diagrams with self energy or vertex corrections adjacent to positron lines, the corresponding modifications are obtained by reflection.
\item A four-fold symmetrization of the integrand (including all diagrams, some of which have been modified) is performed as in Eq.~\eqref{eq:M2symm}.
\end{enumerate}
The outcome is a modified integrand, which is a sum of three contributions, as in Eq.\ (\ref{eq:M0-types}), now written as
\begin{equation}
{\cal M}^{(2)}(k,l)\ =\ \left[  \mathcal M^{(2)}_{\text{\typeI}} + \mathcal M^{(2)}_{\text{\typeII}} \right]_{\rm modified} + \mathcal M^{(2)}_{\text{regular}} \, ,
\end{equation}
with the modified self energy and vertex integrands specified by Eqs.\ (\ref{eq:Smu-symm}) and (\ref{eq:VmuS-average}), respectively.
The symmetrizations of Eq.\ (\ref{eq:M2symm}) are understood.
This will be the starting point for subtraction of infrared and ultraviolet divergences at the integrand level.   The same modifications will
be applied to two-loop form factor subtractions.

Before we proceed with the subtraction of infrared divergences, we
will verify that indeed the new integrand possesses the promised
factorization properties. Specifically, below, we will show  analytically that
in the regions $(H, 1)$ and $(2, H)$,  which were problematic originally,
Ward identities and factorizations are now realized locally in momentum space.   Factorization in
the $(1, H)$ region is self-evident by the absence of
loop-polarizations in Eq.~\eqref{eq:VmuS-average} and \eqref{eq:Smu-symm} and we do not
discuss it any further.  

\subsubsection{Factorization in region $(H,1)$}

  To test for the desired property in region $(H,1)$  ($l$ collinear to $p_1$), we can replace the loop momentum vector $l^\mu$ by $-(1-x)p_1^\mu$ with $1>x>0$ in Eq.\ (\ref{eq:VmuS-average}), except in the singular denominators, $l^2$ and $(p_1+l)^2$.  At fixed, off-shell $k$, we expect the photon carrying momentum $l$ to satisfy a tree-level Ward identity, giving one term that factorizes $l$ and $k$ dependence, and another term in which they are linked, as shown in Fig.~\ref{fig:H1-WI}.  The latter term will cancel against other diagrams that are singular when  $l$ is collinear to $p_1$.    Explicit calculation confirms that in the $l \to -(1-x)p_1$  limit,  singular behavior follows the expected pattern.   
  
 As noted above, the vertex $V_{\rm mod}^\mu$, Eq.\ (\ref{eq:VmuS-average}), is identical to the original form in this region, so we are just exhibiting how this works.
  Applying the Dirac equation and $p_1^2=0$ to the averaged vertex, Eq.\ (\ref{eq:VmuS-average}), the singular part of the vertex diagram is the collection of terms proportional to the symmetric factor $1/l^2(p_1+l)^2$.   These terms (which are even in $l_T$) are given, in detail, by
  \bea
  \lim_{l \to -(1-x)p_1} \left[ (p_1+l)^2 l^2\, V_{\rm mod}^\mu\,
  \right ]  &=& \frac{ 2 i e^3 }{(p_1+k)^2}\,
  \Bigg \{ p_1^\mu\ -\ \frac{ x \left (k^2 + x\psla_1\ksla \right) p_1^\mu}{k^2+2xp_1\cdot k} 
  \nn\\
    &\ & \hspace{-3mm}  -\ \frac{ x \left (k^2 +  (1-x) \psla_1\ksla
      \right) p_1^\mu}{k^2+2 (1-x) p_1\cdot k}   + \frac{  \left (\psla_1 + \ksla  \right) (x \psla_1) \gamma^\mu  \ksla } { k^2 + 2xp_1\cdot k} \  \Bigg \} u(p_1)
  \nn\\
  &\ & \hspace{-10mm} =\  \frac{ -2i e^3 }{(p_1+k)^2}\, \Bigg \{ \frac{ x
    \left( \ksla \psla_1 \ksla \right )}{( k+x p_1)^2} \Bigg \}
  \gamma^\mu\, u(p_1)\, ,
  \label{eq:VH1-1}
  \eea
  where the second form results from the cancellation of all terms proportional to $p_1^\mu$, and where we have observed that the Dirac equation implies $\psla_1\gamma^\mu \ksla\, u(p_1) = - \psla_1 \ksla \gamma^\mu\, u(p_1)$.   A little further rewriting shows that this result equals the sum of terms associated with the Ward identity for a scalar-polarized vector particle, connecting a lowest-order jet with one of the diagrams of the one-loop hard subdiagram,
\begin{align}
\lim_{l \to -(1-x)p_1} \left[ (p_1+l)^2 l^2\, V_{\rm mod}^\mu\,\right ] &\equiv
\lim_{l \to -(1-x)p_1} \left[ (p_1+l)^2 l^2\, V_{0}^\mu\,\right ] \nonumber \\
&= -i e^3 \frac{2x}{1- x}  \left [  \frac{1}{\left(x\psla_1 + \ksla \right)} - \frac{1}{\left ( \psla_1 + \ksla \right )} \right ] \gamma^\mu u(p_1)\,.
   \label{eq:V1-H1-WI}
\end{align}
  This Ward identity is the expected one, and is represented in Fig.\ \ref{fig:H1-WI}.

\subsubsection{Region $(2,H)$ and the $k$ Ward identity}

The final step is to verify that the Ward identity for the exchanged
photon, Fig.~\ref{fig:exchange-WI}, is respected by our
modified integrand, Eq.\ (\ref{eq:VmuS-average}), once it is added to
the appropriately-modified self energy, Eq.\ (\ref{eq:Smu-symm}).
Since we are interested in the behavior of the integral in the $k$
collinear to $p_2$ region, $(2,H)$, we can approximately set $k^2=0$, which
simplifies the algebra considerably. 
Straightforward algebra and the Dirac equation then yield the relatively simple expression
  \bea
   k_\mu V_{\rm mod}^\mu(l,k,p_1)_{k^2=0}\ &=&\  -i e^3 \, \frac{(1-\epsilon)}{(p_1+k)^2}\, \Bigg [ \frac{ (\psla_1 -\slashed l) \ksla}{ (p_1+k+l)^2 l^2 }  + \frac{\slashed l \ksla }{(k-l)^2 (p_1+l)^2 } \Bigg ]_{\langle T \rangle }u(p_1) .
   \nn\\
   \eea
To this expression, we will add the self energy integrand.  Neglecting
terms that vanish at $k^2=0$, we find 
\begin{eqnarray}
&&k_\mu V_{\rm mod}^\mu(l,k,p_1) |_{k^2=0} 
+k_\mu S_{\rm mod}^\mu(l,k,p_1) |_{k^2=0}   = 
  -i e^3 \, \frac{(1-\epsilon)}{(p_1+k)^2}\, 
\nonumber \\
&& \hspace{10mm} \times 
\Bigg[ 
\frac{ (\overline{\slashed l}_{-} - \slashed l) \slashed k}{ (p_1+k+l)^2 l^2 }  +
   \frac{(\slashed l-\overline{\slashed l}_{+} ) \ksla }{(k-l)^2 (p_1+l)^2 }
\Bigg ]_{\langle T \rangle } u(p_1)\, .
\end{eqnarray}
Given that 
\begin{equation}
(\slashed l-\overline{\slashed l}_{\pm} ) \slashed{p}_2  = 0, 
\end{equation}
we verify, 
\begin{equation}
k_\mu V_{\rm mod}^\mu(l,k,p_1) |_{k^2=0} 
+k_\mu S_{\rm mod}^\mu(l,k,p_1) |_{k^2=0}   = 0, \quad k \parallel p_2 \, ,
\end{equation}
and thus the factorization is complete on a point-by-point basis,  
as illustrated in Fig.\ \ref{fig:exchange-WI}.
This takes care of the $(2,H)$ region. Of course, to get the point-by-point Ward identity, we had to use consistent routing of momenta in the vertex and self energy diagrams, 
and that is what our modifications were designed to achieve.

 \subsection{Two-loop ``form factor'' infrared counterterms}
\label{subsec:2-loop-ff}

The amplitude integrands prepared in the previous subsections have the
property that they factorize in {\it all}  infrared limits.  As a result of this factorization,  infrared
singular factors are common to all $e^-(p_1)+ e^+(p_2) \to \gamma^*(q_1)+\ldots +
\gamma^*(q_n)$ amplitudes for arbitrary numbers of photons in the
final state.  In particular, the same singular factors are found in the $2 \to 1$
``form factor'' amplitude, which will enable us to use this simpler process to
organize the infrared structure of the entire class of processes.     
As discussed in Sec.\ \ref{sec:setup}, this universality is known to hold once loop momentum
integrations are carried out.  In this discussion, we have made the 
ingredients for factorization manifest at the level of integrands in
the class of diagrams that we discuss.

To exploit this universality, we begin as in the one-loop case, and generate  integrands for the
form factor, now with modified vertex and self energy integrands,
joining them to the tree level amplitudes for $e^-(p_1)+ e^+(p_2) \to \gamma^*(q_1)+\ldots +
\gamma^*(q_n)$.   
We depict again the one-loop and two-loop form factor functionals,  
following the notations of Eqs.~\eqref{eq:FF2functional_occ1} and \eqref{eq:ffDiagNotation}, as in Figs.\ \ref{fig:F2} and \ref{fig:for F1}, which we
reproduce here as equations for ease of discussion,
\begin{equation}
\label{eq:FF1functional}
  \mathcal F^{(1)}\left[\mathbf P_1 T \mathbf P_1 \right] = \eqs[.20]{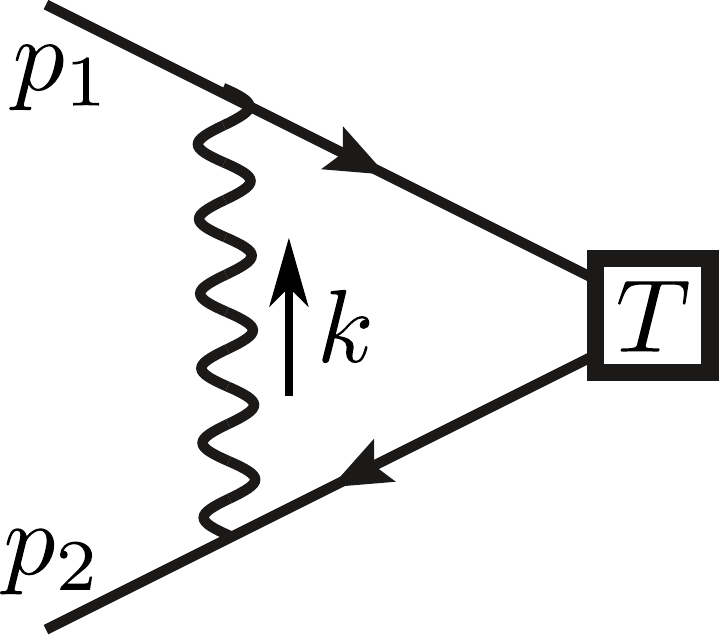}
\end{equation}
\begin{eqnarray}
\label{eq:FF2functional}
  \mathcal F^{(2)}\left[
\mathbf P_1 T \mathbf P_1
  \right]
  &=& 
\eqs[.2]{19a} 
+ \eqs[.2]{19b} 
+ \eqs[.2]{19c}
\nonumber \\
&& + \eqs[.2]{19d} 
+ \eqs[.2]{19e} 
+ \eqs[.2]{19f} \ .
\nonumber \\
\end{eqnarray} 
The loop momenta in the form factor amplitude are assigned and symmetrized with an identical procedure as that described in Subsection \ref{subsec:mod} for the generic $e^-(p_1)+ e^+(p_2) \to \gamma^*(q_1)+\ldots +
\gamma^*(q_n)$ amplitudes.  
For concreteness, we have made specific assignments on the  $k,l$
labels to the virtual photons exchanged among fermion lines in the form factor
diagrams. The averaging over loop momentum routings and
transverse loop momentum reflections of Eq.~\eqref{eq:M2symm} is 
understood implicitly.  
\TypeI\ self energy and \typeII\ vertex subgraphs in the last four diagrams
of Eq.~\eqref{eq:FF2functional} are modified with the extra symmetrizations
explained in subsection~\ref{subsec:mod}, just as the diagrams in Fig.~\ref{fig:typeInII}.

The squared vertex in each figure represents a matrix in spinor space, in between
a pair of $\mathbf P_1\equiv \frac{\slashed p_1 \slashed p_2} {2p_1
  \cdot p_2}$  projectors, Eq.~\eqref{eq:P1projector}. 
$T$ captures contributions from amplitude propagators which remain 
off-shell in infrared limits where either $k$ and/or $l$ vanish. 
As such, and due to factorization, it is independent of the loop momenta
assigned to the virtual photons outside the squared vertex in the above diagrams.  
For our purposes, $T$ will stand for  a tree amplitude integrand,
independent of both $k$ and $l$,
inserted in the one-loop and two-loop form factor of 
Eqs.~\eqref{eq:FF1functional} and \eqref{eq:FF2functional}, or a one-loop
amplitude integrand, depending on the loop momentum $l$, inserted in
the one-loop form factor of Eq.~\eqref{eq:FF1functional}. 

We have explained in Section~\ref{sec:oneloop} how an infrared finite remainder of the one-loop
amplitude integrand is derived with the use of a one-loop form factor
subtraction at the level of the integrand. We recall here the final result  at one loop of
Eqs.~\eqref{eq:M1IRfinite}, equivalent to Eqs.\ \eqref{eq:ffDiagNotation} and \eqref{eq:oneLoopFFCT},
\begin{equation}
 {\cal M}^{(1)}_{{\rm IR-finite}} =
  {\cal M}^{(1)} - \mathcal F^{(1)}
  \left[ {\mathbf P_1} \mathcal{
      \widetilde{M}}^{(0)} {\mathbf P_1}
  \right].
\end{equation}
This result is the form we would like to generalize, including the treatment of its induced ultraviolet singularities.

We are now ready to discuss the subtraction of infrared singularities
for the two-loop amplitude.  
Starting from the modified two-loop integrand $\mathcal M^{(2)}$
constructed in Subsection \ref{subsec:mod}, we will first subtract a
``global'' counterterm that simultaneously cancels all singularities in ``double-IR'' 
limits in which both loop momenta $k$ and $l$ are soft or
collinear to incoming fermion lines, e.g., the double-soft limit with
$k^\mu, l^\mu$ much less than the hard scales, the soft-collinear limit with $k^\mu$ soft, 
$l \parallel p_i$ (here $i=1$ or $2$), the  limit of two collinear pairs
with $k \parallel p_1, \, l \parallel p_2$, and the two-loop collinear
limit with $k , l \parallel p_i$. 
This global double-IR counterterm is precisely the set of diagrams given in Eq.\ (\ref{eq:FF2functional}),
\begin{equation}
  \takelimit_{\text{double-IR}} \mathcal M^{(2)} = \mathcal F^{(2)}
  \left[ {\mathbf P_1} \mathcal{
      \widetilde{M}}^{(0)} {\mathbf P_1}
  \right] \, ,
  \label{eq:doubleIRct}
\end{equation}
where we use the two-loop form factor as a local  approximation of all
double-infrared singularities.   
The subtraction gives
\begin{equation}
  \mathcal M^{(2)}_{\text{double-IR-finite}} = \mathcal M^{(2)} -
  \mathcal F^{(2)} \left[  {\mathbf P_1} \mathcal{
      \widetilde{M}}^{(0)} {\mathbf P_1} \right]\, ,
  \label{eq:twoLoopIRremainder}
\end{equation}
in which we recognize the beginning of the pattern of the two-loop finite remainder, Eq.\ (\ref{eq:M2-fin}), so far with a subtraction
only for the double-IR limits.   The modification of \typeI\ and \typeII\ diagrams is understood in these expressions to match the 
behavior of the modified amplitude, as described above.

\begin{figure}[!h]
  \centering
  \begin{align}
  &\eqs[0.3]{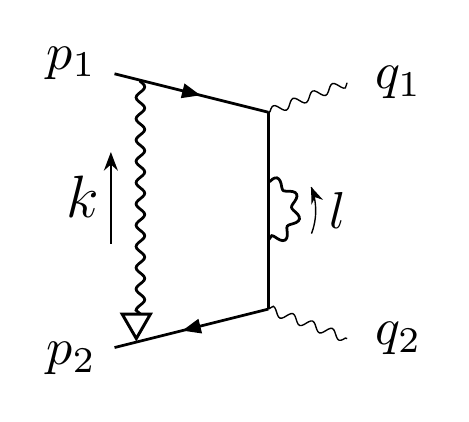} + \eqs[0.3]{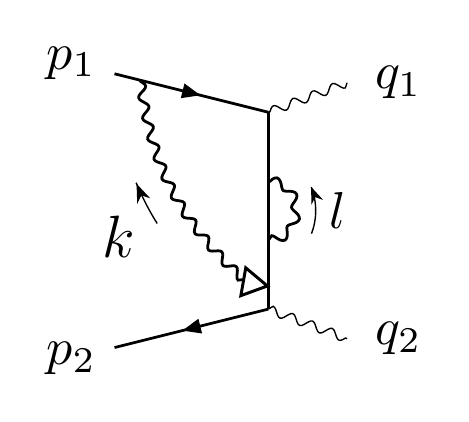}+\eqs[0.3]{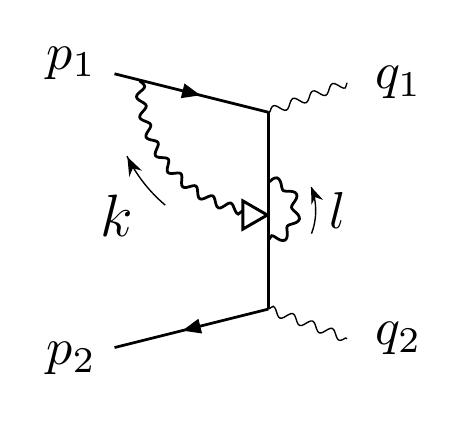} \nonumber \\
  &+ \eqs[0.3]{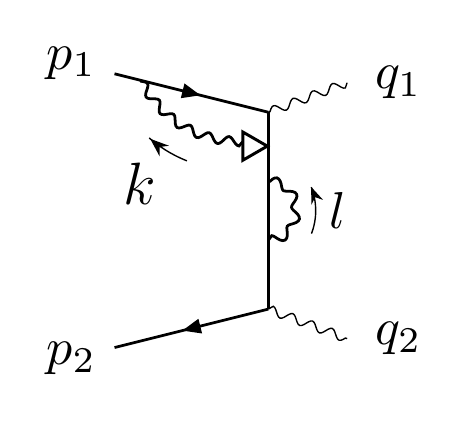} =\eqs[0.3]{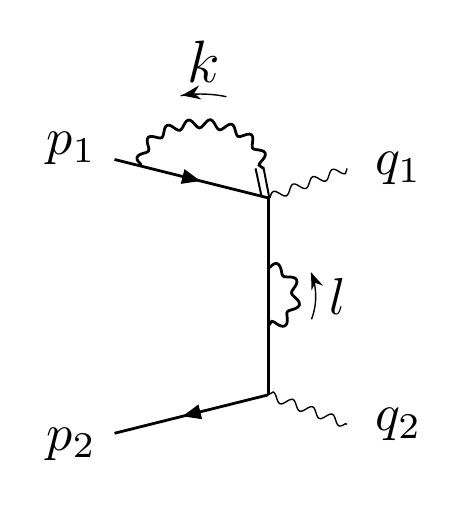}\nonumber
  \end{align}
  \caption{Illustration of factorization of single-collinear singularities of the two-loop amplitude.}
  \label{fig:singleCollFact}
\end{figure}

Next we must subtract a counterterm that cancels all remaining singularities
in ``single-IR'' limits, where one of the two loop momenta becomes
either soft (single-soft) or collinear (single-collinear)
to the incoming electron or positron, and one loop becomes hard while
flowing through the short-distance subdiagram.   (Type S\ and \typeII\ diagrams, in which a potentially
hard loop is disconnected from the hard scattering are already taken care of.)
For this, we are seeking an
approximation of the two-loop integrand remainder after the
subtraction of the double-IR singularities,  
\begin{equation}
  \takelimit_{\text{single-IR}} \, \mathcal
  M^{(2)}_{\text{double-IR-finite}}  
= \takelimit_{\text{single-IR}} \left(\mathcal M^{(2)} - \mathcal F^{(2)} \left[  {\mathbf P_1} \mathcal{
      \widetilde{M}}^{(0)} {\mathbf P_1} \right] \right).
\label{eq:take-single-IR}
\end{equation}
Due to the local factorization properties of our amplitude
integrand, the single-IR singularities of the two-loop amplitude
factorize in terms of a one-loop amplitude with hard momentum and
a universal factor, which is the same for any number of photons in the
final-state and can be approximated with a one-loop form factor
integrand. The factorization can be easily derived by using Ward identities, analogous to the one-loop case. 
For example, Fig.~\ref{fig:singleCollFact} illustrates the factorization of singularities in the single-collinear limit $k \parallel p_1$ in the sum of a class of diagrams.
Factorization implies that the full contribution of regions in the amplitude where one loop is infrared and one hard is given by the product of the one-loop integrand in Eq.\ (\ref{eq:M1-fin}) and the one-loop form factor integrand,
\begin{equation}
\label{eq:F1M2counter}
\takelimit_{\text{single-IR}} \mathcal M^{(2)} 
 = \mathcal F^{(1)} \left[ {\mathbf P_1}   \widetilde {\mathcal M}^{(1)} {\mathbf P_1} \right]\,.
\end{equation}
Similarly, we remove the single-IR limit from the two-loop form factor
counterterm, 
\begin{eqnarray}
 \takelimit_{\text{single-IR}} \, \mathcal F^{(2)} \left[  {\mathbf P_1} \mathcal{
      \widetilde{M}}^{(0)} {\mathbf P_1} \right] 
=\mathcal F^{(1)} \left[ {\mathbf P_1} \mathcal F^{(1)} \left
    [ {\mathbf P_1}  \widetilde{\mathcal M}^{(0)} {\mathbf P_1} \right] 
    {\mathbf P_1} \right]\,.
\label{eq:F1F2M2counter}
\end{eqnarray}
The above compact notation corresponds to  
the following momentum representation
\begin{eqnarray}
&& 
\mathcal F^{(1)} \left[ {\mathbf P_1} \mathcal F^{(1)} \left
    [ {\mathbf P_1}  \widetilde{\mathcal M}^{(0)} {\mathbf P_1} \right] 
    {\mathbf P_1} \right] = \frac {-e^4} 4 
\bar v (p_2) \gamma^\mu \frac{1}{\slashed k - \slashed{p}_2} 
\nonumber \\ 
&& \hspace{30mm} \times \ 
{\mathbf P_1}
\gamma^\nu \frac{1}{\slashed l - \slashed{p}_2}  
 {\mathbf P_1} 
\widetilde{\mathcal M}^{(0)} 
{\mathbf P_1}
\frac{1}{\slashed l + \slashed{p}_1} \gamma_\nu
 \mathbf{ P_1} 
\nonumber \\ 
&& \hspace{30mm} \times \
\frac{1}{\slashed k + \slashed{p}_1} \gamma_\mu u(p_1)
+ \mbox{ (3 symmetric terms of Eq.~\eqref{eq:M2symm})}.
\nonumber\\
\end{eqnarray} 
Combining, Eq.~\eqref{eq:F1M2counter} and Eq.~\eqref{eq:F1F2M2counter} into
Eq.~\eqref{eq:take-single-IR} we have that
\begin{align}
  \takelimit_{\text{single-IR}} \, \mathcal M^{(2)}_{\text{double-IR-finite}} &= \takelimit_{\text{single-IR}} \left(\mathcal M^{(2)} - \mathcal F^{(2)} \left[  {\mathbf P_1} \mathcal{
      \widetilde{M}}^{(0)} {\mathbf P_1} \right] \right) \nonumber \\
   &= \mathcal F^{(1)} \left[ {\mathbf P_1}   \widetilde {\mathcal M}^{(1)} {\mathbf P_1} \right]                                                                                -
     \mathcal F^{(1)} \left[ {\mathbf P_1} \mathcal F^{(1)} \left [ {\mathbf P_1}
      \widetilde {\mathcal M}^{(0)}
{\mathbf P_1}
     \right] {\mathbf P_1}
     \right] \nonumber \\
                                                                              &= \mathcal F^{(1)} \left[ {\mathbf P_1} \left(
 \widetilde{\mathcal M}^{(1)} - \mathcal F^{(1)} \left
    [ {\mathbf P_1}  \widetilde{\mathcal M}^{(0)} {\mathbf P_1} \right]  \right)
    {\mathbf P_1} \right] \nonumber \\
  &= \mathcal F^{(1)} \left[ {\mathbf P_1}  \widetilde{\mathcal M}^{(1)}_{\text{IR-finite}} {\mathbf P_1}\right] \, ,
  \label{eq:singleIRct}
\end{align}
where $\widetilde {\cal M}^{(1)}_{\rm IR-finite}$ is given by Eqs.\ (\ref{eq:M1IRfinite}) using (\ref{eq:oneLoopFFCT}), removing
the external spinors.    This
 provides the second IR subtraction in Eq.\ (\ref{eq:M2-fin}).

We recall from the discussion at the end of Sec.\ \ref{sec:setup} that the $\mathbf P_1$ projectors play an important role in
Eq.~\eqref{eq:singleIRct} in preventing the introduction of spurious
IR singularities. Let us examine their action here, in the momentum representation of the 
above counterterm, which reads
\begin{eqnarray}
&& 
\mathcal F^{(1)} \left[ {\mathbf P_1}  \widetilde{\mathcal
   M}^{(1)}_{\text{IR-finite}} {\mathbf P_1}\right]  
= \frac {-ie^2} 4 
\bar v (p_2) \gamma^\mu \frac{1}{\slashed l - \slashed{p}_2} \ 
{\mathbf P_1} 
\widetilde{\mathcal M}^{(1)}_{\text{IR-finite}}(k) 
{\mathbf P_1}
\frac{1}{\slashed l + \slashed{p}_1} \gamma_\mu u(p_1)
\nonumber \\ [2mm]
&& \hspace{45mm} \
+ \mbox{ 3 symmetric terms of Eq.~\eqref{eq:M2symm}}.
\label{eq:F1M1rep}
\end{eqnarray} 
In a collinear limit such as $k \parallel p_1$ of
Eq.~\eqref{eq:M1IRfinite},  the factor $\widetilde{\mathcal
  M}^{(1)}_{\text{IR-finite}}$, which is defined without external spinors, is actually divergent.  The divergence
  arises from terms of the amplitude that vanish in the Ward identity by the equation of motion for the external
  fermions.  Such non-factoring terms cannot be canceled by a form factor subtraction.    However, given that 
  $\widetilde{\mathcal M}^{(1)}_{\text{IR-finite}}$
is positioned  in  between a pair of projectors which play the role of 
the external spinors, in the sense that $\slashed p_1 \mathbf P_1 = \mathbf P_1 \slashed p_2 = 0$, 
these singularities are guaranteed to vanish.  In the absence of the
projectors, such singularities vanish only in regions where the
other loop momentum $l$ is soft or collinear to one of the incoming
particles.

In conclusion, we have arrived at a compact  preliminary result for the  finite
remainder, subtracted for all infrared limits of our modified two-loop amplitudes,
\begin{equation}
  \mathcal M^{(2)}_{\text{IR-finite}} = \mathcal M^{(2)} 
  - \mathcal F^{(2)} \left[  {\mathbf P_1} \mathcal{
      \widetilde{M}}^{(0)} {\mathbf P_1} \right]
    - \mathcal F^{(1)} \left[  {\mathbf P_1}  \widetilde{\mathcal M}^{(1)}_{\text{IR-finite}} {\mathbf P_1}\right]\,.
    \label{eq:twoLoopIRFiniteRemainder}
\end{equation}
Effectively, this is Eq.\ (\ref{eq:M2-fin}) before
subtractions for UV behavior are carried out.
 Eq.~\eqref{eq:twoLoopIRFiniteRemainder} is valid locally, and the
 right-hand side is free of nonintegrable infrared behavior at all points in
 the integration domain of the loop momenta.   It is worth pointing out that 
the number of IR counterterms in this expression is smaller
than the number of diagrams that contribute to the amplitude.

\subsection{Ultraviolet singularities of the photonic two-loop
  amplitude ${\cal M}^{(2)}$}
  \label{subsec:UV-photon-2}

The integrand ${\cal M}^{(2)}_{\rm IR-finite}$ defined by Eq.~\eqref{eq:twoLoopIRFiniteRemainder}
can be thought of as an unrenormalized two-loop amplitude 
with its infrared singularities subtracted. However, the integral
of this amplitude remainder is still
singular in the ultraviolet limits  $k \to \infty$ and/or $l \to
\infty$. To complete our work, we need to remove these ultraviolet
singularities from the integrand as well.

Our aim will be to construct  ultraviolet counterterms that reflect the structure on the right-hand
side of Eq.~\eqref{eq:twoLoopIRFiniteRemainder}.
As above, we denote them by the action of an operator $\takelimit_{\rm UV}$, so that, schematically,  
\begin{eqnarray}
\label{eq:UVstructure}
\takelimit_{\rm UV} {\cal M}^{(2)}_{\rm IR-finite}
&=& \takelimit_{\rm UV} {\cal M}^{(2)}
- \takelimit_{\rm UV} \mathcal F^{(2)} \left[  {\mathbf P_1} \mathcal{
      \widetilde{M}}^{(0)} {\mathbf P_1} \right]
%\nonumber \\ 
%&& 
%\hspace{30mm} 
- \takelimit_{\rm UV}
\mathcal F^{(1)} \left[  {\mathbf P_1}  \widetilde{\mathcal
    M}^{(1)}_{\text{IR-finite}} {\mathbf P_1}\right]\, .
    \nonumber\\
\end{eqnarray}
The first term in Eq.~\eqref{eq:UVstructure} removes the ultraviolet
singularities of the two-loop amplitude integrand ${\cal M}^{(2)}$ 
\begin{equation}
{\cal M}^{(2)}-\takelimit_{\rm UV} {\cal M}^{(2)}
= 
 {\cal M}^{(2)}_{\rm UV-finite}\,.
\end{equation} 
Similarly, the second term in Eq.~\eqref{eq:UVstructure} cancels the
ultraviolet singularities in the two-loop form factor integrals of $\mathcal F^{(2)}$  which serves as  a counterterm of double infrared
singularities, 
\begin{equation}
\mathcal F^{(2)} \left[  {\mathbf P_1} \mathcal{
\widetilde{M}}^{(0)} {\mathbf P_1} \right]
-\takelimit_{\rm UV} \mathcal F^{(2)} \left[  {\mathbf P_1} \mathcal{
\widetilde{M}}^{(0)} {\mathbf P_1} \right]
= \mathcal F^{(2)}_{\rm UV-finite} \left[  {\mathbf P_1} \mathcal{
\widetilde{M}}^{(0)} {\mathbf P_1} \right].
\end{equation}
The last term in Eq.~\eqref{eq:UVstructure} must remove ultraviolet
singularities from both the one-loop form factor integral of
$\mathcal F^{(1)}$ and its one-loop infrared-subtracted amplitude
kernel $ {\mathbf P_1}  \widetilde{\mathcal M}^{(1)}_{\text{IR-finite}} {\mathbf P_1}$,  
\begin{eqnarray}
&& 
\mathcal F^{(1)} \left[  {\mathbf P_1}  \widetilde{\mathcal
    M}^{(1)}_{\text{IR-finite}} {\mathbf P_1}\right] 
- 
\takelimit_{\rm UV}
\mathcal F^{(1)} \left[  {\mathbf P_1}  \widetilde{\mathcal
    M}^{(1)}_{\text{IR-finite}} {\mathbf P_1}\right]
=\mathcal F^{(1)}_{\rm UV-finite} \left[  {\mathbf P_1}  \widetilde{\mathcal
    M}^{(1)}_{\text{finite}} {\mathbf P_1}\right].
\nonumber \\ 
&&
\end{eqnarray}
We will produce the desired  ultraviolet approximations in
the equations above analogously to the
quantum field theoretical procedure of ultraviolet renormalization, 
by introducing diagrammatic counterterms.  
These will have the effect of removing ultraviolet singularities from
one-loop and two-loop propagator and vertex
subgraphs whenever they appear in diagrams of the 
one-loop ${\cal M}^{(1)}$ and two-loop  ${\cal M}^{(2)}$ amplitudes as 
well as the one-loop $\mathcal F^{(1)}$ and two-loop $\mathcal
F^{(2)}$ form factor integrals. 
Our diagrammatic counterterms will be {\it local} in both momentum and coordinate space.   Their integrands are functions of
the loop momenta, which will combine with corresponding  two-loop integrands to give convergent results.  However, 
upon analytic integration of their loop momenta, their integrated values will also
match the known poles of QED counterterms in the 
dimensional regulation parameter $\epsilon$ (common to any renormalization scheme).  

This local renormalization procedure completes the 
integrand decomposition  of the two-loop bare amplitude into  
\begin{equation}
{\cal M}^{(2)} = {\cal M}^{(2)}_{\rm singular} 
+  {\cal M}^{(2)}_{\rm finite}\,,
\end{equation}
where  all ultraviolet and infrared singularities reside in the first
term, $ {\cal M}^{(2)}_{\rm singular} $, of the right-hand side and the
second term  ${\cal M}^{(2)}_{\rm finite}$ is completely free of
non-integrable singularities in $d=4$ dimensions. 
 The finite remainder is given by  Eq.\ (\ref{eq:M2-fin}), reproduced here,
\begin{equation}
  \mathcal M^{(2)}_{\text{finite}} = \mathcal M^{(2)}_{\text{UV-finite}} 
  - \mathcal F^{(2)}_{\text{UV-finite}} \left[  {\mathbf P_1} \mathcal{
      \widetilde{M}}^{(0)} {\mathbf P_1} \right]
    - \mathcal F^{(1)}_{\text{UV-finite}} \left[  {\mathbf P_1}  \widetilde{\mathcal M}^{(1)}_{\text{finite}} {\mathbf P_1}\right],
    \label{eq:twoLoopFiniteRemainder}
\end{equation}
where all terms on the right-hand side are convergent in the $k \to
\infty$ and/or $l \to \infty$ limits while their sum is free of any
infrared singularities. 
$ {\cal M}^{(2)}_{\rm singular} $  is a sum of  ultraviolet finite
 form factor one-loop and two-loop amplitudes whose analytic
 integration involves well-known 
integrals~\cite{Kramer:1986sg,Kramer:1986sr} and standard reduction 
methods. The integration of  $ {\cal M}^{(2)}_{\rm singular} $  over the
loop momenta in dimensional regularization will be presented in a
separate publication.

 In Section~\ref{sec:oneloop}, we explained in detail the derivation of local
ultraviolet counterterms for the diagrams in ${\cal M}^{(1)}$ and 
$\mathcal F^{(1)}$.  We use these counterterms at two loops as well, 
in order to cancel ultraviolet singularities from one-loop 
propagator and vertex subgraphs in diagrams of ${\cal M}^{(2)}$ and 
$\mathcal F^{(2)}$.  We will elaborate below essential infrared 
properties of their corresponding counterterm diagrams in 
configurations where one of the  loop momenta is in the 
ultraviolet region  and the other is in an infrared singular region. 
We will then extend the derivation of one-loop counterterms of 
one-loop self energy and vertex subgraphs for the modified integrands
as they appear in \typeI\ and \typeII\ diagrams. 
Finally,  we will discuss how to derive counterterms which respect
infrared factorization, when both loop momenta tend to infinity.

\subsubsection{One-loop ultraviolet counterterms in regular diagrams}
\label{subsubsec:subUV1}
In Section~\ref{sec:oneloop}, we derived one-loop local ultraviolet
counterterms for the one-loop QED vertex and electron self energy graphs, Eqs.~\eqref{eq:vtct1} 
and \eqref{eq:bbct1}, respectively.
Our approach for their construction aimed to address the following problem.  
When a loop momentum, for example $k$, is taken to an infinite
value, the second loop momentum $l$ is unrestricted and can assume
values which give rise to infrared singularities.  
For such loop momentum configurations, simultaneous subtractions 
are required for both ultraviolet and soft or collinear
singularities. However, a naive introduction of ultraviolet
counterterms could spoil the factorization of collinear singularities.   

\begin{figure}
  \centering
  \begin{align}
  &\eqs[0.3]{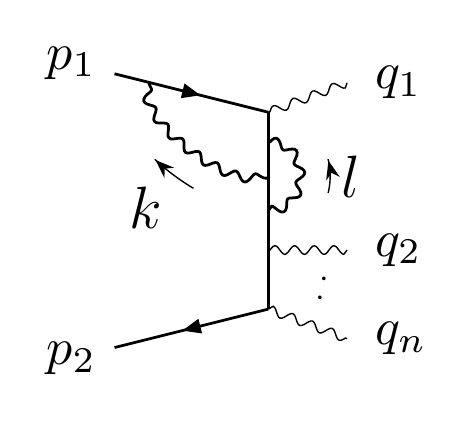} \overset{{l \to \infty}}{\longrightarrow} \eqs[0.3]{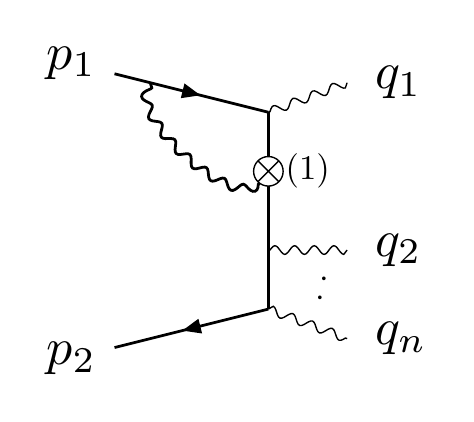}  \nonumber \\
  &\eqs[0.3]{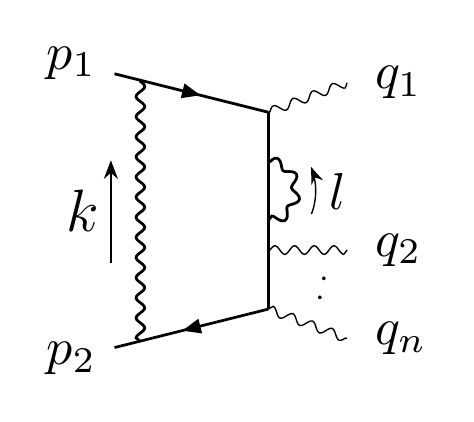} \overset{{l \to \infty}}{\longrightarrow} \eqs[0.3]{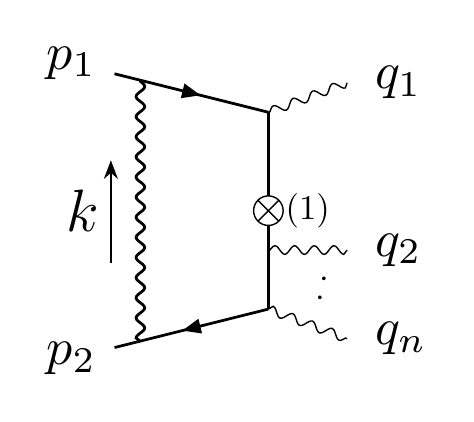} \nonumber
  \end{align}
  \caption{Illustration of ultraviolet counterterms, as defined in
    Eq.~\eqref{eq:vtct1} and \eqref{eq:bbct1}, for one-loop vertex and
       self energy subgraphs in regular diagrams 
(as opposed to \typeI\ and \typeII\ diagrams). 
 \label{fig:hardVertexBubbleUV} }
\end{figure}

\begin{figure}[!h]
  \centering
  \begin{align}
  &\eqs[0.3]{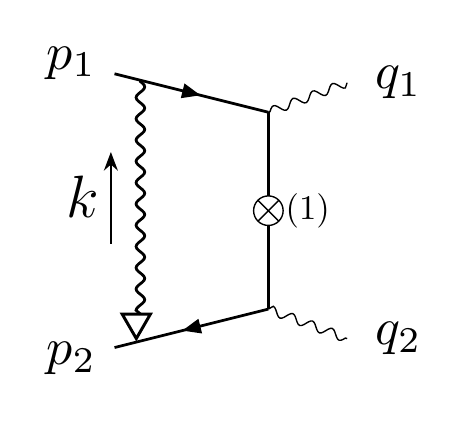} + \eqs[0.3]{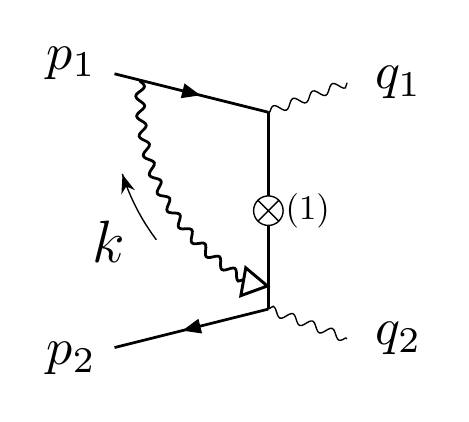} +\eqs[0.3]{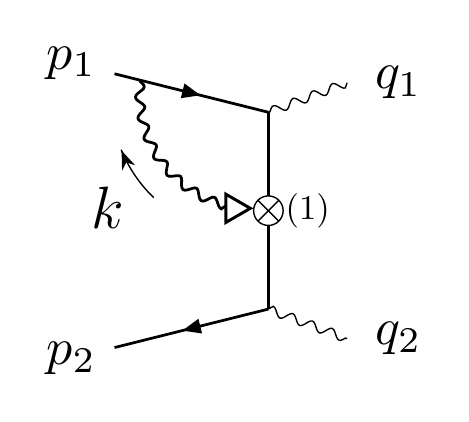} \nonumber \\
  &+\eqs[0.3]{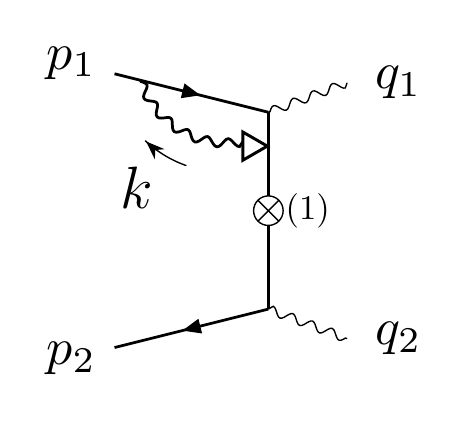} = \eqs[0.3]{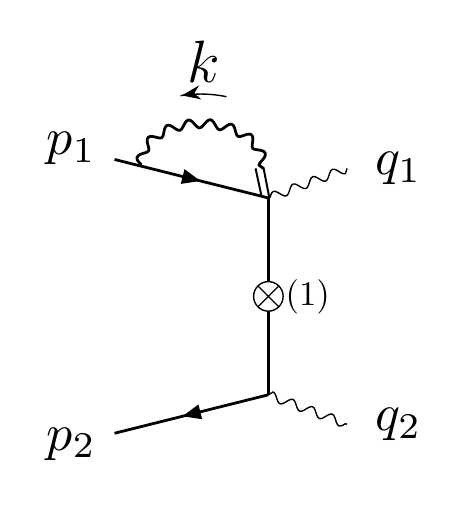} \nonumber
  \end{align}
  \caption{Illustration of a Ward identity that must be satisfied by
    the UV counterterms for one-loop self energy  and vertex
    subgraphs. 
Note that we are referring to counterterms at the integrand level,
    and the counterterm designations in the figure correspond to
    functions whose dependence on the loop momentum $l$ 
is suppressed in the notation.
}
  \label{fig:subUVWardId}
\end{figure}

To address this complication, in Section~\ref{sec:oneloop} we
introduced local ultraviolet counterterms for the one-loop electron
propagator and the one-loop electron-electron-photon vertex that
respect the QED Ward identity. Let us examine the function of the
counterterms of \ Eq.~\eqref{eq:vtct1} and \eqref{eq:bbct1}, for
diagrams in ${\cal M}^{(2)}_{\rm regular} $ such as the ones illustrated 
in Fig.~\ref{fig:hardVertexBubbleUV}.
The corresponding counterterm diagrams,  in which the ultraviolet
limit is approximated by Eqs.~\eqref{eq:vtct1} and \eqref{eq:bbct1},
exhibit the same cancellations  of collinear singularities that occur in the
original diagrams. The pattern of the Ward identity for the
collinear regions of counterterm  diagrams is  illustrated in Fig.~\ref{fig:subUVWardId}.

Therefore, our construction of ultraviolet counterterms ensures that
collinear  singularities associated with the non-UV loop remain in a 
factorized form. This factorized singularity  can then be precisely
canceled by the form factor amplitude $\mathcal
F^{(1)}_{\text{UV-finite}} \left[  {\mathbf P_1}  \widetilde{\mathcal
    M}^{(1)}_{\text{finite}} {\mathbf P_1}\right]$ in
Eq.~\eqref{eq:twoLoopFiniteRemainder}. We remark that for this
cancellation to occur, it is important that  the local vertex of the
form factor counterterm, ${\mathbf P_1}  \widetilde{\mathcal
    M}^{(1)}_{\text{finite}} {\mathbf P_1}$,  is UV regularized
by the use of the same counterterms for one-loop 
self energy and vertex graphs as those we have used in ${\cal
  M}^{(2)}_{\rm regular}$.

\subsubsection{One-loop ultraviolet counterterms specific to  \typeI\  and \typeII\  diagrams}
\label{subsubsec:subUV2}

To complete the derivation of one-loop ultraviolet counterterms for
${\cal M}^{(2)}$ we
have to discuss the ultraviolet limit of the integrand of 
one-loop vertex and self energy subgraphs in \typeII\ and \typeI\
diagrams, which we have cast in a modified form.  

In the large $l$ limit, the modified vertex expression of Eq.~\eqref{eq:VmuS-average} is logarithmically divergent, so we can easily write down its UV subdivergence counterterm by taking the leading large-$l$ limit of the expression,
\begin{equation}
  \takelimit_{l \to \infty} V^\mu_{\rm mod.} =  \frac{2i e^3 \, (1-\epsilon)} {(p_1+k)^2 (l^2-M^2)^2} \left[
    p_1^\mu + k^\mu - (\slashed p_1 + \slashed k) \gamma^\mu  - \frac{ 2 l^\mu \slashed l \slashed k } {l^2 - M^2}
    \right]_{\langle T \rangle } u(p_1) \, .
  \label{eq:Vmu-CT}
\end{equation}
Though the original \typeI\ electron self energy expression
Eq.~\eqref{eq:Smu-original} is linearly divergent in the limit of
large $l$, our equivalent form  of Eq.~\eqref{eq:Smu-symm} is only
logarithmically divergent.  This allows us to easily write down the UV counterterm
\begin{align}
  \takelimit_{l \to \infty} S^\mu_{\rm mod.} &= e^3 \frac{i (1-\epsilon)} {(p_1+k)^2{(l^2-M^2)^2}}
\left[
\slashed p_1 +  \bar{\slashed l}_+ - \bar{\slashed l}_- 
+ \frac{ \left(\bar{\slashed l}_+ + \bar{\slashed l}_-\right)\,2k\cdot l}{(l^2-M^2)}
\right]_{\langle T \rangle } \,\gamma^\mu u(p_1)\,.
  \label{eq:Smu-CT}
\end{align}
The symmetrization of the transverse components of the loop momentum ensures that those momenta remain UV finite
when $k$ is collinear to either $p_1$ or $p_2$.

Importantly, the above counterterms Eqs.~\eqref{eq:Vmu-CT} and
\eqref{eq:Smu-CT} satisfy the necessary Ward identity for
factorization (or rather, cancellation) of the singularity 
in the limit $k \parallel p_2$, analogous to Eq.~\eqref{eq:region2wardid},
\begin{equation}
  k_\mu \takelimit_{l \to \infty} S_{\rm mod.}^\mu\, \Big |_{k \to yp_2}\ =\ -\; k_\mu \takelimit_{l \to \infty} V_{\rm mod.}^\mu\, \Big |_{k \to yp_2}.
\end{equation}
Because of this result, 
 the UV counterterms do not re-introduce unfactorized collinear
singularities when $k \parallel p_2$. It is also easy to check that in
the limit $k \parallel p_1$, the above counterterms do not
re-introduce the ``loop polarization'' problem.  In these results
the transverse symmetrizations in Eqs.\ (\ref{eq:Vmu-CT}) and (\ref{eq:Smu-CT}) 
are necessary.

\TypeI\  and \typeII\
diagrams are also present in the two-loop form factor amplitude ${\cal
  F}^{(2)}$, which we have used as an infrared 
counterterm on  the right-hand side of
Eq.~\eqref{eq:twoLoopIRFiniteRemainder}.  
Consistently, we use the expressions of 
Eq.~\eqref{eq:Vmu-CT}  and Eq.~\eqref{eq:Smu-CT} 
as ultraviolet counterterms in the form factor ${\cal
  F}^{(2)}$ as well. 

\subsubsection{One-loop ultraviolet counterterms specific to the form factor
  ${\cal F}^{(1)}$ and  ${\cal F}^{(2)}$  amplitudes}

In subsections \ref{subsubsec:subUV1} and \ref{subsubsec:subUV2}, we
constructed ultraviolet  counterterms for one-loop vertex and
self energy graphs in regular diagrams as well as \typeI\ and \typeII\  
diagrams. 
Additional ultraviolet counterterms are needed for one-loop vertex
subgraphs in  the form factor
  ${\cal F}^{(1)}$ and  ${\cal F}^{(2)}$  amplitudes  on the 
right-hand side  of Eq.~\eqref{eq:twoLoopIRFiniteRemainder}

\begin{figure}
  \centering
  \includegraphics[width=0.36\textwidth]{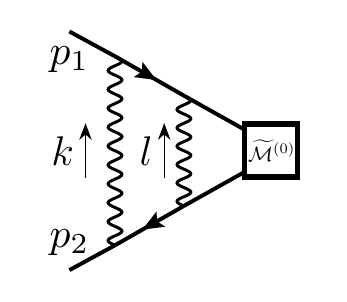}
  \caption{A sample graph that contributes to the second term on the
    right-hand side of Eq.~\eqref{eq:twoLoopIRFiniteRemainder}, using the same notation as Eq.~\eqref{eq:ffDiagNotation}.}
  \label{fig:twoloopFFsample}
\end{figure}

The ultraviolet singularity of  ${\cal F}^{(1)}$ 
 (similar to the usual QED one-loop
vertex) has already been  encountered in Eq.~\eqref{eq:FF1L_UV}. The same singularity appears in 
${\cal F}^{(2)}$, for which 
a sample diagram is shown of Fig.~\ref{fig:twoloopFFsample}. 
The large-$l$ limit in Fig.~\ref{fig:twoloopFFsample} is singular
and the singularity can be also subtracted by a counterterm that is
identical to 
the square bracket of Eq.~\eqref{eq:FF1L_UV}.

Next we construct a one-loop ultraviolet counterterm for  the last term in
Eq.~\eqref{eq:twoLoopIRFiniteRemainder} with a momentum representation~\footnote{As pointed out earlier, the $k \leftrightarrow l$ symmetrization in Eq.~\eqref{eq:fourFoldSym} makes it rather unimportant how we assign $k$ and $l$ to the two photons.}
\begin{equation}
  \mathcal F^{(1)} \left[  {\mathbf P_1}  \widetilde{\mathcal M}^{(1)}_{\text{IR-finite}} {\mathbf P_1}\right] =
  \bar v(p_2) \left[ 
    \left(-i e^2 \right) \frac{\gamma^\mu (\slashed k - \slashed p_2)
      \mathbf P_1
      \widetilde{\mathcal M}^{(1)}_{\text{IR-finite}}(l)
      \mathbf P_1
      (\slashed k + \slashed p_1) \gamma_\mu}{k^2 (k+p_1)^2 (k-p_2)^2} 
    \right]  u(p_1).
\end{equation}
The subtraction of the one-loop singularity as $l$ becomes large,  
by using the counterterms of  Eq.~\eqref{eq:UVM1} and
Eq.~\eqref{eq:FF1L_UV}, 
turns $\widetilde{\mathcal M}^{(1)}_{\text{IR-finite}}(l)$ 
into $\widetilde{\mathcal M}^{(1)}_{\text{finite}}(l)$ of Eq.~\eqref{eq:M1singular}.
A further  counterterm for the one-loop singularity as $k$
becomes large is required and it is similar to
Eq.~\eqref{eq:FF1L_UV}. 
The ultraviolet approximation of $k \to \infty$, following  the $l \to \infty$
subtraction, reads
\begin{equation}
  \takelimit_{k \rightarrow \infty} \mathcal F^{(1)} \left[  {\mathbf P_1}  \widetilde{\mathcal M}^{(1)}_{\text{finite}}(l) {\mathbf P_1}\right] =
  \bar v(p_2) \left[
    \left(-i e^2 \right) \frac{\gamma^\mu\slashed{k}
      \mathbf P_1
      \widetilde{\mathcal M}^{(1)}_{\text{finite}}(l)
      \mathbf P_1
      \slashed{k}\gamma_\mu}{(k^2-M^2)^3} 
    \right]  u(p_1).
\end{equation}

\subsubsection{Double ultraviolet counterterms}

After discussing the subtraction of ultraviolet singularities in single limits
where only one of the loop momenta approaches infinity, we turn our
attention  to double ultraviolet singularities in the region where both loop momenta
$k,l$ become infinitely large.  
As in the standard renormalization procedure at two loops, the subtraction of such singularities is performed \emph{after}
one-loop singularities have already been subtracted. This order of
subtractions for singularities in ultraviolet regions is necessary to
avoid double counting, and leads to counterterms for the
singularities in the double-ultraviolet region which produce integrands that are also convergent 
in single-ultraviolet regions. 

Double-ultraviolet singularities are found in two-loop
one-particle-irreducible (1PI) propagator and vertex graphs or 
subgraphs in the diagrams of ${\cal M}^{(2)}$ and the two-loop form factor amplitude ${\cal F}^{(2)}$ 
of Eq.~\eqref{eq:FF2functional}.  Counterterms for these singularities 
can be constructed straightforwardly, with the following steps. 
\begin{enumerate}
\item We first perform a series expansion of the integrands of the singular
  1PI graphs/subgraphs in the limit that both loop momenta become
  large uniformly, and truncate the series to the order that
  contributes to logarithmic UV singularities. As a result, we obtain
  tadpole-type integrands  whose only denominators are $k^2, \, l^2\,
  (k \pm l)^2$, each raised to some non-negative integer power.
\item We then insert masses into all the denominators to eliminate infrared
  singularities without changing the UV behavior, as was done at one loop
  in Eqs.~\eqref{eq:vtct1} and \eqref{eq:bbct1}. Some of the denominators are
  already massive by this step, because of the previous subtraction of
  one-loop UV sub-divergences, and these denominators are kept massive.
\end{enumerate}

The above procedure leads to counterterms that preserve the
Ward identities of  vertex and propagator graphs at two loops. 
This is a natural consequence of the Taylor expansion around the
ultraviolet limit being truncated  consistently at the same order 
and the introduction of infrared  mass regulators being performed 
in the same manner. The benefit of Ward identity-preserving ultraviolet counterterms is
two-fold.  First, the integration of the counterterms leads to simple
results, making manifest cancellations of double-ultraviolet
counterterms of diagrams with two-loop  vertex subgraphs  
against diagrams with two-loop self energy subgraphs. 
Second, it permits an extension of  our study to three-loop amplitudes
in the future, since our double-ultraviolet counterterms will not
destroy the factorization of collinear singularities associated with a 
third loop.

Taken together with the modified integrands of the two-loop amplitudes, and the
results of Section \ref{sec:fermionloop.tex}, the
UV counterterms described above enable us to construct the fully-subtracted
two-loop amplitude given in Eq.\ (\ref{eq:twoLoopFiniteRemainder}) or equivalently (\ref{eq:M2-fin}).   
The number of UV counterterms is similar to the number for the original
diagrams, because the number of form factor subtractions 
is fixed, independent of the number of final-state photons in the process.
In the following section, we will verify the integrability of our construction
for Eq.\ (\ref{eq:M2-fin}).

\section{Numerical checks of cancellation of singularities for off-shell multi-photon production}
\label{sec:checks}
%\subsection{Setup}

In this section, we numerically check the cancellation of all IR / UV singularities at
the integrand level in $e^-(p_1) + e^+(p_2) \rightarrow \gamma^*(q_1) +
\gamma^*(q_2)$. This is an example from the class of processes for which
our formalism applies.  
Our analysis includes all the terms that make up ${\cal M}_{\rm finite}^{(2)}$ in the
full expression of Eq.\ (\ref{eq:M2-fin}).   

We generate the Feynman diagrams using the program QGRAF
 \cite{Nogueira:1991ex}. Then we use custom Mathematica code to insert
 the Feynman rules, perform loop momentum symmetrizations, and modify certain
 diagram expressions as specified in the preceding text. 
Among the full set of two-loop diagrams, we may summarize the 
following modifications.
\begin{itemize}

\item The finite part of diagrams ${\cal M}^{(2)}_2$, with vacuum polarizations, including both IR
and UV counterterms, is generated as specified by Eq. (\ref{eq:M22}).

\item  Fermion loop diagrams, ${\cal M}^{(2)}_c$ with $c \ge 4$, are evaluated with modified vertices adjacent
to the incoming electron and/or positron lines,  given by Eq. (\ref{eq:gamma-hat}).

\item The diagrams for ${\cal M}^{(2)}_4$, which include light-by-light scattering, have UV subtractions, as 
described in Sec.\ \ref{sec:c=4}, which add to a finite result, but which produce convergence in otherwise divergent
individual diagrams.

\item  For diagrams without fermion loops (photonic contributions), we use the modified two-loop integrands for \typeII\ and \typeI\ subdiagrams, specified by
Eqs.\ (\ref{eq:VmuS-average}) and (\ref{eq:Smu-symm}), respectively.   

\item The \typeI\ and \typeII\ modified integrands are also employed
for the form factor subtractions, as described in Sec.\ \ref{subsec:2-loop-ff}.

\item Ultraviolet counterterms that ensure convergence for two-loop photonic amplitudes and
their IR subtractions are constructed as in Sec.\ \ref{subsec:UV-photon-2}.

\end{itemize}

We check that the integrand has the correct scaling
behavior to be convergent in all possible infrared and ultraviolet
limits at arbitrary phase space points.  For illustration, we present
results for one such point, 
\begin{align}
  p_1^\mu &= (1, 0, 0, 1), \nonumber \\
  p_2^\mu &= (1, 0, 0, -1), \nonumber \\
  q_1^\mu &= (1, 0, 1/3, 1/7), \nonumber \\
  q_2^\mu &= (1, 0, -1/3, -1/7) ,
\end{align}
with polarization vectors (satisfying the transversality condition $\epsilon_i^* \cdot q_i = 0$)
\begin{align}
  \epsilon_1^{* \, \mu} &= (4, -4, 9, 7), \nonumber \\
  \epsilon_2^{* \, \mu} &= (2, -1, -3, -7),
\end{align}
and external spinors in the Weyl basis (satisfying the Dirac equations $\slashed p_1 u(p_1) = \bar v(p_2) \slashed p_2 = 0$)
\begin{align}
  u(p_1) = \begin{pmatrix} 0 \\ -7 \\ 3 \\ 0 \end{pmatrix}, \quad \bar v(p_2) = \begin{pmatrix} 0, & 13, & -9, & 0 \end{pmatrix} \, .
\end{align}

A basis of vectors transverse to the $p_1$-$p_2$ plane is defined as
\begin{equation}
  n_x^\mu = (0, 1, 0, 0), \qquad n_y^\mu = (0,0, 1, 0) \, .
\end{equation}

We also use a parameterization of  the loop momenta
$k^\mu$, $l^\mu$, which depends on a small parameter $\delta$ and
scaling exponents $\omega_{k, \, l}^{i}$, $i=\{+,-,T\}$, starting
from an arbitrary choice of loop momenta at $\delta =1$,
\begin{align}
  k^\mu &= \frac {33} {17} \delta^{\omega_k^+} \, p_1^\mu - \frac {48} {89} \delta^{\omega_k^-} \, p_2^\mu +  \delta^{\omega_k^T} \left( \frac {21} {23}  \, n_x^\mu + \frac {21} {41} \, n_y^\mu \right), \nonumber \\
  l^\mu &= \frac {47} {23} \delta^{\omega_l^+} \, p_1^\mu - \frac {7} {61} \delta^{\omega_l^-} \, p_2^\mu + \delta^{\omega_l^T} \left( -\frac {37} {73} \, n_x^\mu -\frac{39}{67} \, n_y^\mu \right) \, . \label{eq:loopMomentumParam}
\end{align}
By selecting
appropriately the values of the exponents we can approach particular IR / UV
regions. 
The proof of  UV finiteness by testing convergence as loop momenta are scaled to infinity is well established  \cite{Weinberg:1959nj}.   The IR scalings of Eq.\ (\ref{eq:loopMomentumParam}) shown in  in Tables \ref{tab:limits} - \ref{tab:limitsQbox} serve the analogous purpose of testing the behavior of integrals in the neighborhoods of regions where loop momenta are pinched between coalescing poles.  The demonstration of IR finiteness or divergence for wide-angle scattering and production amplitudes by use of these scalings is discussed in Refs.\ \cite{Sterman:1994ce,Collins:2011zzd}.   The explicit tests here demonstrate the adequacy of our IR counterterms in regulating singular behavior in these amplitudes.\footnote{For other processes, typically involving momentum pinches in so-called Glauber regions \cite{Collins:2011zzd}, additional scaling tests become necessary to verify finiteness.}
The mass regulator for the UV counterterms is arbitrarily chosen as $M=5/3$.

The numerical checks explained below have been repeated with various random choices of external / internal momenta, polarization vectors, and spinors, for us to be confident of the cancellation of IR / UV singularities.

\subsection{Photonic contributions}

The ${\cal M}^{(0)}$ tree amplitude for
$e^- + e^+ \rightarrow \gamma^* + \gamma^*$ has a ``$t-$channel'' and
a ``$u-$channel'' Feynman diagram which are symmetric under 
the exchange of the two final state photons. Of course, this exchange
symmetry  is valid for the amplitude at all orders in perturbation theory.  At any given number of loops, we can
identify a diagram as belonging to the $t$ or the $u$ channel by
eliminating all virtual photons from it and mapping it correspondingly
onto one of the two tree Feynman diagrams.  

Our form factor subtractions act on the $t$  and $u$ channels
independently of each other, rendering their corresponding amplitude
contributions  finite separately.  To verify the cancellation of
singularities it is sufficient to consider only one of the channels. 
In the following, we present  results only for one of the two
orderings for the final-state photons. 

To test convergence in a singular region, we assign numerical values
for the exponents
$\omega_k^+,\omega_k^-,\omega_k^T,\omega_l^+,\omega_l^-,\omega_l^T$
which parameterize it appropriately~\cite{Sterman:1978bi,Libby:1978qf,Collins:1989gx}.
For example, in the single-soft limit where $k$ is soft and
$l$ is hard,  we set
$\omega_k^+ = \omega_k^- = \omega_k^T =1, \ \omega_l^+ = \omega_l^- =
\omega_l^T = 0$ in Eq.~\eqref{eq:loopMomentumParam}. 
Now we can perform an analytic Laurent expansion in $\delta$. 
For example, in the same region, we obtain 
\begin{equation}
 \left. \bar v(p_2) \, \widetilde M_{\rm finite}^{(2)} \, u(p_1) 
\right|_{t-channel} = \frac 1 {\delta^3} \left( -630.517 - 24.0761 i \right) + \mathcal O \left( \frac 1 {\delta^2} \right) \, .
\end{equation}
Since all computations are over rational numbers (recall that all
momenta, polarizations, and spinors are chosen to have rational
numerical components), the above numerical value is computed with no
rounding error, and  the final results (involving very large numerators and denominators)
have been converted into floating point numbers here only for
readability.  Therefore we are confident that the expansion coefficients are strictly zero at $\mathcal O(1/\delta^4)$.
Since the integration measure scales as $\int d^4 k \, \int d^4 l \sim \delta^4 \cdot \delta^0 = \delta^4$, the integral scales as $\delta$ and is convergent in the single-soft limit.

We repeat the same procedure for all IR / UV limits and we present the
scaling of the t-channel contribution as $\delta \to 0$ in Table
\ref{tab:limits}.
\begin{table}[H]
  \centering
  \begin{tabular}{|c |c|c|c|c|c|c| p{5em}| p{5em}|}
    \hline
    Limit & $\omega_k^+$ & $\omega_k^-$ & $\omega_k^T$ & $\omega_l^+$ & $\omega_l^-$ & $\omega_l^T$ & amplitude scaling \\ \hline
    $k$ soft & 1 & 1 & 1 & 0 & 0 & 0 & $\delta^{1}$ \\ \hline
    $k \parallel p_1$ & 0 & 2 & 1 & 0 & 0 & 0 & $\delta^{1}$ \\ \hline
    $k, l$ soft & 1 & 1 & 1 & 1 & 1 & 1 & $\delta^{2}$ \\ \hline
    $k$ soft, $l \parallel p_1$ & 2 & 2 & 2 & 0 & 2 & 1 & $\delta^{3}$ \\ \hline
    $k \parallel p_1$, $l \parallel p_2$ & 0 & 2 & 1 & 2 & 0 & 1 & $\delta^{2}$ \\ \hline
    $k, l \parallel p_1$ & 0 & 2 & 1 & 0 & 2 & 1 & $\delta^{2}$ \\ \hline
    $k, l \to \infty$ & -1 & -1 & -1 & -1 & -1 & -1 & $\delta$ \\ \hline
    $k \to \infty$ & -1 & -1 & -1 & 0 & 0 & 0 & $\delta$ \\ \hline
  \end{tabular}
  \caption{The scaling behavior of the photonic contributions to the finite integrand $\widetilde M^{(2)}$ in various limits. The scaling of the integration measure has been taken into account, and the amplitude is convergent if it scales as any positive power of $\delta$. The loop momenta $k$ and $l$ are parameterized as Eq.~\eqref{eq:loopMomentumParam}, with the scaling exponents given in this table. Since the integrals always scale as $\delta$ raised to a positive power, there is no divergence in any of the IR / UV limits.}
  \label{tab:limits}
\end{table}
For brevity, only a set of independent limits is presented. For example,
due to the global symmetrization $k \leftrightarrow l$, the
$k \parallel p_1$ limit is identical to the $l \parallel p_2$
limit. Also, since $p_1$ and $p_2$ are essentially treated in the same
way, the $k \parallel p_2$ limit is not shown separately.

We note that the subtracted integrand converges faster than expected in certain IR limits.
For example, in the double-soft limit, as shown in the 3rd row of Table \ref{tab:limits},
the four-dimensional finite integral is suppressed by $\delta^2$ instead of $\delta$.
This will be beneficial for high-precision numerical integration, and we leave it to
future work to fully understand the unexpected fast convergence properties.

We have also checked that when a loop momentum tends to infinity (the UV limit) while the second
loop momentum becomes soft or collinear, the integrand also scales in a way that guarantees
finiteness of the integral. This is simply a consequence of the Ward-identity preserving nature
of our UV subtraction terms, which guarantees the factorization of infrared divergences with respect
to one of the loop momenta point by point in the space of the value of the other loop momentum.

\subsection{Fermion loop contributions}
Table\ \ref{tab:limitsQbub} shows the relevant IR / UV scaling
behavior of ${\cal M}^{(2)}_{2}$ which receives contributions 
 from Feynman diagrams with a vacuum polarization correction to an  
internal photon propagator. Again, only the t-channel ordering of the two
final-state photons is presented.
\begin{table}[H]
  \centering
  \begin{tabular}{|c |c|c|c|c|c|c| p{5em}| p{5em}|}
    \hline
    Limit & $\omega_k^+$ & $\omega_k^-$ & $\omega_k^T$ & $\omega_l^+$ & $\omega_l^-$ & $\omega_l^T$ & amplitude scaling\\ \hline
    $l$ soft & 0 & 0 & 0 & 1 & 1 & 1 & $\delta$ \\ \hline
    $l \parallel p_1$ & 0 & 0 & 0 & 0 & 2 & 1 & $\delta$ \\ \hline
    $k, l$ soft & 1 & 1 & 1 & 1 & 1 & 1 & $\delta$ \\ \hline
    $k, l \parallel p_1$ & 0 & 2 & 1 & 0 & 2 & 1 & $\delta$ \\ \hline
    $k, l \to \infty$ & -1 & -1 & -1 & -1 & -1 & -1 & $\delta$ \\ \hline
    $k \to \infty$ & -1 & -1 & -1 & 0 & 0 & 0 & $\delta$ \\ \hline
  \end{tabular}
  \caption{The scaling behavior of the vacuum polarization contributions to the finite integrand $\widetilde M^{(2)}_{2}$ in various limits. The loop momenta $k$ and $l$ are parameterized as Eq.~\eqref{eq:loopMomentumParam}, with the scaling exponents given in this table. The scaling of the integration measure has been taken into account. Since the integrals always scale as $\delta$ raised to a positive power, there is no divergence in any of the IR / UV limits.}
  \label{tab:limitsQbub}
\end{table}

Table\ \ref{tab:limitsQbox} shows the relevant IR / UV scaling
behavior of ${\cal M}^{(2)}_{4}$ which receives contributions from six
Feynman diagrams with  fermion box subdiagrams. 
We consider the sum  of all these diagrams in our numerical check 
without deleting those that can be obtained from others by crossing.
\begin{table}[H]
  \centering
  \begin{tabular}{|c |c|c|c|c|c|c| p{5em}| p{5em}|}
    \hline
    Limit & $\omega_k^+$ & $\omega_k^-$ & $\omega_k^T$ & $\omega_l^+$ & $\omega_l^-$ & $\omega_l^T$ & amplitude scaling\\ \hline
    %$l$ soft & 0 & 0 & 0 & 1 & 1 & 1 & $\delta$ \\ \hline
    $l \parallel p_1$ & 0 & 0 & 0 & 0 & 2 & 1 & $\delta$ \\ \hline
    $k, l$ soft & 1 & 1 & 1 & 1 & 1 & 1 & $\delta^3$ \\ \hline
    $k \to \infty$ & -1 & -1 & -1 & 0 & 0 & 0 & $\delta$ \\ \hline
    $k, l \to \infty$ & -1 & -1 & -1 & -1 & -1 & -1 & $\delta$ \\ \hline
    \end{tabular}
  \caption{The scaling behavior of the fermion box contributions to
    the finite integrand $\widetilde M^{(2)}_{4}$ in various
    limits. The loop momenta $k$ and $l$ are parameterized as
    Eq.~\eqref{eq:loopMomentumParam}, with the scaling exponents given
    in this table. The scaling of the integration measure has been
    taken into account. Since the integrals always scale as $\delta$
    raised to a positive power, there is no divergence in any of the
    IR / UV limits. 
}
  \label{tab:limitsQbox}
\end{table}
 
\section{Conclusions}
A striking property of gauge theory scattering amplitudes is  the 
factorization  of their infrared  singularities.    
 In this article, we have taken a first step towards understanding this
 structure better at a practical level, and developing a subtraction algorithm for the
 evaluation of multi-loop amplitudes numerically in exactly four dimensions. We have
worked in an Abelian gauge theory, which we may think of as the perturbative
expansion of massless quantum electrodynamics, and in Feynman gauge.

In loop amplitudes, infrared factorization is generally not manifest point by point in momentum space, or on a diagram by diagram basis.   
Factorization becomes manifest for the amplitude when gauge symmetry 
is exploited, as Ward identities eliminate 
nonfactorizable collinear singularities in combinations of Feynman
diagrams. At the integral level, one can make simplifications that 
permit these group cancellations. First, ``hard''  contributions can be 
explicitly integrated out and UV renormalized. This operation renders
them finite and projects them to point-like tensor structures, which in
turn generate only scalar-polarizations for virtual
photons in collinear regions.
Second, approximations of Feynman integrals  in  the relevant divergent
limits can be manipulated independently of all other limits.  It is therefore
permitted and helpful to shift loop momenta differently for each one
of these limits, which is seemingly essential  in order  to demonstrate the 
cancellations of non-factorizing terms.   These manipulations of the 
integrand are helpful in demonstrating general properties of amplitudes,
such as the factorization of universal infrared functions from process-dependent
hard scattering functions.   They do not, however, provide an algorithmic scheme for 
the numerical calculation of those functions.  

An aim of this work has been to investigate  
whether infrared factorization 
can be implemented  in an alternative way, by exclusively making use of procedures that 
are strictly local, i.e. valid not only for the integral but also for
the integrand.  By committing to implementing factorization
at the integrand level, the two tools in the previous paragraph cannot be employed.
That is, we cannot rely on the results of sub-integrals, or independent shifts of
loop momenta. 
Indeed, integrating out the hard regions leads to a loss of locality,
which prevents the construction of local infrared
counterterms. Locality is also incompatible with 
performing independent shifts of the momenta in diverse singular
regions  of the amplitude.  Rather, a global choice of loop momentum routings should achieve 
the cancellation of non-factorizing divergences in all possible
singular limits simultaneously when diagrams are combined without further shifts in
momentum.

In this article,  we have demonstrated that
this goal  can be achieved through two loops for a large class of
processes in a realistic gauge theory.  In our case study, we
considered a class of QED amplitudes in Feynman gauge, for
the production of multiple off-shell photons in electron-positron annihilation. 

For these amplitudes, we generated a  two-loop integrand in a
form in which factorization is manifest for all infrared singular limits. 
Our integrand is an average over four copies of a usual  Feynman
diagrammatic integrand with symmetric, judiciously chosen,  
loop momenta routings. 

Terms in the integrand originating  from 
one-loop vertex and self energy subgraphs require renormalization and contribute to the
amplitude at ``hard" scales, comparable to the hard momentum transfer
and beyond. Using the symmetries of these integrations, however, we know
that once they are fully carried out in dimensional regularization
they produce only unphysical scalar polarizations on collinear photons that
connect the vertex correction to the hard scattering.  For such photons, QED
Ward identities ensure that non-factorizing contributions cancel at the point-like level
in a sum over diagrams.  This reasoning, however, does not yet tell us how to actually
carry out the relevant integrals numerically.
Indeed, for generic values of the vertex or self energy loop momentum, these subdiagrams 
generate  ``loop'' polarizations on their external photons, in arbitrary directions, as opposed to scalar polarizations in the collinear direction.

In this article, we developed alternative local
representations of these subgraphs,  different than, although of course derived from, those obtained by a direct application of Feynman rules, with properties required for factorization
in all collinear and soft regions.
Our novel representation of the amplitude integrand achieves the goal
of making factorization in all singular limits manifest locally. It allows only
scalar polarizations for those collinear virtual photons that give rise to singular contributions. 
This ensures a simultaneous cancellation of non-factorizing IR divergent contributions in all possible singular limits and preserves, of course, the integrated value of the original amplitude.

With local factorization manifest, we were able 
to remove infrared singularities with local, universal counterterms
for  all $e^+ + e^- \to \gamma_1^* + \ldots
+ \gamma_n^*$ amplitudes, irrespective of the photon multiplicity, $n$, in the final state. 
Specifically, we used a simple form factor amplitude through two loops  
as a ``generating functional" for our infrared counterterms. 
Each form factor integrand is defined by a finite local vertex, either 
 the tree amplitude or the finite remainder of the
one-loop amplitude, between Dirac projection matrices.  The expression for the infrared finite remainder
of the two loop amplitudes in Eq.~(\ref{eq:twoLoopIRFiniteRemainder}) is
an elegant outcome of factorization, and it constitutes a main result
of this article.  We anticipate that this form will generalize to higher orders. We remark that Eq.~(\ref{eq:twoLoopIRFiniteRemainder}) holds generally for final-states where photons are replaced by other $(W,Z,H)$ 
massive bosons.  It is also valid for the Abelian contributions to QCD amplitudes 
for the production of generic colorless final-states with off-shell photons or other massive bosons in quark anti-quark annihilation.

To remove ultraviolet singularities, we introduced
local counterterms for one- and two-loop Green's functions with 
two, three and,  in the case of diagrams with closed fermion loops, four legs.   A
delicate issue in constructing our ultraviolet counterterms has been
the preservation of factorization in singular regions where one of the loop momenta
lies in the ultraviolet regime while the second loop momentum  is
collinear to an incoming particle.   To achieve this, we constructed 
 integrands for ultraviolet counterterms for self energy and vertex
graphs with consistent expansions around the ultraviolet limit, which
preserve  Ward identities and, consequently, local collinear
factorization.  

With the factorization-based method of this article we were able to separate explicitly,
for the first time in an amplitude with arbitrary numbers of external particles, and in a realistic gauge theory, the
finite remainder of two-loop amplitude integrands from all of their 
infrared and ultraviolet singularities. As a test, we computed 
numerically the degree of divergence for all pinch
singularities anticipated  in the two-loop $e^+ + e^- \to \gamma^* + \gamma^*$
amplitude and demonstrated that the subtracted remainder
is indeed integrable, and can in principle be evaluated with
numerical integration in exactly $d=4$ dimensions.
 
We have carried out the analytic integration of the form factor
infrared counterterms and of the ultraviolet counterterms with a 
straightforward application of tensor reduction techniques to well
known master integrals. Our analytic results for the divergent part of
the amplitude 
%\draftnote{Babis: pending confirmation}
%agree with expectations in the literature~\cite{Catani:1998bh,Sterman:2002qn}.
%These results  
will be presented in a forthcoming publication.    

Once local subtractions are in place, numerical integration of the finite remainders is possible, but presents additional challenges. 
An important step in this direction has already been taken in
Ref~\cite{Capatti:2019ypt} and in Ref.~\cite{Capatti:2019edf} where amplitudes, from the class of
processes with equivalent subtractions for pinch singularities as described here, 
were evaluated efficiently at one loop.  Ref.~\cite{Capatti:2019ypt,Capatti:2019edf} presents an
algorithm in which, after a first analytic integration of the energy components of
the loop momenta, a novel deformation of the integration
path for the remaining integrals away from non-pinched singularities
is constructed. The efficiency of this numerical integration method
was tested in Ref.~\cite{Capatti:2019ypt,Capatti:2019edf} with examples of finite
integrals with many external legs and beyond 
two loops.  The application of the same technique for the integration
of the finite remainders of the two-loop amplitudes presented in this
article is an exciting possibility.  

In this article, we formulated 
a novel subtraction method based on local infrared factorization at 
two loops for a class of QED amplitudes for the production of off shell photons.  We are looking forward to
extending our method to generic QCD amplitudes for multi-particle
scattering processes at two loops and higher orders in future work.

\section*{Acknowledgments}

We are grateful to Alexander Ochirov, Alexander Penin, Andrea Pelloni, Armin Schweitzer, Ben Ruijl,
Bernhard Mistlberger, Claude Duhr, Dario Kermanschah, Valentin Hirschi, Vittorio del Duca,
Yao Ma, Zeno Capatti and Zoltan Kunszt for inspiring discussions.  
This research was supported in part by the Pauli Centre for
Theoretical Physics, the National Science Foundation
under Grants PHY-1620628 and PHY-1915093,  by the Swiss National Science Foundation
under contract SNF200021\_179016 and by the European Commission through the ERC grant “pertQCD”.
%\newpage
%\appendix

\newpage
\bibliographystyle{JHEP}
\bibliography{biblio}

\end{document}